\documentclass[fgeqn,usenatbib]{mnras}

\usepackage{newtxtext,newtxmath}
\usepackage{layouts}

\usepackage[T1]{fontenc}

\DeclareRobustCommand{\VAN}[3]{#2}
\let\VANthebibliography\thebibliography
\def\thebibliography{\DeclareRobustCommand{\VAN}[3]{##3}\VANthebibliography}

\usepackage{graphicx}	
\usepackage{amsmath}	
\usepackage{booktabs}
\usepackage[dvipsnames]{xcolor}
\usepackage{multirow}
\usepackage{lscape}
\usepackage[figuresright]{rotating}
\makeatletter
\let\@makecaption=\SFB@maketablecaption
\makeatother
\usepackage{adjustbox}

\newcommand{\Msun}{\mathrm{M_{\odot}}}
\newcommand{\fable}{\textsc{fable}}

\newcommand{\NoAGNHighSN}{\textit{NoAGNHighSN}}
\newcommand{\NoAGN}{\textit{NoAGN}}
\newcommand{\BondiFidDutyFromZFourAccOnly}{\textit{BondiZ4AccOnly}}
\newcommand{\BondiFidDutyFromZFour}{\textit{BondiZ4}}
\newcommand{\BondiBoostFidDutyFromZFour}{\textit{BondiBoostZ4}}
\newcommand{\BondiFidDutyFromZSixAccOnly}{\textit{BondiZ6AccOnly}}
\newcommand{\BondiFidDutyFromZSix}{\textit{BondiZ6}}
\newcommand{\BondiBoostFidDutyFromZSix}{\textit{BondiBoostZ6}}
\newcommand{\BondiExtraBoostFidDutyFromZSix}{\textit{BondiExtraBoostZ6}}
\newcommand{\EddLowDutyLightFromZSix}{\textit{SupplyLimLowDutyLightZ6}}
\newcommand{\EddLowDutyLightFromZSixToZTwo}{\textit{SupplyLimLowDutyLightZ6-2}}
\newcommand{\EddNoDutyFromZFour}{\textit{SupplyLimNoDutyZ4}}
\newcommand{\EddNoDutyFromZFourToZTwo}{\textit{SupplyLimNoDutyZ4-2}}

\newcommand{\NoAGNHighSNRes}{\textit{NoAGNHighSNRes}}
\newcommand{\NoAGNRes}{\textit{NoAGNRes}}
\newcommand{\EddNoDutyFromZFourRes}{\textit{SupplyLimNoDutyZ4Res}}

\definecolor{HighSNColour}{RGB}{23.0, 100.2,171.0}
\definecolor{SNColour}{RGB}{147.8, 196.4, 222.8}
\definecolor{SNBondi1e2Colour}{RGB}{157.529411765, 153.545098039, 199.780392157}
\definecolor{SNBondi1e3Colour}{RGB}{114.454901961, 97.9098039216, 171.839215686}
\definecolor{SNLightLowDutyZ2Colour}{RGB}{243.77397945, 182.1436389, 30.5171913}
\definecolor{SNLightLowDutyZ0Colour}{RGB}{240.21459,139.26739065,3.34461315}
\definecolor{SNHeavyNoDutyZ2Colour}{RGB}{75.2, 176.0, 98.0}
\definecolor{SNHeavyNoDutyZ0Colour}{RGB}{0.0, 100.156862745, 40.3333333333}

\definecolor{HighSNResColour}{RGB}{0, 0, 255}
\definecolor{SNResColour}{RGB}{0, 255, 255}
\definecolor{SNHeavyNoDutyZ0ResColour}{RGB}{0, 255, 0}

\title[Constraining the role of AGN in dwarfs]{Two can play at that game: constraining the role of supernova and AGN feedback in dwarf galaxies with cosmological zoom-in simulations}

\author[Koudmani et al.]{
Sophie Koudmani,$^{1,2}$\thanks{E-mail: skoudmani@ast.cam.ac.uk}
Debora Sijacki$^{1}$
and Matthew C. Smith$^{3,4}$
\\
$^1$ Institute of Astronomy and Kavli Institute for Cosmology, University of Cambridge, Madingley Road, Cambridge CB3 0HA, UK \\
$^{2}$ St Catharine's College, University of Cambridge, Trumpington Street, Cambridge CB2 1RL, UK \\
$^{3}$ Universit\"at Heidelberg, Zentrum f\"ur Astronomie, Institut f\"ur theoretische Astrophysik, Albert-Ueberle-Str. 2, 69120 Heidelberg, Germany \\
$^{4}$ Max-Planck-Institut f\"ur Astronomie, K\"onigstuhl 17, 69117 Heidelberg, Germany
}

\date{MNRAS, in press}

\pubyear{2022}

\begin{document}
\label{firstpage}
\pagerange{\pageref{firstpage}--\pageref{lastpage}}
\maketitle

\begin{abstract}
There is growing observational evidence for dwarf galaxies hosting active galactic nuclei (AGN), including hints of AGN-driven outflows in dwarfs. However, in the common theoretical model of galaxy formation, efficient supernova (SN) feedback is the tool of choice for regulating star formation in the low-mass regime. In this paper, we present a suite of high-resolution cosmological dwarf zoom-in simulations relaxing the assumption of strong SN feedback, with the goal to determine whether more moderate SN feedback in combination with an efficient AGN could be a suitable alternative. Importantly, we find that there are sufficient amounts of gas to power brief Eddington-limited accretion episodes in dwarfs. This leads to a variety of outcomes depending on the AGN accretion model: from no additional suppression to moderate regulation of star formation to catastrophic quenching. Efficient AGN can drive powerful outflows, depleting the gas reservoir of their hosts via ejective feedback and then maintaining a quiescent state through heating the circumgalactic medium. Moderate AGN outflows can be as efficient as the strong SN feedback commonly employed, leading to star formation regulation and HI gas masses in agreement with observations of field dwarfs. All efficient AGN set-ups are associated with overmassive black holes (BHs) compared to the (heavily extrapolated) observed BH mass – stellar mass scaling relations, with future direct observational constraints in this mass regime being crucially needed. Efficient AGN activity is mostly restricted to high redshifts, with hot, accelerated outflows and high X-ray luminosities being the clearest tell-tale signs for future observational campaigns.
\end{abstract}

\begin{keywords}
methods: numerical -- galaxies: active -- galaxies: star formation -- galaxies: dwarf -- galaxies: evolution -- galaxies: formation.
\end{keywords}



\section{Introduction}

Dwarf galaxies are the smallest building blocks of structure formation in the $\Lambda$-Cold Dark Matter ($\Lambda$CDM) Universe with stellar masses of less than $3 \times 10^{9} \ \Msun$ (or approximately the mass of the Large Magellanic Cloud). As such they play an important role within several areas of astronomy.

One of these is near-field cosmology with \textit{Gaia}, DES, VISTA and, in the future, the Vera Rubin Observatory all working towards a complete and unbiased census of the Local Group. Low-mass galaxies also play a crucial role in reionization, with \textit{JWST} set to constrain the faint end of the high-redshift luminosity function to be combined with observational probes of reionization and its evolving patchiness from LOFAR, HERA and SKA.

Moreover, dwarf galaxies offer an ideal testbed to study galactic feedback mechanisms as they are highly susceptible to these processes due to their shallow potential wells, in particular galactic winds play a crucial role in regulating the baryon cycle of dwarfs \citep[see e.g.][for a recent perspective]{collins_observational_2022}. Future X-ray missions such as \textit{Athena}, \textit{AXIS} or \textit{Lynx} will provide much improved constraints on energetic feedback processes and hot outflows as well as the presence of active black holes (BHs) in these low mass systems. Furthermore, radio facilities such as SKA or ngVLA will be able to constrain the radio power evolution of dwarf galaxies through cosmic time as well as map the properties of cool outflows \citep[see e.g.][for reviews on galactic winds and the role of upcoming observational missions in constraining outflow properties]{veilleux_galactic_2005,veilleux_cool_2020}. Also note that the future gravitational observatory LISA will provide complementary constraints on BH merger rates which will also give insights into the non-active BH population at the low-mass end \citep[e.g.][]{amaro-seoane_astrophysics_2022}.

Overall, there are multi-faceted, multi-messenger efforts to study dwarfs observationally throughout cosmic time. Dwarfs have also drawn significant interest from the theoretical community due to a number of apparent discrepancies when comparing dwarf galaxies to predictions from dark matter (DM) only simulations. These include a mismatch between the number of observed Milky Way satellites and the predicted number of DM haloes that may be hosting these systems, the so-called missing satellites problem \citep[e.g.][]{kauffmann_formation_1993,klypin_where_1999,moore_dark_1999}. Furthermore, there is an apparent lack of observed massive satellites, the too-big-to-fail problem \citep[e.g.][]{boylan-kolchin_too_2011}, as the inferred central masses of the most massive observed satellites do not match the central masses of the most massive simulated subhaloes. Another apparent discrepancy concerns the inferred DM halo profiles of observed
dwarfs, which do not seem to agree with the cuspy DM profile predicted by the
$\Lambda$CDM model -- the cusp versus core problem \citep[e.g.][]{flores_observational_1994,moore_evidence_1994}.

This has prompted some to question the validity of CDM and investigate alternatives such as warm DM \citep[e.g.][]{lovell_haloes_2012}, self-interacting DM \citep[e.g.][]{vogelsberger_dwarf_2014} or fuzzy DM \citep[e.g.][]{marsh_model_2014}. On the other hand, there is also a large body of theoretical work that has identified a number of important baryonic processes that could largely alleviate the aforementioned discrepancies \citep[see][for a recent review]{sales_baryonic_2022}. In particular, reionization suppression \citep[e.g.][]{efstathiou_suppressing_1992,okamoto_mass_2008,fitts_fire_2017,katz_how_2020} and stellar feedback \citep[e.g.][]{navarro_cores_1996,governato_bulgeless_2010,parry_baryons_2012,hopkins_galaxies_2014,hopkins_fire-2_2018,kimm_towards_2015,emerick_stellar_2018,smith_supernova_2018,smith_cosmological_2019,gutcke_lyra_2021}. 

In recent years, it has also been suggested that active galactic nuclei (AGN) could play a role in the resolution of the dwarf galaxy problems \citep[e.g.][]{silk_feedback_2017}, which represents a paradigm shift as previously stellar feedback mechanisms, and in particular supernovae (SNe), had been hypothesised to be responsible for regulating the low-mass end. AGN feedback, on the other hand, had been thought to only play an important role in high-mass galaxies.

This common theoretical model has been called into question by the mounting observational evidence that at least some dwarf galaxies do host AGN, with detections of AGN activity in dwarfs spanning the whole electromagnetic spectrum from X-ray \citep[e.g.][]{schramm_unveiling_2013,baldassare_50000_2015,baldassare_x-ray_2017,lemons_x-ray_2015,miller_x-ray_2015,mezcua_population_2016,mezcua_intermediate-mass_2018,pardo_x-ray_2016,aird_x-rays_2018,birchall_x-ray_2020,birchall_incidence_2022,latimer_agn_2021} to optical \citep[e.g.][]{greene_active_2004,greene_new_2007,reines_dwarf_2013,chilingarian_population_2018,molina_sample_2021-1,polimera_resolve_2022} to IR \citep[e.g.][]{satyapal_discovery_2014,sartori_search_2015,marleau_infrared_2017,kaviraj_agn_2019} and to radio observations \citep[e.g.][]{greene_radio_2006,wrobel_radio_2006,wrobel_steep-spectrum_2008,nyland_intermediate-mass_2012,nyland_multi-wavelength_2017,reines_parsec-scale_2012,reines_candidate_2014,mezcua_extended_2018,mezcua_radio_2019,reines_new_2020,davis_radio_2022}.

What triggers an AGN episode in the low-mass regime is still unclear, with observational studies finding no clear correlation between cosmological environment and AGN activity in dwarfs \citep[e.g.][]{kristensen_environments_2020,davis_radio_2022}. Furthermore, the impact of AGN on their dwarf hosts (if any) is still hotly debated, but there is a growing number of tentative  observational hints.

For example, \citet{penny_sdss-iv_2018} examined six quenched low-mass galaxies from the MaNGA survey with emission line ratios indicative of AGN, with five of these displaying a kinematically offset ionized gas component, which they interpret as a sign of AGN-driven outflows. \citet{manzano-king_agn-driven_2019} observe 29 dwarf galaxies with AGN signatures using Keck LRIS long-slit spectroscopy and find that a third have high-velocity ionized gas outflows. \citet{liu_integral_2020} and \citet{bohn_near-infrared_2021} obtained integral-field spectroscopy and infrared observations of a subset of these dwarfs with putative AGN-driven outflows confirming that the outflows are primarily photoionized by the AGN with similar coronal line detections and (relative) energetics to more massive galaxies. Similarly, \citet{mezcua_radio_2019} have identified radio jets in dwarf galaxies with jet efficiencies as powerful as those of massive galaxies. These observations suggest that AGN could play an important role in the evolution of their dwarf hosts. 

This is further supported by \citet{dickey_agn_2019} who present long-slit spectroscopy for 20 isolated, quiescent dwarfs and find that 16 of these have AGN-like line ratios. However, they caution that only five of these dwarfs were classified as AGN hosts in SDSS due to lower resolution and signal-to-noise. Similarly, \citet{mezcua_hidden_2020} highlight the power of resolved observations, and in particular IFU surveys, for identifying hidden AGN in dwarfs.

These tantalizing observational hints at AGN feedback in dwarfs have motivated theorists to look into this largely unexplored regime -- so far with mixed results. Whilst analytical models and general AGN energetics look promising \citep[e.g.][]{silk_feedback_2017,dashyan_agn_2018}, numerical simulations find that strong SN feedback can hinder BH growth \citep[e.g.][]{dubois_black_2015,angles-alcazar_black_2017,habouzit_blossoms_2017,trebitsch_escape_2018}. Furthermore, due to the shallow potential wells of dwarf galaxies, the BH may not necessarily be at the centre of the dwarf, exacerbating the issue of gas availability \citep[e.g.][]{bellovary_origins_2021,sharma_hidden_2022}.

Though if conditions in hydrodynamical simulations are such that BHs in dwarf may grow efficiently, then the AGN can contribute to the regulation of star formation \citep[e.g.][]{barai_intermediate-mass_2019,sharma_black_2020,koudmani_little_2021,wellons_exploring_2022} and significantly boost outflow temperatures and velocities \citep[e.g.][]{koudmani_fast_2019}. In cosmological simulations, minor mergers can trigger this AGN activity in dwarfs whilst high-density environments are detrimental \citep[e.g.][]{kristensen_merger_2021}. 

Theoretical studies have also shown that multimessenger signatures will be a crucial tool for constraining the massive BH population in dwarfs \citep[e.g.][]{tamfal_formation_2018,bellovary_multimessenger_2019,pacucci_separating_2020,volonteri_black_2020}. Similarly tidal disruption event (TDE) studies will deliver valuable additional constraints on this population \citep[e.g.][]{zubovas_tidal_2019,pfister_enhancement_2020}. 

The active BH occupation fraction in dwarfs is already relatively well constrained for X-ray bright AGN \citep[e.g.][]{aird_x-rays_2018,mezcua_extended_2018,birchall_x-ray_2020}. However, simulations struggle to match these constraints with some simulations (primarily with heavy seeds) such as Romulus or IllustrisTNG overproducing bright AGN \citep{haidar_black_2022,sharma_hidden_2022} whilst other simulations (primarily with lighter seeds) such as EAGLE, Illustris or \fable \ significantly underproduce bright AGN \citep{koudmani_little_2021,haidar_black_2022}. This is at least partly connected to the widely used Bondi prescription \citep{hoyle_effect_1939,bondi_mechanism_1944,bondi_spherically_1952} for BH accretion which due to its square dependency on the BH mass suppresses the growth (and thereby the luminosities) of low-mass BHs. Moreover, the aforementioned strong SN feedback, commonly employed by cosmological simulations to regulate the low-mass end of the galaxy stellar mass function, further acts to suppress the BH growth in dwarf galaxies.

In this paper, our aim is to investigate the impact of AGN feedback in a cosmological environment, relaxing the assumption of strong SN feedback. For this investigation we use the \fable \ galaxy formation model \citep{henden_fable_2018} but vary the strength of the SN feedback and the BH accretion prescription. We want to assess whether a more moderate SN feedback parametrization in combination with AGN feedback at the low-mass end can have the same success in regulating star formation as the fiducial strong SN feedback in \fable. To this end, we run high-resolution cosmological zoom-in simulations of one low-mass dwarf galaxy reducing the strength of the SN feedback compared to \fable \ and we carry out several numerical experiments including varying the boost factor of the Bondi prescription and supply-limited accretion based on local gas availability.

In Section~\ref{sec:methods}, we outline the simulation set-up and feedback models employed. We then introduce our simulation suite with visualisations of the large-scale inflow -- outflow structures in Section~\ref{subsec:overview_vis} before analysing a set of representative zoom-in simulation runs with a focus on the cosmic evolution of star formation and BH accretion (see Section~\ref{ResultsSubSec:CosmicEvMainRuns}), outflow properties (see Section~\ref{subsec:outflow_props}), stellar and BH assembly compared to observational constraints (see Sections~\ref{subsec:stellar_assembly}~and~\ref{subsec:bh_assembly}, respectively), central DM distributions (see Section~\ref{subsec:dm_props}) and a comparison with observed properties of dwarf AGN and their hosts (see Section~\ref{subsec:comp_obs}). Finally, we discuss our results in Section~\ref{sec:discussion} and present our conclusions in Section~\ref{sec:conclusion}.

\section{Methodology} \label{sec:methods}

\subsection{General set-up}

We perform zoom-in simulations of a dwarf galaxy in a low-density environment using the \textsc{arepo} code \citep{springel_e_2010}, where the equations of hydrodynamics are solved on a moving unstructured mesh defined by the Voronoi tessellation of a set of discrete points which (approximately) move with the velocity of the local flow.

The initial conditions are generated with the \textsc{music} code \citep{hahn_multi-scale_2011} at $z=127$ and, as in \fable, we use the cosmological parameters from \citet{planck_collaboration_planck_2016}.

The target dwarf halo is selected at $z=0$ from a $10 \ \mathrm{cMpc} \, h^{-1}$ box with $256^{3}$ particles, i.e. the coarse resolution is $7.47 \times 10^{6} \ \Msun$. In the coarse simulation, the virial mass is $1.04 \times 10^{10} \ \Msun$ and the virial radius is 62.0~kpc at $z=0$. This halo is then re-simulated at higher resolution with the selection region at $z=0$ set to a sphere of radius 736~kpc. 

The resolution within this zoom-in region is increased by a factor of $16^{3}$ so that the high-resolution DM mass is $m_\mathrm{dm}= 1536 \ \Msun$ and the high-resolution target gas mass is $\bar{m}_\mathrm{gas} = 287 \ \Msun$. For the high-resolution DM particles, gas cells, and star particles, we employ comoving softening lengths of 0.129 ckpc which are held constant from $z=2$. We also perform even higher resolution runs for three selected set-ups with target gas mass resolution $\bar{m}_\mathrm{gas} = 35.9 \ \Msun$ and DM particle mass $m_\mathrm{dm}= 192 \ \Msun$.

Note that this is the same system as `Dwarf 1' in \citet{smith_cosmological_2019} which focused on investigating the multi-phase interstellar medium (ISM) model from \citet{smith_supernova_2018} in a cosmological context. Here we use the same initial conditions but evolve the system based on the \fable \ physics, varying the SN energetics and AGN feedback prescriptions.

\subsection{Galaxy formation model and parameters}

Our aim is to determine whether AGN feedback could be a viable alternative to the strong SN feedback that is being employed by the majority of large-scale cosmological simulations to regulate the low-mass end of the galaxy population \citep[e.g.][]{vogelsberger_properties_2014,schaye_eagle_2015,henden_fable_2018,pillepich_simulating_2018}. The \fable \ galaxy formation model, which forms the basis for our investigations, has been described in detail in \citet{henden_fable_2018} and is based on the Illustris galaxy formation model \citep{vogelsberger_model_2013}, but with updated SN and AGN feedback prescriptions. In the following, we only recap the most important details and outline the modifications made to the \fable \ set-up.

\subsubsection{Star formation and ISM model}
The star formation and ISM model in \fable \ is unmodified from Illustris and we just briefly summarise the most important aspects, see \citet{vogelsberger_model_2013} for details.

The ISM is modelled using an effective equation of state (eEOS) which represents the global dynamical behaviour of condensed clouds and ambient hot gas \citep[][]{springel_cosmological_2003}. These two phases are governed by star formation, cloud evaporation arising from SNe and cloud growth caused by cooling. Stars are formed stochastically above a density threshold of $n_\mathrm{H} \sim 0.1 \ \mathrm{cm}^{-3}$ following the empirically defined Kennicutt–Schmidt relation and assuming a Chabrier \citep{chabrier_galactic_2003} initial mass function.

Within this subgrid model the ejecta from SNe are returned as hot gas, directly heating the ambient hot phase, so that the eEOS model implicitly contains thermal feedback from SNe. The ambient hot gas can cool radiatively down to temperatures of approximately $10^4$~K, though adiabatic cooling may allow the gas to reach even lower temperatures (note that the temperature assumed for the clouds is $10^3$~K, however, the model is not too sensitive to this value as long as this temperature is chosen $\ll 10^4$~K). The absence of metal-line cooling below $10^{4}$~K or molecular cooling means that the gas will be much ``puffier'' than in simulations that include cooling channels down to low temperatures. Therefore the star formation and stellar feedback will likely be considerably smoother in time and less spatially concentrated than for a resolved multi-phase ISM model.

Finally, for the eEOS model of star formation, it is explicitly assumed that the cold clouds and the hot surrounding medium remain coupled at all times. Consequently, the high entropy of the supernova remnants is trapped in the potential wells and it is very difficult to drive galactic winds. To circumvent this issue, the eEOS is coupled with a galactic wind model for stellar feedback as described in the following Section~\ref{subsubsec:stellar_feedback_zooms}.

\subsubsection{Stellar-feedback-driven winds}\label{subsubsec:stellar_feedback_zooms}
The galactic wind model for stellar feedback in \fable \ is based on the Illustris model \citep{vogelsberger_model_2013} with a few modifications to the parameters that govern the wind energetics.

Wind particles are stochastically launched from star-forming regions driven by the available energy from core collapse SNe with the SNe rates provided by the eEOS model. The wind energy factor $\epsilon_\mathrm{W,SN}$ (which gives the fraction of energy available from each core collapse SN) is set to $\epsilon_\mathrm{W,SN}=1.09$ in Illustris and $\epsilon_\mathrm{W,SN}=1.5$ in \fable, i.e. a very strong energy normalisation ($> 100$ per cent).

This strong coupling may be assumed due to the presence of additional stellar feedback processes other than SNe (such as photoionization or stellar winds) which would contribute to the driving of galactic outflows but are not explicitly included in the modelling. However, it could also be reasoned that the value of this coupling should be chosen smaller than 100 per cent given that there are further uncertainties as to when and how the feedback energy should be coupled to the ISM (and the radiative losses associated with this).

In fact, as an extreme choice, the wind particles in Illustris and \fable \ are temporarily decoupled from hydrodynamic interactions until they have left the ISM (based on either a density threshold or a limit on the elapsed travel time), so that cooling losses are minimised. Hence the $\epsilon_\mathrm{W,SN}=1.5$ parametrization in \fable \ likely represents an upper limit. Also note that whilst the Illustris winds are purely kinetic, one third of the wind energy in \fable \ is injected thermally which reduces overcooling and thereby further strengthens the SN feedback.

With this parametrization the SN feedback is the dominant process regulating star formation at the low-mass end \citep[see][]{henden_fable_2018}. Furthermore, we found that with the fiducial \fable \ set-up, AGN activity in low-mass galaxies is potentially artificially suppressed: the bright X-ray AGN observed in local dwarfs are not reproduced by the \fable \ model and the BHs in the \fable \ low-mass galaxies are potentially undermassive compared to the observed relations \citep{koudmani_little_2021}.

Here we would like to assess whether AGN feedback in tandem with more moderate SN feedback parameters could be a plausible alternative to the strong SN feedback set-up. To this end, we significantly reduce the energy imparted to the SN-driven winds by a factor of three for our dwarf AGN simulations\footnote{Note that all other parameters pertaining to star formation or stellar feedback are the same as in \fable.}: in addition to the fiducial \fable \ set-up with $\epsilon_\mathrm{W,SN}=1.5$, the \NoAGNHighSN \ run, we also run a simulation with $\epsilon_\mathrm{W,SN}=0.5$, the \NoAGN \ run, and then add various AGN feedback prescriptions to this set-up (see Table~\ref{tab:zoomruns}). All main simulation runs, which we focus on in this paper, are highlighted in bold in this summary table .

\begin{table*}
\caption{Overview of the cosmological zoom-in simulation runs, where the main simulation runs are highlighted in bold and in colour. We list the wind energy factor, $\epsilon_\mathrm{W, SN}$, which gives the fraction of SNII energy available to drive galactic winds (see Section~\ref{subsubsec:stellar_feedback_zooms} for details on this parameter), the BH accretion prescription which is either based on the Bondi prescription (see Section~\ref{subsubsec:BH_acc_bondi}) or limited by the gas supply (see Section~\ref{subsubsec:BH_acc_edd}) in the central resolved region of the galaxy defined by the spatial resolution of the simulation ($3 \times$ gravitational softening length which equals 0.387 and 0.194 ckpc for the two resolutions employed, respectively) -- note that BH accretion is Eddington-limited at all times for all set-ups and accretion-only runs without AGN energy injection are labelled as `AccOnly', the BH seed mass $M_\mathrm{seed}$, the seeding redshift $z_\mathrm{seed}$, the time interval between AGN bursts here referred to as `duty cycle' $\Delta t_\mathrm{AGN}$ (no duty cycle implies continuous injection of AGN energy proportional to accretion rate, whilst with a duty cycle the AGN energy is accumulated over $\Delta t_\mathrm{AGN}$ and then released in one AGN burst leading to more intermittent and energetic AGN feedback, see Section~\ref{MethodsSubsubsec:BH_feedback_zooms} for details on the AGN feedback injection), the impact of the BH on star formation (SF) at $z=1$ compared to the \NoAGN \ simulation set-up (see Fig.~\ref{fig:SfProperties} and Fig.~\ref{fig:SFHiResRuns}), the agreement with the stellar mass -- halo mass (SMHM) relation (see Fig.~\ref{fig:StellarAssembly}), and resolution of the simulation as indicated by the target gas cell mass $\bar{m}_\mathrm{gas}$.}
\begin{center}
\begin{tabular}{@{}lcccccccc@{}}
\toprule
\textbf{Simulation name} & \textbf{Wind energy} & \textbf{BH accretion} $\dot{M}_\mathrm{BH}$  &  \textbf{BH seed} & \textbf{BH} & \textbf{`Duty cycle'}  & \textbf{BH impact}  & \textbf{Agrees} & \textbf{Reso-}  \\ 
 & \textbf{factor} $\epsilon_\mathrm{W, SN}$  & \textit{Bondi (with $\alpha$ boost}   & \textbf{mass} & \textbf{seeding}   & $\Delta t_\mathrm{AGN}$ [Myr] &  \textit{SF at} $z=1$ & \textbf{with} & \textbf{lution}   \\ 
 & \textit{Fraction of SNII} & \textit{ factor) or SupplyLim}   & $M_\mathrm{seed}$& \textbf{redshift} & \textit{Time between} & \textit{compared to}  & \textbf{SMHM}  & $\bar{m}_\mathrm{gas}$ \\ 
 & \textit{energy for winds} & \textit{(may limit to high-z)}  &$[\Msun]$  & $z_\mathrm{seed}$   & \textit{AGN bursts} & \textit{NoAGN sim.} & \textbf{relation} & $[\Msun]$  \\ \toprule
\multicolumn{8}{@{} l}{\underline{\textbf{SN-ONLY SIMULATIONS}}} \vspace{0.1cm} \\
        \textcolor{HighSNColour}{\bf NoAGNHighSN} & 1.5 & - & - &- & - & - & Yes & 287  \\ 
        \textcolor{SNColour}{\bf NoAGN} & 0.5 & - & - & - & - & - & No & 287  \\ 
        \textcolor{HighSNResColour}{\bf NoAGNHighSNRes} & 1.5 & - & - & - & - & - & Yes & 35.9  \\ 
        \textcolor{SNResColour}{\bf NoAGNRes} & 0.5 & - & - & - &  - & - & No & 35.9  \\ \midrule
\multicolumn{8}{@{} l}{\underline{\textbf{INEFFICIENT AGN SIMULATIONS}}} \vspace{0.1cm} \\
        BondiZ6AccOnly & 0.5 & $\dot{M}_\mathrm{Bondi}(\alpha = 10^{2})$ & $10^{3}$ & 6 & - & No impact & No & 287  \\ 
        BondiZ6 & 0.5 & $\dot{M}_\mathrm{Bondi}(\alpha = 10^{2})$ &  $10^{3}$ & 6 & 25 & No impact & No & 287 \\
        BondiBoostZ6 & 0.5 & $\dot{M}_\mathrm{Bondi}(\alpha = 10^{3})$ & $10^{3}$ & 6 & 25 & No impact & No & 287  \\ 
        BondiZ4AccOnly & 0.5 & $\dot{M}_\mathrm{Bondi}(\alpha = 10^{2})$ & $10^{4}$ & 4 & - & No impact & No & 287  \\ 
        \textcolor{SNBondi1e2Colour}{\bf BondiZ4} & 0.5 & $\dot{M}_\mathrm{Bondi}(\alpha = 10^{2})$  & $10^{4}$ & 4 & 25 & No impact & No & 287  \\ 
        SupplyLimNoDutyLightZ6-2 & 0.5 & $\dot{M}_\mathrm{supply}(z\geq2)$ & $10^{2}$ & 6 & - & No impact & No & 287 \\
        SupplyLimNoDutyZ6-2 & 0.5 & $\dot{M}_\mathrm{supply}(z\geq2)$ & $10^{3}$ & 6 & - & No impact & No & 287\\  \midrule
 \multicolumn{8}{@{} l}{\underline{\textbf{MODERATELY-EFFICIENT AGN SIMULATIONS}}} \vspace{0.1cm}  \\
        \textcolor{SNLightLowDutyZ2Colour}{\bf SupplyLimLowDutyLightZ6-2} & 0.5 & $\dot{M}_\mathrm{supply}(z\geq2)$   & $10^{2}$ & 6 & 10 & Suppressed & Yes & 287 \\ 
        SupplyLimLightZ6-2 & 0.5 & $\dot{M}_\mathrm{supply}(z\geq2)$  & $10^{2}$ & 6 & 25 & Suppressed & Yes & 287  \\
        SupplyLimVeryLowDutyZ6-2 & 0.5 & $\dot{M}_\mathrm{supply}(z\geq2)$ & $10^{3}$ & 6 & 5 & Suppressed & Yes & 287 \\
        SupplyLimLowDutyZ6-2 & 0.5 & $\dot{M}_\mathrm{supply}(z\geq2)$ & $10^{3}$ & 6 & 10 & Suppressed & Yes & 287 \\ 
        \textcolor{SNHeavyNoDutyZ2Colour}{\bf SupplyLimNoDutyZ4-2} & 0.5 & $\dot{M}_\mathrm{supply}(z\geq2)$  & $10^{4}$ & 4 & - & Suppressed & Yes & 287  \\ 
        \textcolor{SNHeavyNoDutyZ0ResColour}{\bf SupplyLimNoDutyZ4Res} & 0.5 & $\dot{M}_\mathrm{supply}$  & $10^{4}$ & 4 & - & Suppressed & Yes & 35.9 \\  \midrule
 \multicolumn{8}{@{} l}{\underline{\textbf{EFFICIENT AGN SIMULATIONS}}} \vspace{0.1cm}  \\
        BondiExtraBoostZ6 & 0.5 & $\dot{M}_\mathrm{Bondi}(\alpha = 10^{4})$ & $10^{3}$ & 6 & 25 & Quenched & Yes & 287  \\
        \textcolor{SNBondi1e3Colour}{\bf BondiBoostZ4} & 0.5 & $\dot{M}_\mathrm{Bondi}(\alpha = 10^{3})$  & $10^{4}$ & 4 & 25 & Quenched & Yes & 287 \\ 
        \textcolor{SNLightLowDutyZ0Colour}{\bf SupplyLimLowDutyLightZ6} & 0.5 & $\dot{M}_\mathrm{supply}$  & $10^{2}$ & 6 & 10 & Quenched & Yes & 287\\ 
        SupplyLimZ6-2 & 0.5 & $\dot{M}_\mathrm{supply}(z\geq2)$ & $10^{3}$ & 6 & 25 & Quenched & Yes & 287\\ 
        \textcolor{SNHeavyNoDutyZ0Colour}{\bf SupplyLimNoDutyZ4} & 0.5 & $\dot{M}_\mathrm{supply}$  & $10^{4}$ & 4 & - & Quenched & Yes & 287 \\
        SupplyLimZ4-2 & 0.5 & $\dot{M}_\mathrm{supply}(z\geq2)$  & $10^{4}$ & 4 & 25 & Quenched & Yes & 287 \\ 
\bottomrule
\end{tabular}
\end{center}
\label{tab:zoomruns}
\end{table*}

These main simulations were performed until $z=0$ for the fiducial resolution of $\bar{m}_\mathrm{gas} = 287 \ \Msun$ and until $z=1$ for the very high resolution of $\bar{m}_\mathrm{gas} = 35.9 \ \Msun$. The additional simulation set-ups at the fiducial resolution were also performed to $z=1$.

We do not explore other values of the wind energy factor or vary alternative parameters related to the SN feedback efficiency as this would be beyond the scope of this work. Here we focus on determining whether weaker SN feedback can be compensated for by AGN feedback in terms of overall star formation suppression at the low-mass end of the galaxy population. We note that the impact of AGN feedback is going to be further degenerate with other stellar feedback processes and additional physical components such as cosmic rays or a multi-phase ISM, so our AGN results have to be interpreted in the context of this more simplistic stellar feedback model, with different physical processes potentially changing the conclusions from our work.

\subsubsection{BH seeding} \label{subsubsec:BH_seeding}
BH seeding mechanisms are still largely unconstrained \citep[see][for a recent review of BH assembly mechanisms]{inayoshi_assembly_2020}. Hence we just seed the BH particle at a target redshift $z_\mathrm{seed}$ with BH subgrid mass $M_\mathrm{seed}$ on-the-fly, by turning the gas cell with highest density\footnote{Note that this can lead to small stochastic fluctuations between runs with similar set-ups as different gas cells may have the highest density at the time of seeding so that the BHs get seeded at slightly different locations in the centre of the galaxy.} at the time of seeding into a BH sink particle. In addition to the redshift criterion, we also use a halo mass criterion, matching the mass of the main halo at $z_\mathrm{seed}$, to ensure that we only seed one BH per simulation. Hence we do not (explicitly) explore BH -- BH mergers with these set-ups.

We focus on light to intermediate-mass seeds with $M_\mathrm{seed} =10^{2}$--$10^{4} \ \Msun$, which could result from Population III star remnants or gravitational runaway in star clusters. We do not consider very massive BH seeds from direct gas collapse with $M_\mathrm{seed} \gtrsim 10^{5} \ \Msun$ as such high seed masses would be unrealistic for our low-mass systems.

We test two different seeding redshifts, $z_\mathrm{seed}=6$ and $z_\mathrm{seed}=4$. These seeding redshifts are chosen such that there has been sufficient stellar mass build-up, with $M_\mathrm{stellar} > 10^{5} \ \Msun$, for the galaxy to host a massive BH, yet the galaxy has only assembled a small fraction of its final stellar mass, with $M_\mathrm{stellar}(z=z_\mathrm{seed}) < 0.02 M_\mathrm{stellar}(z=0) \sim 10^{8} \ \Msun$. For the $z_\mathrm{seed}=6$ case, we employ light seeds with masses $M_\mathrm{seed} = 10^{2} \ \Msun$ or $M_\mathrm{seed} = 10^{3} \Msun$, whilst for seeding at $z_\mathrm{seed}=4$ we set the seed mass to $M_\mathrm{seed} = 10^{4} \ \Msun$, see Table~\ref{tab:zoomruns}. 

The set-up with $M_\mathrm{seed} = 10^{2} \ \Msun$ represents a conservative choice which allows us to explore the growth of light seeds, whilst we use the massive seed set-up with $M_\mathrm{seed} = 10^{4} \ \Msun$ to investigate the maximum AGN impact. Seeding the massive seed at an even lower redshift would yield better agreement with the $z=0$ BH -- galaxy scaling relations, however, seeding at such a late point in the galactic assembly would make it difficult for the AGN to have a significant influence on the stellar mass evolution. Given that the scaling relations are heavily extrapolated in this mass regime we therefore insert the massive seeds at $z=4$.

Note that whilst the value of the BH subgrid mass, which is used to calculate the BH accretion rates, is set to $M_\mathrm{seed}$, we set the initial dynamical mass, $M_\mathrm{dyn}$, which is used to calculate the gravitational potential of the BH, to $M_\mathrm{dyn}=10^{4} \ \Msun$ for all set-ups. We found that this was necessary to ensure that the BH dynamics are properly resolved, avoiding the BH scattering out of the central region. For the light seed runs, we then only allow the dynamical BH mass to increase once the subgrid BH mass has caught up. This also allows us to dispense with the repositioning scheme that was used in \fable \ to keep the BHs at the potential minimum of their host haloes.

\subsubsection{BH accretion: Bondi-based runs} \label{subsubsec:BH_acc_bondi}

In \fable, the BHs accrete at a Bondi-Hoyle-Lyttleton-like rate \citep{hoyle_effect_1939,bondi_mechanism_1944,bondi_spherically_1952}, $\dot{M}_\mathrm{Bondi}$, boosted by a factor of $\alpha=100$ (and limited by the Eddington rate). The BH accretion rate, $\dot{M}_\mathrm{BH}$, is then given by:

\begin{equation}
    \dot{M}_\mathrm{BH} = \alpha \dot{M}_\mathrm{Bondi} = \alpha \frac{4 \pi \mathrm{G}^{2} M_\mathrm{BH}^{2} \rho}{c_\mathrm{s}^{3}},
    \label{eq:ZoomsBondiRate}
\end{equation}
where $\mathrm{G}$ is the gravitational constant, $M_\mathrm{BH}$ is the BH mass, $\rho$ is the density and $c_\mathrm{s}$ is the sound speed in the vicinity of the BH\footnote{The traditional Bondi-Hoyle-Lyttleton model also includes a relative velocity term in the denominator, though we neglect this here for simplicity (also note that our relative velocities may not always be accurate given the increased dynamical BH mass for some of our runs).}. The boost factor $\alpha$ was introduced in the Bondi accretion prescription to account for the unresolved multi-phase nature of the ISM \citep[e.g.][]{springel_modelling_2005,booth_cosmological_2009,johansson_evolution_2009,sijacki_gravitational_2011}. We estimate the density and the sound speed in the vicinity of the BH based on the gas properties within the BH smoothing length, which is defined as the radius enclosing the 32 nearest gas cells. These neighbouring gas cells are then also used as the reservoir for the gas draining, weighted by the density kernel and the volume of the gas cells. We note that, different from \fable, we impose a maximum accretion radius which limits the BH smoothing length to the resolved gas region as defined by three times the (comoving) gravitational softening length, $r_\mathrm{max} = 3 \times \epsilon_\mathrm{soft} = 0.387$~ckpc, to avoid artificially enhancing accretion rates by continually increasing the accretion region. Due to the self-regulatory nature of the Bondi rate, which is proportional to the gas density and inversely proportional to the cube of the sound speed, BH accretion and feedback are naturally lowered as the central region is rarefied so that we find that a sufficient number of neighbours is always available within $r_\mathrm{max}$ for the Bondi-based simulation set-ups.

Firstly, we set up runs with the fiducial $\alpha=100$ set-up from \fable, labelled as `Bondi' to test whether the reduction of the SN feedback strength is sufficient to allow for high-luminosity AGN in dwarfs. We also set up simulations with AGN accretion but without AGN feedback (labelled as `AccOnly') to assess the impact of feedback self-regulation on the BH growth.

The strong dependency of the Bondi model on $M_\mathrm{BH}$ suppresses the growth of low-mass BHs so that it is difficult to explore efficient AGN accretion in dwarf galaxies within the constraints of this framework.

To circumvent these constraints, we also set up simulations where we increase the boost parameter $\alpha$ as a cheap way to explore higher accretion rates. We emphasise that these boost factors should then be thought of as numerical experiments rather than within the Bondi framework. We run simulations with $\alpha=10^{3}$, labelled as `BondiBoost', for both seed masses, see Table~\ref{tab:zoomruns}. Again due to the significant cost of the simulation, we can only explore a limited number of set-ups and hence we simply increase this parameter by an order of magnitude to explore the high-accretion regime.

For the intermediate-mass seeds ($M_\mathrm{seed} = 10^{3} \ \Msun$), we also trial an even more extreme parameter choice with $\alpha=10^{4}$, labelled as `BondiExtraBoost', as lower seed masses are even more severely affected by the mass-dependent suppression from the Bondi model. Given the difficulty of growing intermediate-mass seeds with the Bondi model, we do not trial light seeds of mass $M_\mathrm{seed} = 10^{2} \ \Msun$ with this accretion prescription.

Note that for all of these simulation runs, we impose the Eddington limit as a cap on the accretion rates.

\subsubsection{BH accretion: Supply-limited accretion runs} \label{subsubsec:BH_acc_edd}

In addition to the boost factors, we also utilise a supply-limited accretion scheme to explore the high-accretion regime (labelled as `SupplyLim'). With this set-up, if the gas availability in the central resolved region allows, the BH accretes at 100 per cent of the Eddington rate, $\dot{M}_\mathrm{Edd}$, which is given by:

\begin{equation}
    \dot{M}_\mathrm{Edd} = \frac{4 \pi \mathrm{G} M_\mathrm{BH} \mathrm{m_{p}}}{\epsilon_\mathrm{r} \mathrm{\sigma_{T} \mathrm{c}}},
\end{equation}
where $\mathrm{m_{p}}$ is the proton mass, $\epsilon_\mathrm{r}$ is the radiative efficiency, $\mathrm{\sigma_{T}}$ is the Thomson cross section and $\mathrm{c}$ is the speed of light.

Accretion is halted when the number of gas cell neighbours, $N_\mathrm{ngb}$, within the resolved region around the BH, which again we define as three times the (comoving) gravitational softening length $r_\mathrm{max} = 3 \times \epsilon_\mathrm{soft} = 0.387$~ckpc (or $r_\mathrm{max} =0.194$~ckpc for the very high resolution run \EddNoDutyFromZFourRes), falls below a critical value of 16 neighbours.

We also test several set-ups where the supply-limited accretion is further restricted to the high-redshift regime only ($z \geq z_\mathrm{end} = 2$), both to investigate the impact of continued AGN activity versus high-redshift AGN activity and to assess whether high-redshift AGN could imprint signatures on their hosts that are still observable after the AGN has shut down at lower redshifts. The supply-limited accretion rate is then given by
\begin{equation}
    \dot{M}_\mathrm{supply}=
    \begin{cases}
      \dot{M}_\mathrm{Edd}, & \text{if}\ N_\mathrm{ngb}(3 \epsilon_\mathrm{soft}) \geq 16 \ \text{and} \ z \geq z_\mathrm{end}\\
      0, & \text{otherwise.}
    \end{cases}
  \end{equation}
This leads to heavily self-regulated BH growth due to feedback decreasing the density around the BH (see Section~\ref{ResultsSubSec:CosmicEvMainRuns}). Note that this feedback-regulated growth also means that the length of the duty cycle becomes crucial, see Section~\ref{MethodsSubsubsec:BH_feedback_zooms}.

For these simulations, we explore set-ups with all of the three different seed masses, as detailed in Table \ref{tab:zoomruns}, since low-mass seeds are no longer heavily suppressed in their growth with gas availability setting the BH accretion rates instead.

\subsubsection{BH feedback}\label{MethodsSubsubsec:BH_feedback_zooms}
The BH luminosity is obtained by multiplying the BH accretion rate with the radiative efficiency, which is here assumed to have the value $\epsilon_\mathrm{r}=0.1$, as in \fable. For simplicity, we then always couple a fraction $\epsilon_\mathrm{f}=0.1$ of this luminosity as thermal feedback, rather than distinguishing between a quasar and a radio mode as in the original \fable \ implementation or in Illustris \citep{sijacki_illustris_2015}. This feedback energy is then distributed over the gas cell neighbours within the BH smoothing length (which is limited to the central resolved region given by $r_\mathrm{max}$), weighted by the gas cell masses.

For the runs based on the Bondi prescription, we inject the thermal energy following a duty cycle of $\Delta t_\mathrm{AGN}=25$~Myr, as in \fable, to avoid artificial overcooling \citep[also see][]{booth_cosmological_2009}. Here duty cycle refers to the time interval between AGN bursts; with this implementation the AGN energy is stored up over $\Delta t_\mathrm{AGN}$ and then released in one burst which leads to more effective AGN feedback as the relative energy content change of the surrounding gas cells will be larger, increasing the cooling time. Furthermore, the more intermittent nature of the feedback means that the gas reservoir may recover in-between the AGN feedback bursts and therefore also leading to more efficient AGN accretion.

This is especially true for the supply-limited accretion set-ups, where we found that the presence of a duty cycle and its duration can significantly affect the AGN accretion rates due to strong dependence on the central gas densities. Hence, we additionally test shorter duty cycle durations of $5$ and $10$~Myr as well as runs without any duty cycle for the supply-limited accretion runs.

\section{Results}

\subsection{Overview of the simulation suite} \label{subsec:overview_vis}

\begin{figure*}
    \centering
    \includegraphics[width=\textwidth]{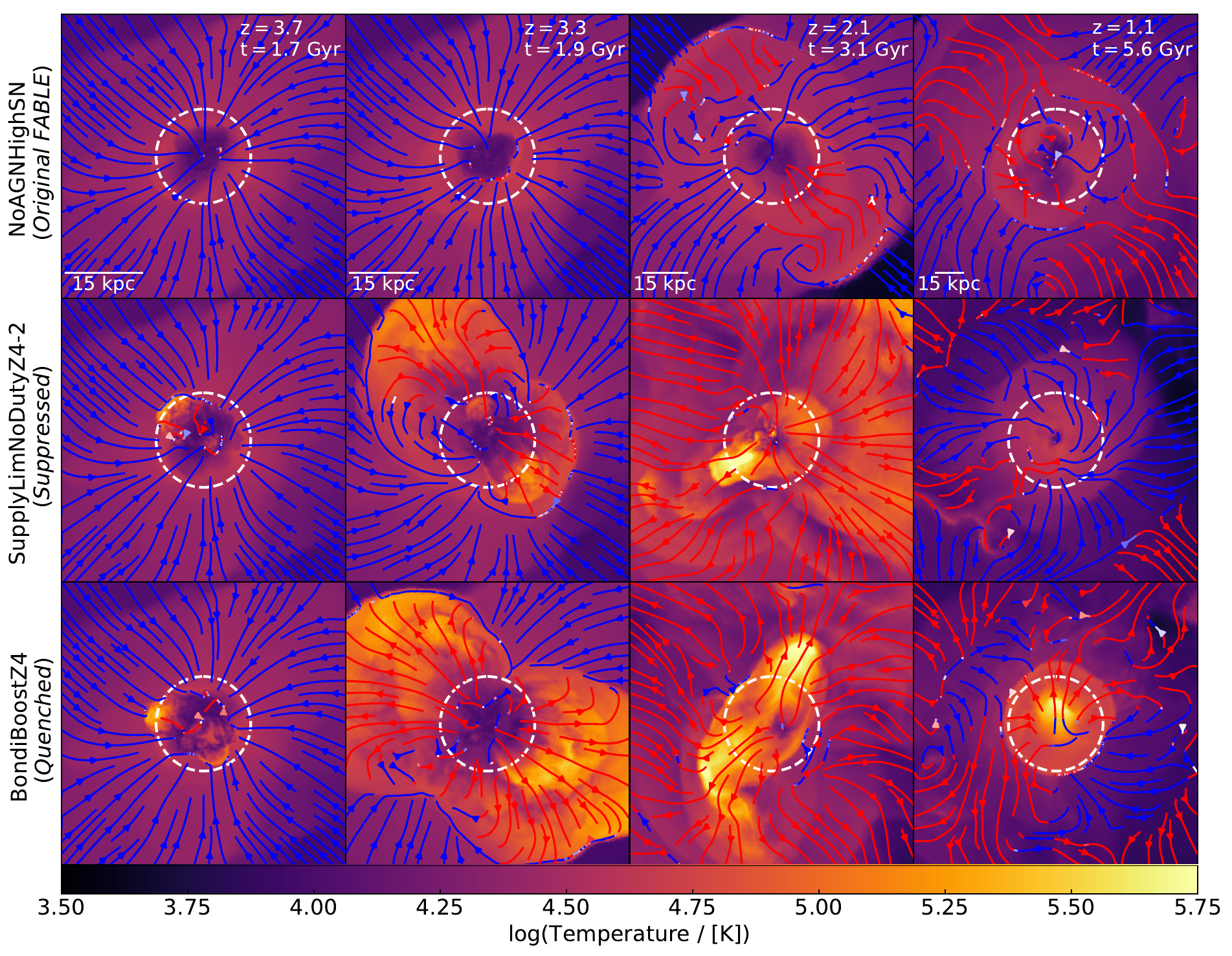}
    \caption{Large-scale gas temperature projections and streamlines of representative dwarf zoom-in simulations. The projection dimensions are $6 R_\mathrm{vir} \times 6 R_\mathrm{vir} \times 10~\mathrm{kpc}$. The colour-coding of the velocity streamlines indicates the sign of the gas radial velocity with red streamlines for outflowing gas and blue streamlines for inflowing gas. The region enclosed by the virial radius is shown as a white dashed circle (note that the virial radii at the redshifts shown here approximately correspond to $R_\mathrm{vir} \sim 9, 10, 15, 25$~kpc). The AGN feedback configurations (middle and bottom rows) shown here drive hotter and more powerful outflows than the \NoAGNHighSN \ run. This allows the AGN to efficiently regulate star formation via the suppression of cosmic inflows.}
    \label{fig:LargeScaleProj}
\end{figure*}

We begin our investigation with inspecting the large-scale temperature maps of selected zoom-in simulation runs. 

Fig.~\ref{fig:LargeScaleProj} shows $6 R_\mathrm{vir} \times 6 R_\mathrm{vir} \times 10~\mathrm{kpc}$ temperature projections for the \NoAGNHighSN, \EddNoDutyFromZFourToZTwo, and \BondiBoostFidDutyFromZFour \ runs at four different redshifts. The streamlines of the outflows are also plotted, with the colour-coding according to the sign of the radial velocity (red streamlines for outflowing and blue streamlines for inflowing gas). Furthermore, we indicate the region enclosed by the virial radius as a white dashed circle.

The \NoAGNHighSN \ run (first row) experiences significant gas inflows, in particular at early times. By $z \sim 2.0$, large-scale SN-driven outflows have developed which are able to partially suppress the cosmic inflows. The temperature maps of the \NoAGN \ run (not shown here) look very similar, however the outflows are slightly weaker due to the lower SN energetics and hence the inflow suppression at late times is also a little bit weaker. The \NoAGNHighSN \ run is equivalent to the original \fable \ set-up from \citet{henden_fable_2018}, as this low-mass system would not meet the halo mass threshold employed in \fable \ for BH seeding for the duration of the simulation and therefore only be regulated by stellar feedback.

The \EddNoDutyFromZFourToZTwo \ run (second row) drives powerful outflows which build up to higher temperatures and larger scales during the active AGN phase. At $z=2$, inflows are virtually completely suppressed by the AGN -- significantly more so than in the \NoAGNHighSN \ run. After the AGN is switched off (by construction) at redshift $z=2$, the outflows die down and by $z \sim 1$, cosmic inflows are able to reach the galaxy supplying fresh fuel and rejuvenating star formation, so that the total suppression of star formation is only short-lived. 

In the last row, we show the temperature projections of the \BondiBoostFidDutyFromZFour \ run. Note that with this run, the AGN is active from $z_\mathrm{seed}=4$ to the end of the simulation, however, due to the depletion of the gas supply, the main phase of activity is restricted to $z>2$. During this phase, the AGN drives powerful hot outflows that propagate far beyond the virial radius. For lower redshifts, the AGN activity is restricted to infrequent bursts triggered by a temporary build-up of dense gas in the centre of the dwarf. These bursts are sufficient to maintain the hot outflow at the scale of the virial radius so that with this AGN set-up the gas supply is not replenished and star formation remains suppressed until $z=0$.

In the following sections, we explore the impact of different AGN feedback configurations on star formation and outflow properties more systematically.

\subsection{Cosmic evolution of the main simulation runs} \label{ResultsSubSec:CosmicEvMainRuns}

\begin{figure*}
    \centering
    \includegraphics[width=0.498\textwidth]{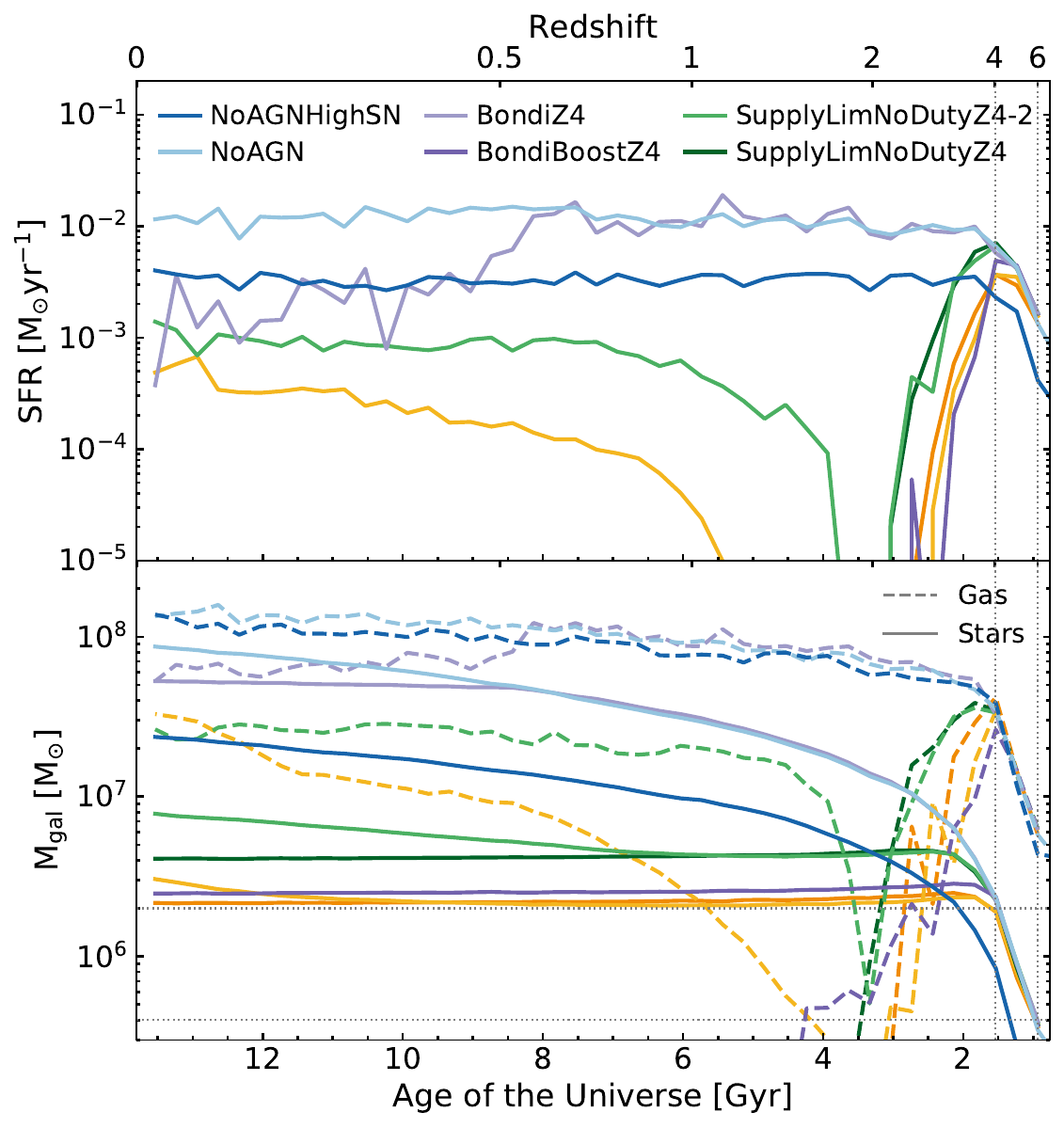} 
    \includegraphics[width=0.498\textwidth]{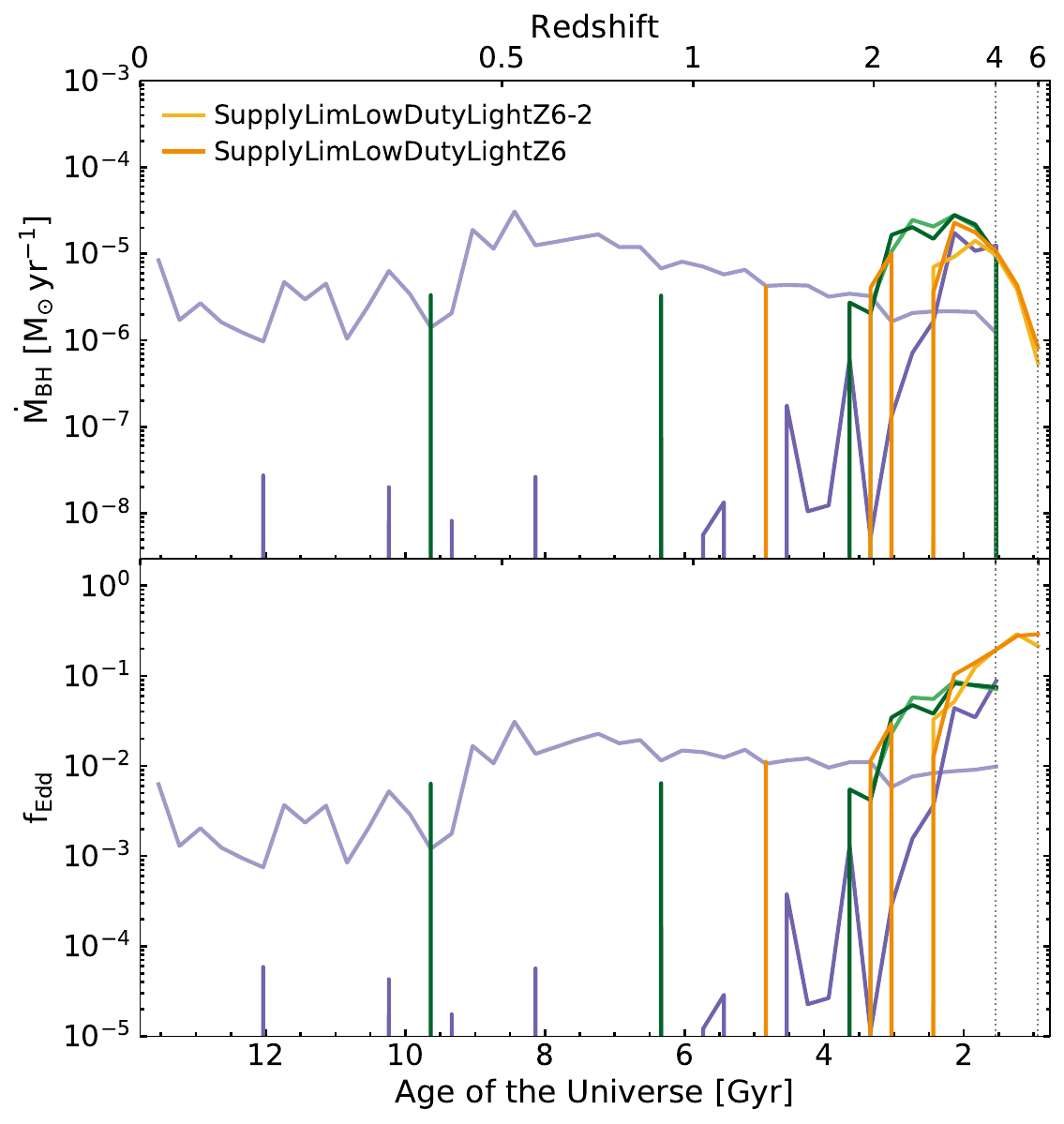}
    \caption{Star formation and accretion properties of the dwarf zoom-in simulations as a function of cosmic time. The Bondi-based runs with $M_\mathrm{seed}=10^{4} \ \Msun$ are shown in purple and the supply-limited accretion runs with $M_\mathrm{seed}=10^{4} \ \Msun$ and $M_\mathrm{seed}=10^{2} \ \Msun$ are shown in green and orange, respectively. For reference the star formation properties of the base \NoAGN \ run and the \NoAGNHighSN \ run are also plotted in light blue and dark blue, respectively (see Table~\ref{tab:zoomruns} for an overview of the zoom-in simulation set-ups). \textit{Left panels}: The upper left panel shows the SFRs within twice the stellar half-mass radius and the lower left panel shows stellar masses (solid lines) and gas masses (dashed lines) within twice the stellar half-mass radius. The seeding redshifts and stellar masses at seeding are indicated as grey dotted lines. The efficient AGN runs are able to drive the gas out of the galaxy, however, if the AGN is switched off at $z=2$, the gas reservoir and star formation may (partially) recover. \textit{Right panels}: The upper right panel shows the BH accretion rate and the lower right panel shows the Eddington fraction. The Bondi prescription needs to be boosted by $\alpha=10^{3}$ so that AGN feedback can significantly suppress star formation compared to the base \NoAGN \ run. Otherwise the growth of low-mass BHs is suppressed by the Bondi model and the AGN feedback is inefficient by construction. Supply-limited accretion schemes, on the other hand, can reach high Eddington fractions, demonstrating that there are sufficient amounts of gas available in these dwarfs for efficient AGN accretion. Note that for all of the quantities plotted here there is small-scale variability which has been smoothed out over time bins of $\Delta t = 300$~Myr to highlight trends. }
    \label{fig:SfProperties}
\end{figure*}

Firstly, we provide an overview of the redshift evolution of the main simulation runs in terms of star formation and BH accretion activity. 

Fig.~\ref{fig:SfProperties} shows the star formation rates (SFRs, upper left panel), stellar and gas masses (lower left panel) as well as BH mass accretion rate (upper right panel) and Eddington fraction (lower right panel) of the main dwarf simulation runs as a function of time. These simulation runs have been chosen as representative outcomes of the different physical set-ups explored (see Table~\ref{tab:zoomruns}). For clarity, we show the temporal evolution averaged over time bins of $\Delta t = 300$~Myr to smooth out fluctuations and highlight the general trends. Furthermore, we only show gas or stellar masses larger than $3 \times 10^{5} \ \Msun$. However, we note that the efficient AGN runs retain a small gas reservoir below this limit ($M_\mathrm{gas} \lesssim 10^{5} \ \Msun$).

Focusing on the left panel, we can see that the \NoAGNHighSN \ set-up (dark blue lines) is significantly more efficient at suppressing star formation than the \NoAGN \ run (light blue lines), with the final stellar masses at $z=0$ differing by a factor of $\sim 5$. In contrast, the difference in gas masses is negligible, suggesting that the increased SN energetics preferentially lead to additional gas heating rather than ejective feedback.  

For the set-ups with AGN activity, we find that there are three distinct branches of star formation evolution: no significant impact from the BH, brief shutdown of star formation during the peak of AGN activity followed by suppressed SFRs, and sustained star formation suppression leading to a quenched state. Here we choose six AGN runs at the fiducial resolution as representative examples (highlighted in Table~\ref{tab:zoomruns}): the two Bondi-based runs ($z_\mathrm{seed}=4$) shown in purple, and the four supply-limited-accretion runs shown in green for the late seeds ($z_\mathrm{seed}=4$) and in yellow for the early seeds ($z_\mathrm{seed}=6$), respectively.

For the Bondi-based runs, we employ the same duty cycle duration as was used for the original \fable \ simulation ($\Delta t_\mathrm{AGN}=25$~Myr, which gives the time interval in-between AGN energy injection episodes) and then explore the impact of varying the boost factor in the Bondi rate (see Section~\ref{subsubsec:BH_acc_bondi}) as well as the seeding masses and redshifts (see Section~\ref{subsubsec:BH_seeding}). 
The Bondi-based AGN runs with the fiducial \fable \ boost factor ($\alpha=100$) do not have a significantly different stellar mass evolution from the \NoAGN \ run. This even applies to the \BondiFidDutyFromZFour \ run which employs a relatively high seed mass of $M_\mathrm{seed}=10^{4} \ \Msun$, shown in light-purple. This run has only moderate accretion rates due to the efficient self-regulation by the AGN thermal feedback with $f_\mathrm{Edd} \sim 10^{-3 }$ -- $10^{-2}$ (for comparison in the equivalent run with accretion only and no feedback injection, BH growth is extremely efficient, reaching the Eddington limit at $t\sim 5$~Gyr, see Appendix~\ref{appsec:BondiAccretionRates}). These moderate accretion rates persist until $z=0$, with AGN feedback leading to a mild suppression in gas masses and SFRs at $z < 1$.

However, if the boost parameter $\alpha$ in the Bondi rate is enhanced by a factor of ten (so that $\alpha=10^{3}$), the AGN has a crucial effect on the evolution of the dwarf, as can be seen for the \BondiBoostFidDutyFromZFour \ run shown in dark purple. From $z_\mathrm{seed}=4$ to $z\sim 1$, the gas mass decreases by over two orders of magnitude from $M_\mathrm{gas} \sim 3 \times 10^{7} \ \Msun$ to $M_\mathrm{gas} \sim 10^{5} \ \Msun$. This simulation results in a significantly lower stellar mass than both of the no-AGN runs, with the final stellar mass about one order of magnitude lower than the \NoAGNHighSN \ set-up\footnote{Also note that the stellar mass of this run slightly decreases once the galaxy has been quenched. This is due to our definition of the stellar mass which we measure within twice the stellar half-mass radius; as the stellar distribution slightly contracts the stellar mass is then nominally reduced. Furthermore, the mass loss due to stellar evolution contributes as well.}.

With the increased boost factor the \BondiBoostFidDutyFromZFour \ simulation has a sufficiently large accretion rate with $f_{\rm Edd} > 0.01$ at $z > 3$, to drive a hot outflow, expelling the majority of the gas reservoir as well as preventing further accretion from the cosmic web. The long-term suppression of SFRs and the gas reservoir is particularly significant given the minimal AGN activity at low redshifts. The BH accretion rates decline dramatically after the initial peak at $z \sim 3$, with only occasional low-energy bursts for $z<2$. However, these outbursts are sufficient to maintain a hot bubble at the scale of the virial radius, as can be seen in Fig.~\ref{fig:LargeScaleProj}.

We note that for Bondi-based runs with lighter seeds ($M_\mathrm{seed}=10^{3} \ \Msun$ and $z_\mathrm{seed}=6$, not shown here), the BH does not affect star formation, even with the $\alpha=10^{3}$ boost parameter. The growth of the light seeds is suppressed by the quadratic dependency on $M_\mathrm{BH}$ in the Bondi prescription so that the impact of the AGN feedback is only minimal, i.e. the limitations of the Bondi model are so severe that one cannot explore high accretion onto low-mass seeds using this model. To test this explicitly, we set up an additional run as an experiment where we boosted the Bondi rate by a further factor of a 100, i.e. $\alpha=10^{4}$. This run (\BondiExtraBoostFidDutyFromZSix, see Table~\ref{tab:zoomruns} and Appendix~\ref{appsec:BondiAccretionRates}) produces a notable suppression in the SFR and gas mass, with the stellar mass suppression compared to the no-AGN runs even more severe since the galaxy gets quenched earlier (not shown here).

With BH accretion limited to the central region in our simulated dwarfs (see Section~\ref{subsubsec:BH_acc_bondi} for details), this explicitly demonstrates that there is in principle enough gas available in these dwarf galaxies for efficient AGN accretion and feedback (modulo uncertainties in the stellar feedback regulation of the gas reservoir), though this efficient AGN feedback can only be obtained when departing from the Bondi model by modifying the boost parameter or basing AGN accretion on an alternative model altogether. 

Motivated by these findings, we also performed several simulations where the target Eddington fraction of the accreting BHs is set to 100 per cent -- provided that sufficient amounts of gas are available in the central region to accrete at the Eddington limit (see Section~\ref{subsubsec:BH_acc_edd} for details). For these supply-limited accretion runs, the effectiveness of the AGN feedback depends on the BH mass but also, crucially, on the time interval between AGN energy injection episodes, the AGN `duty cycle'. With the supply-limited accretion set-ups, the BH accretes whenever the central resolved region (defined by three times the comoving gravitational softening length, $\epsilon_\mathrm{soft}=0.129$~ckpc for the fiducial resolution) contains at least 16 gas cells (which translates to a minimum gas mass reservoir of approximately $4600 \ \Msun$ for the fiducial resolution). A longer duty cycle allows for the gas reservoir in the central region of the galaxy to recover in between feedback injections, fuelling further accretion activity. Furthermore, the injection of the feedback energy stored up over the length of the duty cycle is more impactful and reduces overcooling effects.

\begin{figure*}
    \centering
    \includegraphics[width=\textwidth]{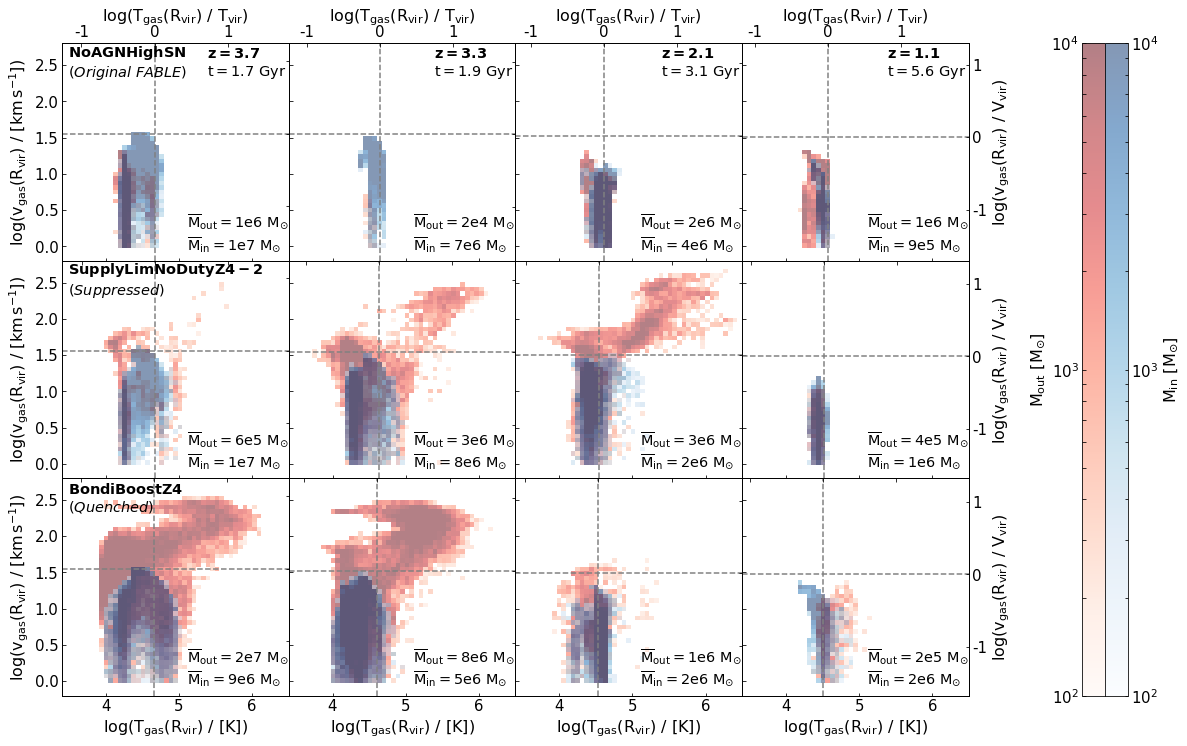}
    \caption{Outflow properties of representative dwarf zoom-in simulations. The blue-shaded and red-shaded histograms show the distribution of the inflowing and outflowing gas, respectively, at the virial radius, $R_\mathrm{vir}$, (shell width is 3 kpc) in temperature -- velocity space. The simulation names are given in the upper left-hand corner and the redshifts/simulation times are specified in the upper right-hand corner. Furthermore, the total outflowing and inflowing mass at $R_\mathrm{vir}$ is indicated in the lower right-hand corner. For reference, the virial temperature and virial velocity are shown as dashed lines. With AGN feedback the outflows are more powerful and have a significant fast and hot component which may escape the halo.}
    \label{fig:OutflowProperties}
\end{figure*}

Indeed we find that without a duty cycle, massive seeds with $M_\mathrm{seed}=10^{4} \ \Msun$ are required for the supply-limited accretion set-up to be effective, whilst the lighter variants with $10^{2} \ \Msun$ and $10^{3} \ \Msun$ have no effect when there is not any duty cycle employed (see Table~\ref{tab:zoomruns}). We test two different variations of the supply-limited accretion runs with massive seeds and without a duty cycle: in the first scenario the BH is active from $z_\mathrm{seed}=4$ to the end of the simulation (see \EddNoDutyFromZFour, dark green lines), whilst for the other set-up we switch off accretion and feedback at $z=2$ (see \EddNoDutyFromZFourToZTwo, light green lines). The latter allows for a recovery of the gas supply once the AGN has shut down, resulting in only a brief shutdown of star formation (though the SFRs remain suppressed until $z=0$). For the \EddNoDutyFromZFour \ run, on the other hand, the gas reservoir cannot recover due to the continued AGN activity. Though again we note that the AGN activity is mainly restricted to high redshift and the quenched state at low redshift is maintained with only very occasional AGN bursts.

With an AGN duty cycle, all of the supply-limited AGN accretion parameter configurations explored have at least a temporary impact on star formation. In some cases, the duty cycle leads to excessive BH growth (see Section~\ref{subsec:bh_assembly}), in particular using the fiducial \fable \ duty cycle of $\Delta t_\mathrm{AGN} = 25$~Myr results in extreme BH masses for all seed masses explored.

Here we present two representative set-ups that have a duty cycle yet the BH growth is (relatively) moderate, with $M_\mathrm{seed}=10^{2} \ \Msun$ and $\Delta t_\mathrm{AGN} = 10$~Myr. Again we investigate two configurations: in the first set-up AGN activity is switched off at $z=2$ (see \EddLowDutyLightFromZSixToZTwo, yellow lines) whilst the other set-up allows AGN activity to continue until $z=0$, subject to gas availability (see \EddLowDutyLightFromZSix, orange lines). As with the other pair of supply-limited-accretion runs, we find that star formation and the gas reservoir are suppressed rapidly once the BH is seeded. If the AGN is switched off at $z=2$ the gas supply and star formation activity may recover, otherwise the dwarf remains quenched and gas-poor -- in this case with just a single additional AGN burst at $z \gtrsim 1$.

\subsection{Outflow properties} \label{subsec:outflow_props}

\subsubsection{Outflow kinematics and energetics}

\begin{figure*}
    \centering
    \includegraphics[width=\textwidth]{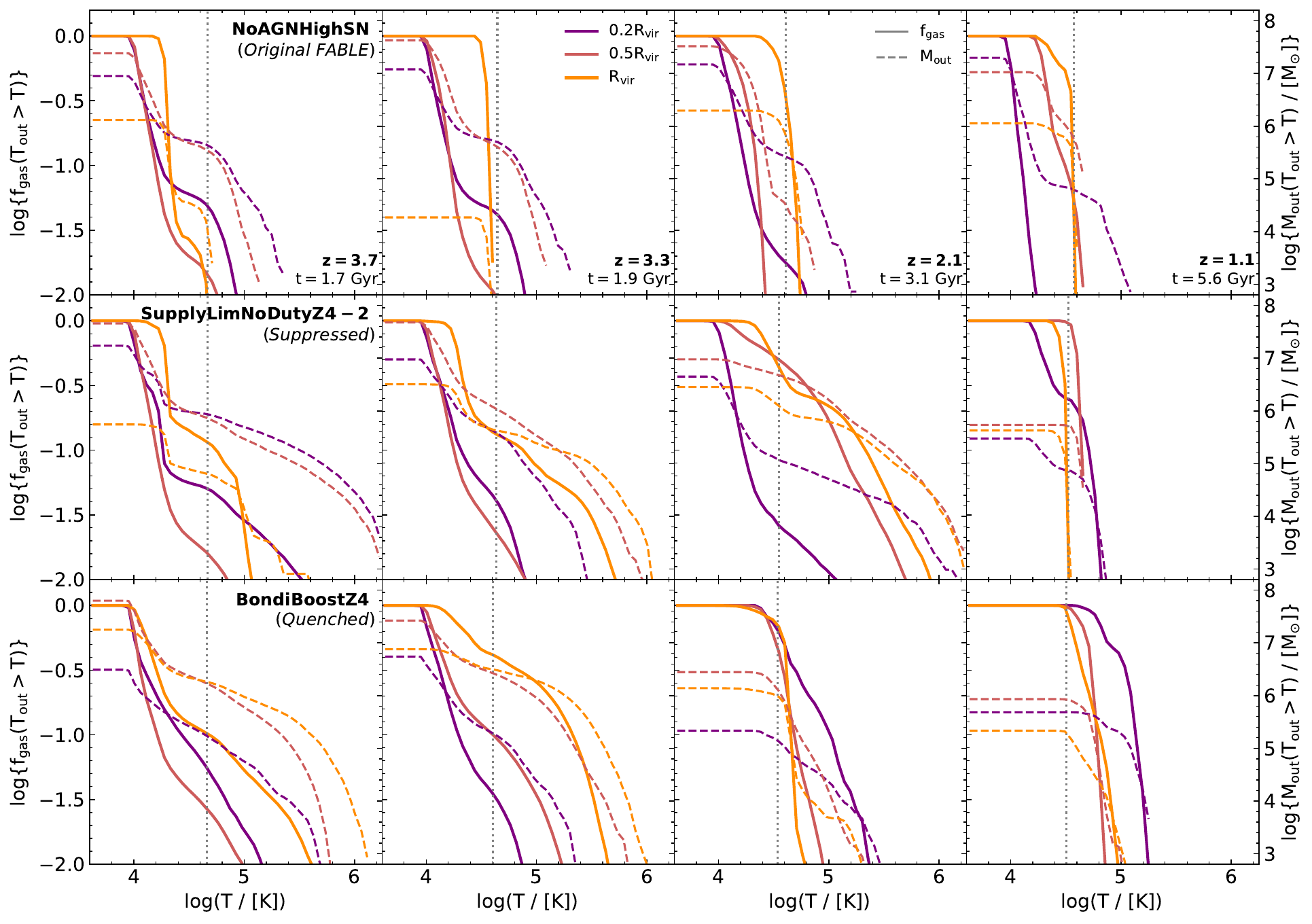}
    \caption{Thermal properties of the outflowing gas for a selection of dwarf zoom-in simulation set-ups, as indicated by the row labels. The solid lines show the fraction of outflowing gas above temperature $T$, whilst the dashed lines show the outflowing gas mass above temperature $T$. The colour-coding indicates where the outflow properties are measured with orange at $R_\mathrm{vir}$, dark red at $0.5R_\mathrm{vir}$ and purple at $0.2R_\mathrm{vir}$. The spherical shell widths for these radii are $3$~kpc, $2$~kpc and $1$~kpc, respectively. The dotted grey line indicates the virial temperature. The most significant differences in the temperature distribution lie at $R_\mathrm{vir}$, with SN feedback unable to push the hot gas to large scales.}
    \label{fig:OutflowTempFrac}
\end{figure*}

\begin{figure*}
    \centering
    \includegraphics[width=\textwidth]{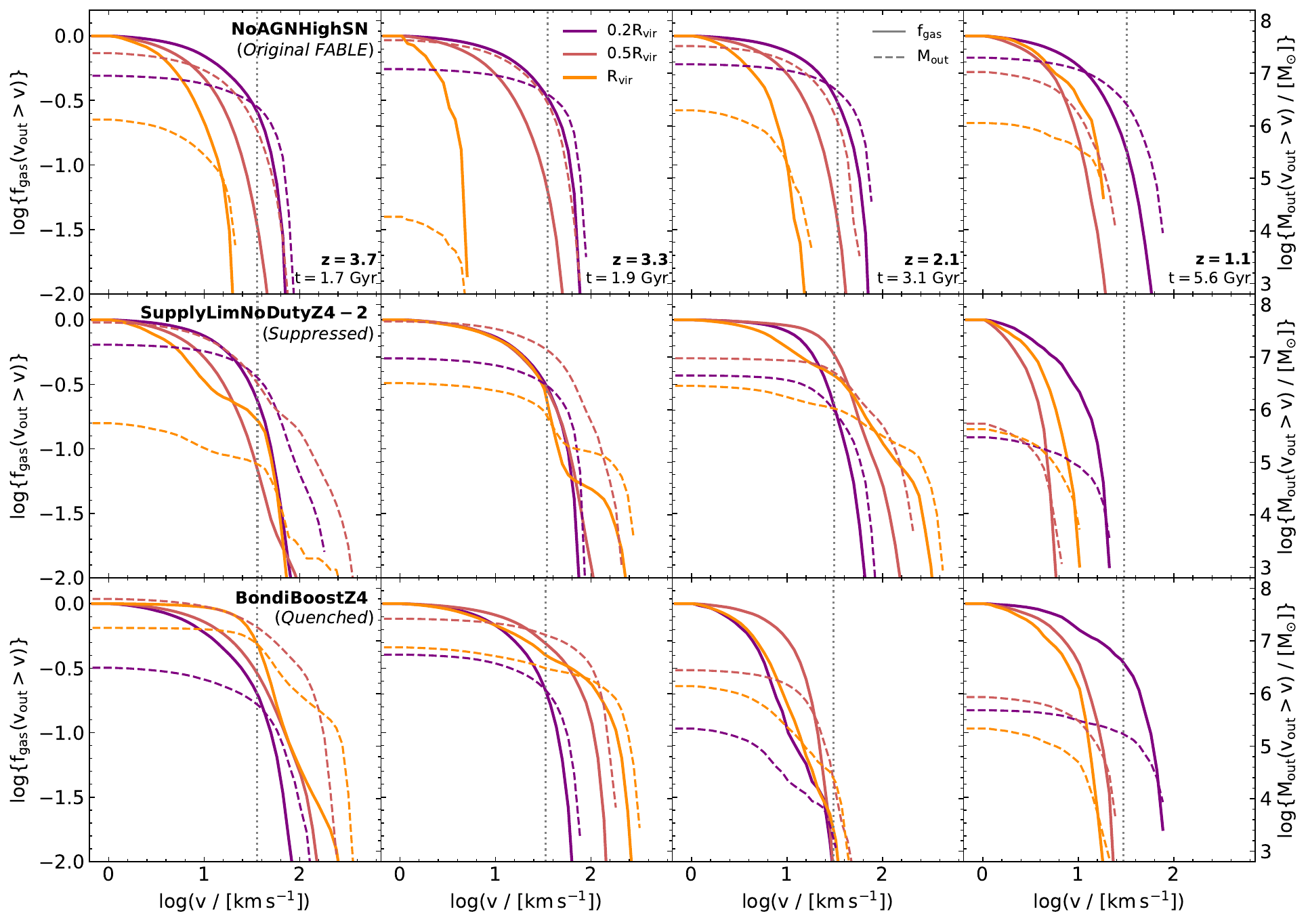}
    \caption{Kinematic properties of the outflowing gas for a selection of dwarf zoom-in simulation set-ups, as indicated by the row labels. The solid lines show the fraction of outflowing gas above velocity $v$, whilst the dashed lines show the outflowing gas mass above velocity $v$. The colour-coding indicates where the outflow properties are measured, with orange at $R_\mathrm{vir}$, dark red at $0.5R_\mathrm{vir}$ and purple at $0.2R_\mathrm{vir}$. The spherical shell widths for these radii are $3$~kpc, $2$~kpc and $1$~kpc, respectively. The dotted grey line indicates the virial velocity. Similarly to the temperature distributions, the most significant differences lie at $R_\mathrm{vir}$, with SN-driven outflows decelerating towards large-scales, whilst high-redshift efficient AGN feedback is able to accelerate the outflows towards $R_\mathrm{vir}$.}
    \label{fig:OutflowVelFrac}
\end{figure*}

In this section, we examine how AGN feedback influences the baryon cycle of the simulated dwarf galaxies by investigating the kinematic and energetic properties of inflows and outflows.

Fig.~\ref{fig:OutflowProperties} shows 2D mass histograms of the gas velocities and temperatures of the inflowing gas (in blue) and outflowing gas (in red) in a spherical slice of width $\Delta r = 3$~kpc centred at $R_\mathrm{vir}$. Note that we only consider gas moving with at least $v_{r}=1 \ \mathrm{km \, s^{-1}}$.

For reference, the virial temperature and virial velocity are shown as dashed lines. Furthermore, the total inflowing and outflowing gas mass is given in the lower right-hand corner of each panel. We choose the same redshifts and simulation runs as for the temperature projections in Fig.~\ref{fig:LargeScaleProj}: \NoAGNHighSN \ run in the top row, \EddNoDutyFromZFourToZTwo \ in the middle row and \BondiBoostFidDutyFromZFour \ in the bottom row. The redshift (and corresponding cosmic time) is given in the upper right-hand corner of the top row panels. In Appendix~\ref{appsec:Outflows}, we present plots analogous to Fig.~\ref{fig:OutflowProperties} for the inflow and outflow properties at $0.2R_\mathrm{vir}$ and $0.5R_\mathrm{vir}$. Note that the virial radii at the redshifts shown here approximately correspond to $R_\mathrm{vir} \sim 9, 10, 15, 25$~kpc.

Focusing firstly on the \NoAGNHighSN \ run, at high redshifts ($z>2$), the inflows dominate over the outflowing material. The outflows only move at low velocities and are therefore unable to escape the halo. Note that the SN feedback does produce a fast and energetic outflow component closer to the galaxy centre (see Fig.~\ref{fig:OutflowProperties0_5-0_2Rvir}), however, due to gas cooling and deceleration, this component cannot propagate to the virial radius. At $z \sim 1$, the outflow component becomes more prominent than the inflow component in terms of overall mass budget; though given the low velocities, these outflows form a galactic fountain and the overall gas reservoir of the galaxy remains relatively steady (see Fig.~\ref{fig:SfProperties}).

The \EddNoDutyFromZFourToZTwo \ simulation has significant outflow activity at early times, which peaks at $z \sim 2$, with the outflowing mass exceeding the inflowing mass. For $z>2$, there is a hot and fast component present which is able to escape the host galaxy, with the escaping outflow comprising a total gas mass of $\sim 10^{5}$~--~$10^{6} \ \Msun$. Therefore the outflows regulate the gas reservoir both by preventive and ejective feedback, heating the circumgalactic medium and driving the gas out of the galaxy. After the AGN has switched off (by construction) at $z=2$, the outflow component is significantly reduced with the inflowing component almost three times more massive than the outflowing gas. This allows for a recovery of the gas reservoir and renewed star formation at low redshifts as seen in Fig.~\ref{fig:SfProperties}.

Next, we turn our focus to the \BondiBoostFidDutyFromZFour \ run in the bottom row. The AGN duty cycle employed in this run reduces overcooling effects and therefore leads to a much more effective coupling between the AGN energy and the ISM. This results in the rapid development of a powerful outflow which has a prominent component at high velocities that would be expected to escape the galaxy, with total gas mass between $\sim 10^{6}$~--~$10^{7} \ \Msun$ for $z>3$. This fast component also has a significant amount of hot gas which then cools back down again to just below the virial temperature (see horizontal feature at $\log(v_\mathrm{gas} / [\mathrm{km \, s^{-1}}]) \sim 2.5$). At late times ($z \leq 2$), as the AGN activity reduces, there is only a minor lingering outflow component which is thermodynamically distinguished from the inflowing component by its higher temperatures but nevertheless, due the relatively low velocities, unlikely to escape the halo.

The mass distribution in the temperature -- velocity phase space for these three different feedback set-ups can be distinguished more quantitatively in Fig.~\ref{fig:OutflowTempFrac} and Fig.~\ref{fig:OutflowVelFrac}. These figures show the (mass-weighted) fraction as well as the total gas mass of outflowing gas moving above a certain temperature and velocity, respectively. The redshifts shown match the redshifts selected for Fig.~\ref{fig:OutflowProperties} and Fig.~\ref{fig:LargeScaleProj}. The colour coding indicates the radius at which the outflow properties are measured, with orange for $R_\mathrm{vir}$, dark red for $0.5R_\mathrm{vir}$ and purple for $0.2R_\mathrm{vir}$. The spherical shell widths for these radii are $3$~kpc, $2$~kpc and $1$~kpc, respectively.

For the \NoAGNHighSN \ run, Fig.~\ref{fig:OutflowTempFrac} shows that gas outflows with temperatures significantly above $T_\mathrm{vir}$ are only produced locally at $r \sim 0.2 R_\mathrm{vir}$ and even for these small scales, significant fractions of hot gas can only be found at early times ($z \gtrsim 3$) with our SN-only set-up. These small-scale hot outflows do not propagate to larger scales as the outflows significantly decelerate from the inner regions to the virial radius (see Fig.~\ref{fig:OutflowVelFrac}). Considering the cumulative gas mass, we can see a decrease in the gas mass of the fast outflow component from $0.2 R_\mathrm{vir}$ to $R_\mathrm{vir}$ at early times ($z \gtrsim 3$). The overall mass budget of outflows with velocities exceeding $V_\mathrm{vir}$ is very low compared to the AGN runs, especially at early times when the SN feedback still has to build up and only has a negligible impact on cosmic inflows. Overall, this then reinforces the picture of the \fable \ SN feedback acting in a `dispersive' rather than `ejective' manner to regulate star formation.

\begin{figure*}
    \centering
    \includegraphics[width=\textwidth]{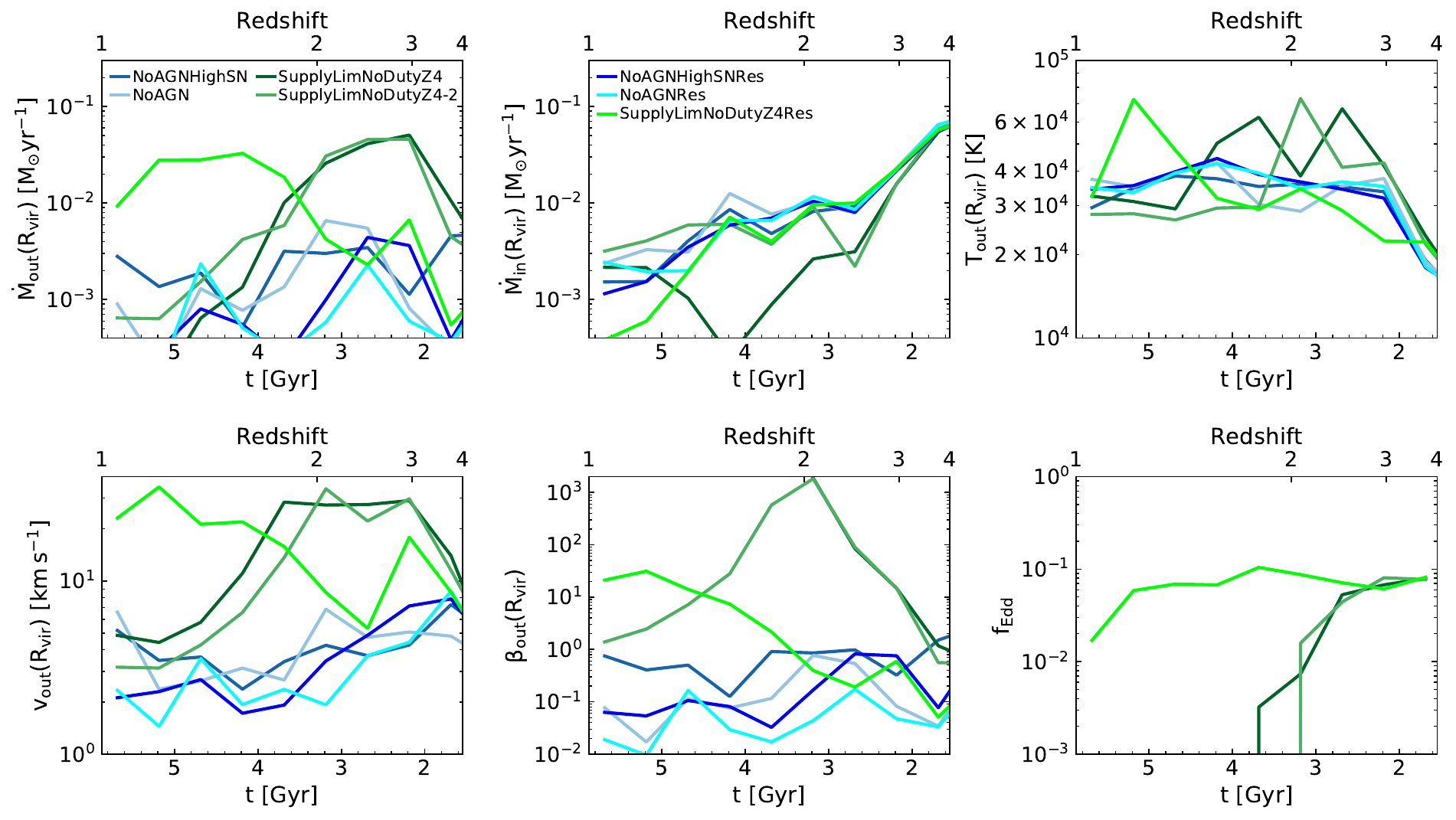}
    \caption{Time evolution of the outflow characteristics (mass outflow rate $\dot{M}_\mathrm{out}$, mass inflow rate $\dot{M}_\mathrm{in}$, outflow temperature $T_\mathrm{out}$, outflow velocity $v_\mathrm{out}$ and mass loading factor $\beta_\mathrm{out}$) at the virial radius, and the Eddington fraction $f_\mathrm{Edd}$. Here we contrast a selection of the main simulation runs at the fiducial resolution ($\bar{m}_\mathrm{gas} = 287 \ \Msun$) with equivalent zoom-in simulations with eight times higher mass resolution ($\bar{m}_\mathrm{gas} = 35.9 \ \Msun$). At higher resolution BH accretion levels are higher and more sustained. However, the impact is also more localised so that it takes significantly longer to establish large-scale AGN-driven outflows. For all quantities shown here, there is small-scale variability which has been smoothed out over time bins of $\Delta t = 500$~Myr to highlight trends.}
    \label{fig:OutflowsTimeEvol}
\end{figure*}

The \EddNoDutyFromZFourToZTwo \ set-up, on the other hand, already displays a large-scale hot outflow generated by the AGN at early times, with approximately 10 per cent of the outflowing gas exceeding $T_\mathrm{vir}$ at the virial radius. At $z \sim 2$, over 50 per cent of the outflowing gas is above $T_\mathrm{vir}$ at $0.5R_\mathrm{vir}$ and at $R_\mathrm{vir}$. However, after the AGN is switched off at $z=2$, this hot bubble quickly dissipates with no very hot gas remaining at the scale of $R_\mathrm{vir}$. The AGN feedback still has a lasting impact though, in particular at $0.2 R_\mathrm{vir}$ where the warm gas component ($T \sim 10^{4} \ K$) is notably diminished, demonstrating that the inner gas remains heated and dispersed (which also increases the efficiency of the stellar feedback) so that star formation remains suppressed with this set-up even after the AGN is switched off.

The \BondiBoostFidDutyFromZFour \ behaves similarly to the \EddNoDutyFromZFourToZTwo \ run at early times, though the impact of the AGN is even stronger, partly due to the AGN duty cycle. There is a clear increase in outflow velocities from small to large scales at early times ($z \gtrsim 3$) when the AGN is at peak activity, indicating that the outflows are accelerating. At late times ($z>2$), as the AGN activity significantly decreases due to the depleted gas supply, the large-scale fast and energetic AGN-driven outflow can no longer be maintained. However, the hot gas fraction remains above 50 per cent in the inner region for the outflowing gas mass, demonstrating that the occasional low-level AGN bursts (see right-hand panel of Fig.~\ref{fig:SfProperties}) are sufficient to drive a hot bubble dispersing the ISM and shielding the galaxy from cosmic inflows, hence keeping the dwarf galaxy in a quenched state.

\subsubsection{Outflow resolution dependence} \label{subsec:ResDependence}

Next, we investigate the resolution dependence of our results, in particular for the supply-limited-accretion simulations. These set-ups are directly tied to the resolved scale due to our criterion for only accreting gas from within three times the gravitational softening length (with the accretion rate set to zero if this region contains fewer than 16 gas cells). In this section, we outline how resolution affects the ability of the BH to efficiently accrete, drive outflows and regulate star formation in the host galaxy. For general resolution effects related to the stellar feedback, see Appendix~\ref{appsec:HiresStellar}.

To test the resolution dependence, we repeat a selection of the main zoom-in simulation runs at eight times higher mass resolution (two times higher spatial resolution): the two SN-only runs (\NoAGNHighSNRes \ and \NoAGNRes) and the supply-limited-accretion run with a massive BH seed formed at $z=4$ (\EddNoDutyFromZFourRes). Note that we perform the high-resolution runs only until $z=1$.

Fig.~\ref{fig:OutflowsTimeEvol} shows the cosmic evolution of the outflow properties at the virial radius (with slice width $3$~kpc) and Eddington fraction of these high-resolution zoom-in simulations as well as their equivalent fiducial-resolution set-ups. For comparison, we also show the properties of the \EddNoDutyFromZFourToZTwo \ fiducial-resolution simulation.

Focusing first on the Eddington fraction, we can see that at higher resolution the BH is able to accrete even more efficiently, maintaining $f_\mathrm{Edd} \sim 0.1$ for the majority of the simulation time (with a decrease to $f_\mathrm{Edd} \sim 0.02$ near $z=1$). This significant difference in BH accretion behaviour at higher resolution has various reasons. Whilst the accretion region within the BH smoothing length (defined by the 32 nearest gas cell neighbours and limited by three times the comoving gravitational softening length) is significantly smaller, similarly the injection region which is also set by the smoothing length corresponds to a significantly lower gas mass. Consequently, the AGN energy injection is much more spatially concentrated (though note this also means the temperature change of the affected gas cell will be larger) and it takes significantly longer for the AGN feedback to take effect and drive an outflow. 

This delayed outflow allows for the gas reservoir to remain intact until much lower redshifts, providing ample of gas for BH accretion. Crucially the hot AGN-driven outflow bubble also only takes effect much later, so that inflow suppression with the high-resolution \EddNoDutyFromZFourRes \ set-up does not occur until $t \sim 4.7$~Gyr. At this point in cosmic evolution, the inflow rates onto the dwarf halo are diminished anyway due to stellar feedback taking effect, so that the inflow suppression from the AGN only plays a negligible role.

Notably, the outflow enhancement, inflow suppression as well as temperature and velocity boosts are at similar absolute levels for the high-resolution and fiducial-resolution simulations. The AGN-boosted outflow is merely delayed at higher resolution and therefore less impactful.

In addition to the changed inflow -- outflow dynamics, the higher resolution also promotes the formation of high-density regions which enhance BH accretion rates and result in higher overall BH masses. Note that similar BH mass trends in low-mass galaxies with stellar masses $\lesssim 10^{10} \ \Msun$ are seen for increasing the resolution with the IllustrisTNG galaxy formation model \citep[][]{pillepich_simulating_2018}.

Crucially, the catastrophic quenching and gas depletion effects observed for the fiducial-resolution simulation are produced by a combination of efficiently driving out the gas and then preventing the gas reservoir from being replenished at high redshift, when cosmic inflow rates are high. The delayed AGN-driven outflow with the higher-resolution simulation yields a much more moderate outcome, with the gas reservoir merely diminished rather than completely depleted. This is also more in line with the observations of lower gas masses in dwarfs with AGN activity.

Similarly, due to the catastrophic suppression of star formation, the fiducial-resolution runs reach unrealistically high mass-loading factors ($\beta \gtrsim 10^{3}$), whilst the mass loading factors for the \EddNoDutyFromZFourRes \ simulation are much more moderate ranging from $\beta \sim 0.1$ to $\beta \sim 10$.

Fig.~\ref{fig:OutflowsTimeEvol} also illustrates how the difference between star formation in the \EddNoDutyFromZFour \ and \EddNoDutyFromZFourToZTwo \ set-ups arises from differing mass inflow rates (rather than the mass outflow rates). Whilst the mass outflow rates are very similar across cosmic time (with \EddNoDutyFromZFourToZTwo \ having slightly higher mass outflow rates at late times due to higher star formation and therefore higher stellar feedback activity), the mass inflow rates for the \EddNoDutyFromZFourToZTwo \ are only very briefly suppressed and quickly recover to the inflow rate of the SN-only reference run once the AGN has been switched off at $z=2$. For the \EddNoDutyFromZFour \ simulation, on the other hand, the inflow rates keeps decreasing until $t \sim 4$~Gyr when gas inflows are completely suppressed. As the gas reservoir has been completely depleted at this point, and therefore the AGN has shut off, the inflows can slowly recover reaching the inflow rates of the SN-only reference run at $t \sim 5$~Gyr. However, as discussed previously, at this point inflow rates are generally very low so that the gas reservoir cannot recover with this set-up.

Overall, we find that the high-resolution set-up results in an intermediate case in terms of star formation suppression, with delayed outflows leading to only mild gas depletion so that the BH can continue accreting throughout the duration of the simulation until $z=1$. Subsequently, the BH ends up being more overmassive with $\log(M_\mathrm{BH} (z=1) / \Msun) = 4.6$, however due to the higher stellar mass this results in a similar offset from the extrapolated scaling relations. 

The AGN feedback models we have investigated in this paper cover a large parameter space resulting in a diversity of outcomes in terms of star formation suppression and it should be noted that the very-high-resolution \EddNoDutyFromZFourRes \ simulation is not necessarily more realistic than the fiducial-resolution runs. Rather these runs likely bracket the range of possible AGN feedback outcomes.

It will be imperative to explore more sophisticated AGN accretion and feedback models and combine these with results from current and upcoming surveys to elucidate the role of AGN in dwarfs. In particular, \textit{JWST} will play a key role in constraining the multiphase outflow properties of dwarf galaxies across cosmic time through sensitive measurements of nebular lines and of (near and mid)-IR emission. 

\subsection{Stellar assembly} \label{subsec:stellar_assembly} 

\begin{figure*}
    \centering
    \includegraphics[width=\textwidth]{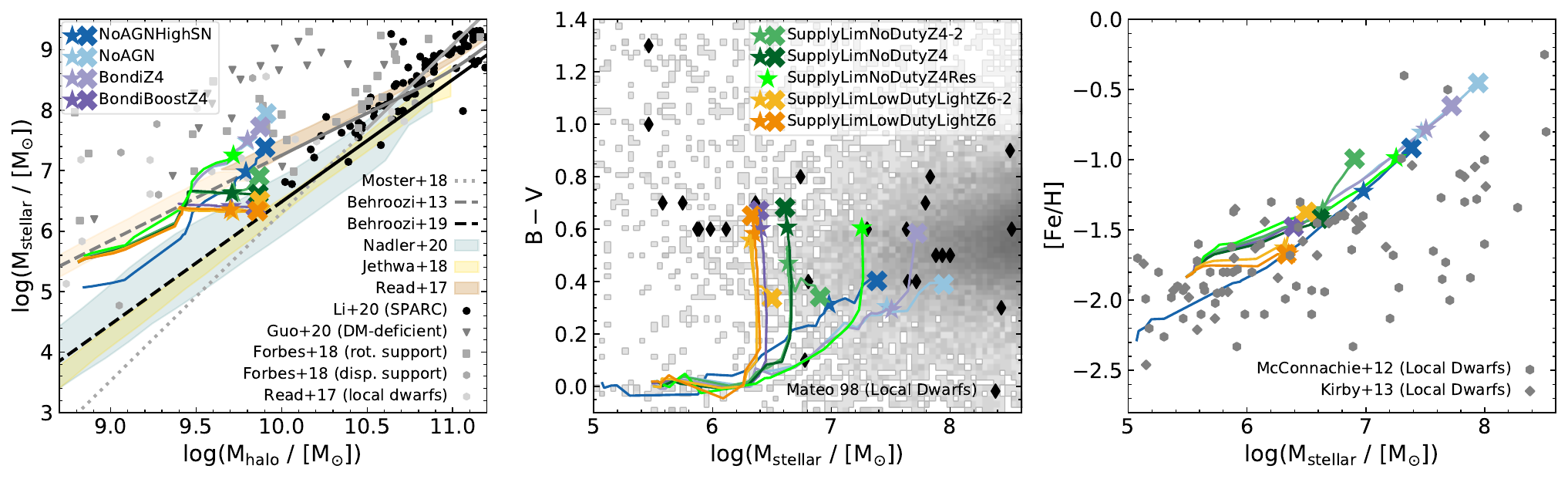}
    \caption{Integrated stellar and halo properties in comparison with observational constraints. We show the redshift evolution of the main dwarf zoom-in simulations from $z=6$~--~$0$, with the $z=1$ and $z=0$ data points highlighted by stars and crosses, respectively. The colour-coding for the different simulation set-ups is indicated in the legend. \textit{Left panel:} Stellar mass -- halo mass (SMHM) relation. SMHM relations based on empirical models as well as individual observed galaxies are plotted for comparison as indicated by the legend. \textit{Middle panel:} Galaxy colours against stellar mass. The (converted to B -- V) colours from the SDSS galaxies are indicated as a grey histogram and integrated photometry of the local dwarfs is plotted as black diamonds \citep{mateo_dwarf_1998}. \textit{Right panel:} Stellar metallicity against stellar mass. We also plot the stellar metallicities of the local dwarfs from \citet{mcconnachie_observed_2012} and \citet{kirby_universal_2013}, note that there is some overlap between these two data sets. \textit{Overall}, we find that AGN feedback in dwarfs cannot be ruled out with these observational constraints on stellar assembly, however, there are some marked differences. Dwarfs with AGN feedback tend to be more quenched, redder, and more metal-poor.}
    \label{fig:StellarAssembly}
\end{figure*}

In this section, we aim to assess whether the integrated stellar properties of our different simulation set-ups are realistic compared with observations, and if this allows us to constrain the nature of AGN feedback in dwarf galaxies. To this end, we show the stellar mass -- halo mass (SMHM) relation, colour -- stellar mass relation, and metallicity -- stellar mass relation in Fig.~\ref{fig:StellarAssembly}.

\subsubsection{Stellar mass -- halo mass relation}

Firstly, we investigate whether the simulated dwarfs are in line with observational expectations for the SMHM relation. The left-hand panel of Fig.~\ref{fig:StellarAssembly} shows the redshift evolution of the main simulation runs in $M_\mathrm{halo}$ -- $M_\mathrm{stellar}$ space\footnote{There are different definitions of halo mass and we here employ the virial mass as defined by the mass enclosed in a sphere whose mean density is 200 times the critical density of the Universe, at the time the halo is considered.}, with the data points at redshifts $z=1$ and $z=0$ marked by stars and crosses, respectively. We also show the redshift evolution of the high-resolution AGN run \EddNoDutyFromZFourRes, which was completed to $z=1$.

For comparison, we plot several SMHM relations based on empirical models from the literature, including \citet{behroozi_average_2013,behroozi_universemachine_2019} and \citet{moster_emerge_2018}. The low-mass end of the SMHM relation is still largely unconstrained -- as evident by the order of magnitude differences between these models. With $M_\mathrm{halo}(z=0) \sim 10^{10} \ \Msun$ our dwarf system sits at the very edge of the validity of these relations. The \citet{behroozi_average_2013,behroozi_universemachine_2019} models extend down to $\log(M_\mathrm{halo} / \Msun) = 10.0$ and the \citet{moster_emerge_2018} model is valid down to $\log(M_\mathrm{halo} / \Msun) = 10.5$. Below these masses we show the extrapolated relations as dashed lines and dotted lines, respectively.

Note that there are several pitfalls when extrapolating SMHM relations below the minimum mass considered for the modelling and validation \citep[e.g. see discussion in][]{moster_emerge_2018}, in particular with regards to physical processes that are only effective at the low-mass end, such as reionization suppression, and therefore not included in the modelling. Furthermore, due to more bursty star formation histories at the low-mass end, the scatter in the SMHM relation would also be expected to increase as we move to lower masses. 

It can therefore be more instructive to compare dwarf simulations to SMHM relations that have been specifically constructed for low-mass galaxy samples. We therefore also show the SMHM relations derived for dwarf galaxy samples by \citet{read_stellar_2017}, \citet{jethwa_upper_2018} and \citet{nadler_milky_2020} as beige-shaded\footnote{The light beige region indicates where the SMHM relation relies on a power law extrapolation of the SDSS stellar mass function below $M_\mathrm{stellar} = 10^{7} \ \Msun$.}, yellow-shaded and blue-shaded regions,  respectively. \citet{read_stellar_2017} is in better agreement with the extrapolated relation from \citet{behroozi_average_2013}, whilst \citet{jethwa_upper_2018} and \citet{nadler_milky_2020} align better with the extrapolated relation from \citet{behroozi_universemachine_2019}. This is not surprising since the \citet{read_stellar_2017} relation is based on the SDSS stellar mass function from \citet{behroozi_average_2013}, leading to a similar normalization. 

Note that \citet{behroozi_universemachine_2019} caution that the stellar mass function in their previous work \citep{behroozi_average_2013} assumed a strong surface-brightness incompleteness correction that is no longer observationally supported, pointing to the need for a lower normalization. Though it should be emphasised that the scatter at the low-mass end of the SMHM is significant. To this end, we also show various observed data points colour-coded on a grey-scale according to the sample with the SPARC data from \citet{li_comprehensive_2020} as black filled circles, the dark-matter-deficient dwarf galaxy sample from \citet{guo_further_2020} as dark grey inverted triangles (these points should be seen as upper limits), nearby dwarf galaxies from \citet{forbes_extending_2018} as grey squares and hexagons for rotationally and dispersion supported systems, respectively. Finally, we plot the local dwarfs from \citet{read_stellar_2017} as light-grey hexagons -- provided that they have not already been included in the aforementioned samples. Note that whilst these data points give an indication of the scatter, they are biased towards higher stellar masses due to the completeness limits of the surveys.

Given the high scatter and uncertainty in the SMHM relations, it is difficult to draw firm conclusions, however, the lower stellar mass normalizations seem to be favoured by more recent studies. To assess whether the stellar masses of the different simulation runs are realistic, we therefore take the \citet{moster_emerge_2018} and \citet{behroozi_average_2013} relations as lower and upper limits, respectively. Note that the evolution of the SMHM relation between $z=1$ and $z=0$ is minimal \citep[see e.g.][figure 9]{behroozi_universemachine_2019}, so we do not plot the $z=1$ relations separately here. For both the $z=1$ and $z=0$ simulated data points we then find that the runs with efficient AGN feedback produce realistic stellar masses given the observational constraints and expectations from empirical models. The \NoAGNHighSN \ is more towards the upper limit of the predicted relations but still within the constraints. The stellar masses of the inefficient AGN runs (such as \BondiFidDutyFromZFour) and \NoAGN run are likely too high (though still within the scatter of observed dwarfs). Similarly, the more gentle \EddNoDutyFromZFourRes \ simulation has a final stellar mass at $z=1$ that is around the upper limit for the predicted stellar mass at $z=0$, though it still falls within the significant scatter.

\subsubsection{Galaxy colours}

Next, we turn towards comparing the galaxy colours of our simulated dwarfs with observational constraints. The middle panel of Fig.~\ref{fig:StellarAssembly} shows the $B - V$ colour against stellar mass. As before, we show the evolutionary tracks of the simulated dwarfs from $z=6$ to $z=0$, with the $z=1$ and $z=0$ data points highlighted by star and cross symbols, respectively. Again, we also include the \EddNoDutyFromZFourRes \ which was completed to $z=1$. Furthermore, we plot the colours of the SDSS galaxies from the NASA Sloan Atlas (NSA) as a grey histogram as well as the galaxy colours of local dwarfs from \citet{mateo_dwarf_1998} as black diamonds.

We use the up-to-date \citet{bruzual_stellar_2003} stellar population synthesis model, assuming a \citet{chabrier_galactic_2003} IMF, to calculate the galaxy colours as a function of stellar age and metallicity, including all star particles in the main halo. For the NSA galaxies, we convert from the SDSS magnitudes using the transformation equations from \citet{jester_sloan_2005}\footnote{There are no transformation equations explicitly for galaxies currently available, however the stellar transformation equations should provide a good approximation provided there are no strong emission lines present.}.

One important caveat with the galaxy colour comparison is that we have not accounted for dust attenuation when calculating the colours of the simulated runs. However, since the gas mass and metallicity are very low for the majority of the AGN runs, dust attenuation would also be expected to be low for these set-ups.

Our simulated galaxy colours are well within the scatter of the observational data, with the efficient AGN runs resulting in significantly redder galaxies than the inefficient AGN runs or no-AGN runs at $z=1$. For the simulation set-ups where the AGN is switched off at $z=2$, the galaxies move towards blue colours due to the rejuvenated star formation, resulting in a similar $z=0$ galaxy colour as the two no-AGN runs. For the runs where star formation continues to be suppressed by the (infrequent) AGN bursts, the galaxy colour remains firmly in the red region.

Interestingly, the late-time star formation suppression in the \BondiFidDutyFromZFour \ run, whilst having a negligible effect on the final stellar mass, significantly reddens the host galaxy compared to the no-AGN runs, suggesting that whilst stellar masses may not necessarily differ for dwarf galaxies with late-time AGN activity, galaxy colours (and star formation histories) could be a more distinguishing feature. This is also supported by the more gentle AGN feedback in the \EddNoDutyFromZFourRes \ set-up, which results in a similar level of reddening as the strong AGN runs at $z=1$, despite the significantly higher stellar mass.

\subsubsection{Stellar metallicities}

Finally, we compare the integrated stellar metallicities of our simulated dwarfs with constraints from the Local Group \citep{mcconnachie_observed_2012, kirby_universal_2013}. As before, we show the evolutionary tracks of the main simulations runs from $z=6$ to $z=0$ and for the high-resolution \EddNoDutyFromZFourRes \ run from $z=6$ to $z=1$, with the $z=1$ and $z=0$ data highlighted by star and cross markers, respectively. 

The simulated stellar metallicities are calculated within twice the stellar half mass radius as the mean of the logarithmic metallicity values in solar units. The simulations are broadly within the scatter of the observed metallicities, however, the simulated metallicity values are somewhat higher than the observed ones, suggesting that the feedback might not be ejective enough, even with AGN feedback reinforcing SN feedback \citep[see also][]{agertz_edge_2020}. In particular early and sustained AGN feedback may be necessary to obtain good agreement, as the best agreement is obtained for the \EddLowDutyLightFromZSix \ run (see orange line). To ensure consistency with the observations we here assume a solar metallicity value of $Z_\mathrm{\odot} = 0.019$ in accordance with the solar composition values employed by \citet{kirby_universal_2013} which uses the stellar population model from \citet{kirby_multi-element_2010}\footnote{The \citet{kirby_multi-element_2010} model assumes solar composition values from \citet{anders_abundances_1989}, except for the iron abundance where a small adjustment is made with $\mathrm{Fe}=7.52$, though the resulting metal mass fraction is still $Z_\mathrm{\odot} = 0.019$ to 2 significant figures.}. Though note that this is significantly higher than the solar metallicity value assumed for \fable \ ($Z_\mathrm{\odot} = 0.0127$). We make this adjustment here to allow for a more direct comparison, however, we note that with the \fable \ solar metallicity, whilst the simulated metallicities are still within the scatter of the observed data, the offset towards the higher end of the distribution is more significant.

\subsubsection{Dwarf stellar assembly constraints on AGN feedback}
The comparisons with observational constraints on the stellar assembly of dwarf galaxies do not disfavour AGN feedback, however, there are several distinguishing features of AGN activity. There is a tendency for dwarfs with AGN feedback to be more quenched, redder and more metal poor. These findings are in agreement with tentative observational trends (albeit at higher dwarf masses), e.g. \citet{dickey_agn_2019} find that 16 out of 20 quiescent dwarf galaxies in their sample have AGN-like line ratios, though larger sample sizes as well as extending these samples to lower galaxy masses will be necessary to confirm these trends.

\subsection{BH assembly} \label{subsec:bh_assembly}
\begin{figure*}
    \centering
    \includegraphics[width=\textwidth]{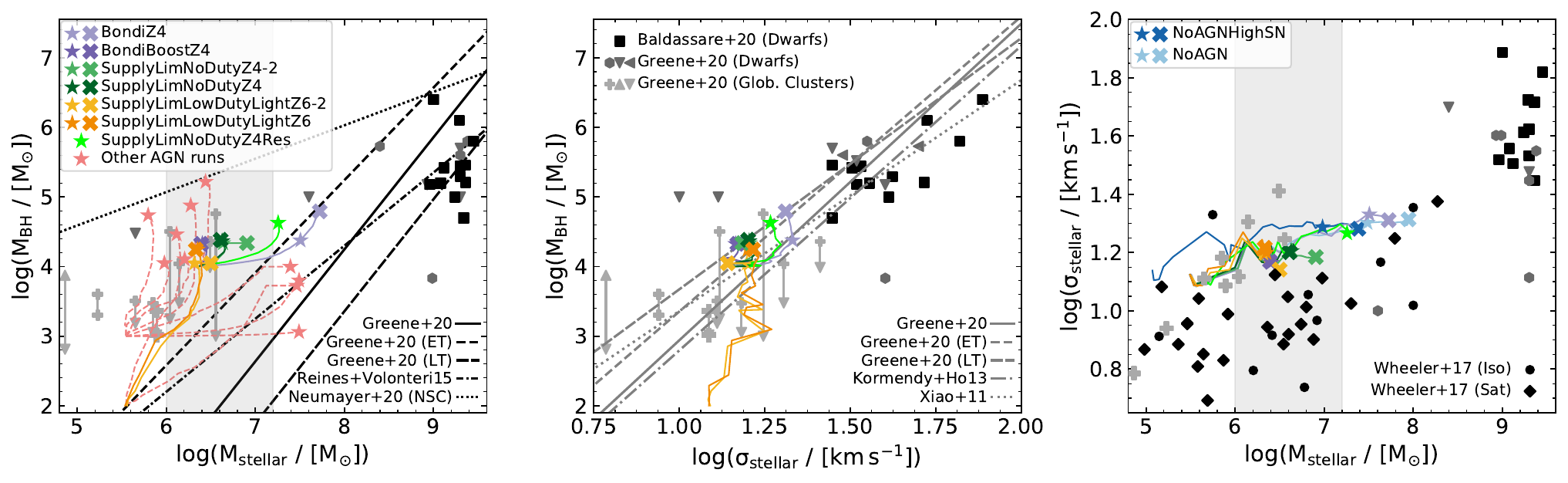}
    \caption{BH -- galaxy co-evolution and stellar velocity dispersions compared to observational constraints. We show the redshift evolution of the main dwarf zoom-in simulations from $z=6$ to $z=0$, with the $z=1$ and $z=0$ data points highlighted by stars and crosses, respectively. The vertically shaded grey band in the left- and righthand panels indicates the stellar mass range expected from SMHM relations for our system at $z=0$. We also show data from observational studies as indicated by the figure legends. The BH masses (and their presence) in globular clusters are still contested so we show a range spanning the minimum and maximum detected BH mass values in the literature as compiled by \citet{greene_intermediate-mass_2020}. \textit{Left panel:} BH mass against stellar mass (measured within twice the stellar half mass radius). In addition to the main simulation runs, we also plot all other dwarf zoom-ins with AGN feedback from Table~\ref{tab:zoomruns} as coral dashed lines (note that these additional simulations were only performed to $z=1$). For comparison we also show the observational (and extrapolated) BH mass -- stellar scaling relations from the literature by \citet{reines_relations_2015} and \citet{greene_intermediate-mass_2020} as well as the NSC mass -- stellar mass relation from \citet{neumayer_nuclear_2020}, which can be taken as an upper limit. \textit{Middle panel:} BH mass against stellar velocity dispersion (measured within a fixed aperture of radius $0.3$~kpc). Again we show observational (and extrapolated) scaling relations for comparison, as indicated by the legend. Here the simulations are in much better agreement with the observational expectations than for the $M_\mathrm{BH}$ -- $M_\mathrm{stellar}$ relation. \textit{Right panel:} A potential explanation for this is the flattening of the stellar velocity dispersion -- stellar mass relation at the low-mass end. However, we also note that our simulated velocity dispersions are at the upper end of the observed distributions and lower dispersions could again lead to the BHs being categorised as overmassive with respect to the extrapolated observed relations.}
    \label{fig:BHMassStellarMassScaling}
\end{figure*}

In this section, we investigate the BH assembly and the level of agreement with observational constraints on BH -- galaxy co-evolution.

Fig.~\ref{fig:BHMassStellarMassScaling} shows the BH mass against stellar mass (left panel), BH mass against stellar velocity dispersion (middle panel) as well as stellar velocity dispersion against stellar mass (right panel) for the main simulation runs and the high-resolution run \EddNoDutyFromZFourRes. In all cases, the simulated stellar mass is calculated within twice the stellar half mass radius and the (1D) stellar velocity dispersion is calculated within a fixed aperture of radius $0.3$~kpc. Observational data and scaling relations are also plotted for comparison\footnote{Note that we do not make adjustments for the different IMFs that have been assumed in converting between stellar luminosities and masses.}. In the following, we discuss these different indicators of co-assembly and whether these can rule out any of the AGN models and parametrizations. 

\subsubsection{BH mass -- stellar mass evolution}

Firstly, we focus on the left-hand panel which shows the BH mass -- stellar mass scaling relations. The evolutionary tracks of the main simulation runs are plotted from $z=6$ to $z=0$ as solid lines colour-coded according to the simulation set-up, with the $z=1$ and $z=0$ data points highlighted as stars and crosses, respectively. For reference, we also show the redshift evolution of all other AGN feedback runs (excluding accretion-only runs) from Table~\ref{tab:zoomruns} as dashed coral lines and the high-resolution \EddNoDutyFromZFourRes \ run as a bright green line, with the $z=1$ data points highlighted as stars (note that these simulations were only completed to $z=1$).

The grey-shaded region denotes the $10^{6} \ \Msun \lesssim M_\mathrm{stellar} \lesssim 10^{7.2} \ \Msun$ range which corresponds to the stellar mass range spanned by the SMHM relations from Fig.~\ref{fig:StellarAssembly}. Galaxies to the right or left of this band are likely under- or over-regulated, respectively.

For comparison, we also show some observed scaling relations. There is still significant uncertainty about the shape and scatter of the low-mass end of the BH -- galaxy scaling relations, so we have selected a few representative examples from the literature that include dwarf galaxies in their data samples. 

The \citet{greene_intermediate-mass_2020} BH masses are based on the dynamical BH masses from \citet{kormendy_coevolution_2013} supplemented with additional dynamical BH mass measurements published since then (particularly in the dwarf regime), including upper limits. We plot the relation based on the whole galaxy sample as a solid black line as well as the relations based on early-type and late-type galaxies only as loosely and densely dashed black lines, respectively. Note that the late-type relation has a significantly lower normalisation than the early-type relation and the relation based on the overall sample. This difference in normalisations could be attributed to the much weaker correlation between disc mass and BH mass \citep[e.g.][]{kormendy_inward_1995}.

For our stellar mass range of interest ($10^{6} \ \Msun \lesssim M_\mathrm{stellar} \lesssim 10^{7} \ \Msun$), these relations are still heavily extrapolated with the lowest mass galaxy in this observational sample (excluding upper limits) having a stellar mass of $M_\mathrm{stellar} = 2.5 \times 10^{8} \ \Msun$. If we do take the extrapolations at face value, the \citet{greene_intermediate-mass_2020} relation for the whole galaxy sample would suggest BH masses of $M_\mathrm{BH} \lesssim 10^{2} \ \Msun$ for our stellar mass range of interest, i.e. lower than the lowest seed mass explored. However, it is most likely not appropriate to extrapolate the scaling relations into this mass regime as we would expect a flattening of the relations, with the transition mass dependent on the seeding mechanism \citep[see e.g.][]{greene_intermediate-mass_2020}.

We also show the scaling relations from \citet{reines_relations_2015}, which is based on a sample of local AGN including dwarfs, as a dashed-dotted black line. This relation is significantly shallower than the \citet{greene_intermediate-mass_2020} relations leading to larger BH masses than the overall \citet{greene_intermediate-mass_2020} relation in the low-mass dwarf regime. Nevertheless, we find that simulations which are in agreement with this extrapolated relation have already formed too many stars at $z=1$ to match the expected stellar mass range from the $z=0$ SMHM relations. It is worth stressing the little constraining power of these observed relations, however, as the \citet{reines_relations_2015} relation is also heavily extrapolated, with the least massive dwarf in this sample having a stellar mass of $M_\mathrm{stellar} = 4.3 \times 10^{8} \ \Msun$.

Overall, the extrapolated BH mass -- galaxy mass scaling relations could then be possibly regarded as lower limits for the BH masses in the dwarf regime. To obtain an upper limit, we use the nuclear star cluster (NSC) -- galaxy scaling relations. NSCs and massive BHs are found to co-exist in many galaxies, with the former being the dominant central mass component in dwarfs\footnote{The scatter for the $M_\mathrm{BH}/M_\mathrm{NSC}$ ratio at a given galaxy mass is over three orders of magnitude, however, there is a trend of increasing $M_\mathrm{BH}/M_\mathrm{NSC}$ ratio with galaxy mass and BHs only starting to dominate at stellar masses above a few times $10^{10} \ \Msun$.} and therefore providing a good indicator of BH mass upper limits. \citet{neumayer_nuclear_2020} present a compilation of NSC mass measurements which extends down to low-mass dwarfs with stellar masses $\lesssim 10^{6} \ \Msun$, i.e. the relation is valid for the whole $z=0$ galaxy stellar mass range considered here. The region bounded by the lower limit of the extrapolated BH -- galaxy scaling relations and the upper limit of the BH -- NSC scaling relation and the stellar mass limits derived from the SMHM relation (grey-shaded region) can then be regarded as the most plausible region of parameter space.

Finally, we also show some of the individual observed dwarf data, with the virial BH mass estimates for active dwarf galaxies from \citet{baldassare_populating_2020} as black squares and the dynamical measurements from \citet{greene_intermediate-mass_2020} as dark-grey hexagons and triangles for mean values and upper limits, respectively. Errorbars are omitted for clarity. These individual observed data points demonstrate that there are virtually no BH mass measurements for classical dwarfs (with $M_\mathrm{stellar} \sim 10^{7} \ \Msun$). To extend our observational data comparison to lower-mass systems, we also include some intermediate-mass black hole (IMBH) candidates from globular clusters around the Milky Way and M31, as compiled by \citet{greene_intermediate-mass_2020}. Note some of these globular clusters are hypothesised to be remnants of dwarf galaxies disrupted by the tidal field of the Milky Way, such as $\omega$ Cen \citep[e.g.][]{freeman_globular_1993,bekki_formation_2003,meza_accretion_2005}. In any case, the presence of IMBHs in globular clusters is still very controversial and the uncertainty in the BH mass values is significant. Hence for each globular cluster, we show a range corresponding to the lowest and highest BH mass value reported in the literature \citep[as compiled by][]{greene_intermediate-mass_2020}, with crosses indicating mean values and triangles indicating lower/upper limits.

The observed BH masses of globular clusters lie significantly above the extrapolated scaling relations for galaxies. This may be partly driven by an observational bias towards detecting more massive BHs but it could also hint at a flattening of the relation in this mass regime. Another contributing factor for systems such as $\omega$ Cen may be that the stellar mass is significantly reduced due to the stripping by the Milky Way disc. Nevertheless, the globular clusters may provide an indication for reasonable BH masses in this stellar mass regime.

Inspecting the evolution of our simulated dwarf galaxies in $M_\mathrm{stellar}$ -- $M_\mathrm{BH}$ space, we find that the AGN runs fall into roughly two categories. We either have galaxies with efficient BH growth that (mostly) match the stellar mass expected from SMHM relations, but have potentially overmassive BHs -- in some cases even reaching the local constraints from NSCs at $z=1$. Alternatively, our simulations admit solutions that are in agreement with the extrapolated BH -- galaxy scaling relations, which however have under-regulated galaxies, with the stellar mass exceeding the expectations SMHM relations based on empirical models. 

The under-regulated simulation runs\footnote{Note that the accretion-only runs are not plotted here.} are listed in Table~\ref{tab:zoomruns} as the `No impact' category, with all five runs exceeding the expected stellar mass range for $z=0$ at $z=1$. These runs are either based on the Bondi model with insufficient boost factors for their seed masses or based on the supply-limited accretion scheme but without a duty cycle and starting from light to intermediate-mass seeds so that efficient accretion becomes difficult.

Our main simulation runs with AGN feedback have mostly been selected from the set-ups with efficient BH growth (with the exception of \BondiFidDutyFromZFour), yet among these overmassive solutions we have picked the runs with comparatively more modest BH growth and star formation regulation. Whilst all of the efficient main simulation runs form overmassive BHs with respect to the extrapolated scaling relations, the BH masses are still one dex below the NSC constraints and well within the scatter of the putative BH masses for globular clusters, suggesting that these set-ups are physically plausible given the observational uncertainties.

The high-resolution \EddNoDutyFromZFourRes \ run falls in-between these two classes, straddling the upper limit for the expected stellar mass and being overmassive with respect to the extrapolated scaling relations yet in good agreement with inferred BH masses of globular clusters as well as upper limits for BHs in dwarfs.

Overall, we find that, with our galaxy formation model, all effective AGN runs end up being overmassive compared to the extrapolated $M_\mathrm{BH}$ -- $M_\mathrm{stellar}$ scaling relations. Next-generation extremely large telescopes and ALMA at full capacity may be able to extend dynamical BH mass measurements to BH masses $M_\mathrm{BH} < 10^{5} \ \Msun$, though supplementary surveys are needed to identify suitable targets, in particular with regards to dwarf galaxy centres and molecular gas content \citep[see][]{greene_intermediate-mass_2020}.

\subsubsection{BH mass -- stellar velocity dispersion evolution}

Next, we compare our simulations with observational predictions for the BH mass -- stellar velocity dispersion relation (see middle panel). We plot the scaling relations for the full dynamical sample from \citet{greene_intermediate-mass_2020} as well as for early-type and late-type galaxies only as solid, loosely dashed and densely dashed grey lines, respectively. Note that the difference between these three relations is much smaller than for the equivalent BH mass -- stellar mass relations, in the regime where there is data. For reference, we also show the original scaling relation from \citet{kormendy_coevolution_2013} as well as the scaling relation from \citet{xiao_exploring_2011}, which is based on low-mass Seyfert 1 galaxies.

In addition to the (extrapolated) scaling relations, we also plot various observed data. As in the left-hand side panel, we include the dwarfs from \citet{baldassare_populating_2020} as well as the dwarfs and globular clusters from \citet{greene_intermediate-mass_2020}. For the latter the uncertainties in BH mass (and presence) are significant, so again we indicate the whole range of BH masses from the literature as collated by \citet{greene_intermediate-mass_2020}.

For the simulations, we calculate the (mass-weighted) 1D stellar velocity dispersion, as the root mean square value of the $x$-, $y$- and $z$-directions, within a fixed 2D-aperture of radius $0.3$~kpc, which is well-matched to the 1 arcsecond resolution from \citet{baldassare_populating_2020}, as well as typical dwarf galaxy bulge sizes relied on in other studies \citep{schutte_black_2019}. We also varied our aperture size between $0.1$~kpc and $1.1$~kpc \citep[the latter corresponds to the resolution from][at the maximum redshift of the sample]{baldassare_populating_2020} and found no significant difference as the central stellar velocity dispersion profiles are relatively flat.

Interestingly, while the main simulation runs lie above the extrapolation of the BH mass -- stellar mass relation, all of these runs are in good agreement with the BH mass -- stellar velocity dispersion relation. The same applies to the globular cluster data. This does not necessarily suggest an inconsistency, e.g. \citet{baldassare_populating_2020} find that for their observed dwarf sample deviations from the BH mass -- stellar velocity dispersion relation do not depend on the $M_\mathrm{BH}/M_\mathrm{stellar}$ ratio (with all of the dwarfs having similar stellar masses).

To investigate this more closely, we inspect the evolution of the simulation runs in stellar velocity dispersion -- stellar mass space. In addition to the dwarf and globular cluster data from \citet{greene_intermediate-mass_2020} and \citet{baldassare_populating_2020} we also show the Local Group dwarfs as collated by \citet{wheeler_no-spin_2017} which fall into the same mass range as our simulated system, with isolated dwarfs shown as black circles and satellite dwarfs as black diamonds. We checked how these observational datasets compare with the stellar velocity dispersions measured for the NSA dwarf galaxies, though given the instrumental resolution, velocity dispersion measurements below $70 \ \mathrm{km \ s^{-1}}$ are unreliable, preventing a direct comparison between our simulated systems and the NSA dwarfs. However, we find that where the SDSS dwarf measurements become more accurate they are in good agreement with the higher-resolution measurements from \citet{baldassare_populating_2020} (not shown here).

The stellar velocity dispersions of the simulated dwarfs (as well as the observed globular clusters) have systematic positive offsets from the mean of the observed dwarf galaxies but are still within the scatter. Though it should be noted that if the simulated data were shifted by $\sim 0.2$ dex towards the centre of the observed dwarf distribution then this would be sufficient for the simulated BHs to be overmassive with respect to the extrapolated BH mass -- stellar velocity dispersion relations. This indicates that the discrepancy between the two relations could be explained by the simulated velocity dispersions being slightly too high. Similarly, the globular clusters may have velocity dispersions corresponding to an overall more massive systems if they are the remnant cores of dwarf galaxies.

\begin{figure*}
    \centering
    \includegraphics[width=\textwidth]{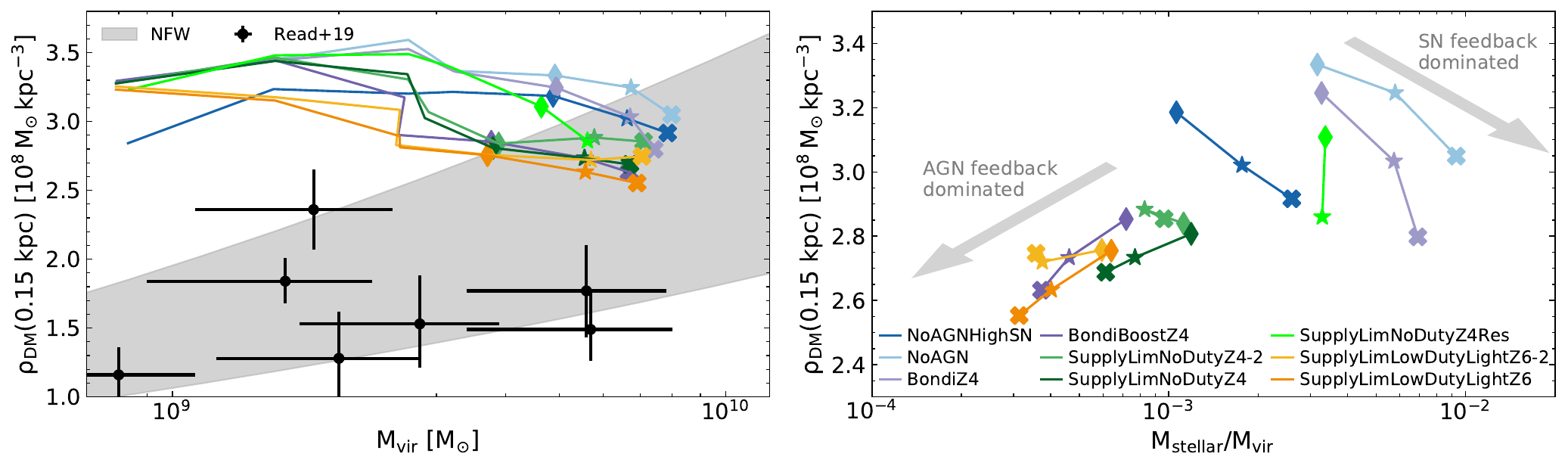}
    \caption{Central DM densities for the simulated dwarfs with observed data from \citet{read_dark_2019} plotted for comparison. We show the binned redshift evolution of the main dwarf zoom-in simulations, with the $z=2$, $z=1$ and $z=0$ data points highlighted by diamonds, stars and crosses, respectively. \textit{Left panel:} Central DM densities from $z=6$~--~$0$ as a function of virial mass. The grey-shaded band shows the central densities expected for an NFW profile, with the band width corresponding to a $1\sigma$ scatter in halo concentration. \textit{Right panel:} Central DM densities from $z=2$~--~$0$ as function of stellar mass to virial mass ratio. Here we restrict the dynamical range of the vertical axis to the densities covered by the simulated data to show the small yet systematic trend towards lower central densities for lower stellar mass to virial mass ratios for the AGN feedback dominated set-ups.}
    \label{fig:DMProperties}
\end{figure*}

To constrain the role of AGN feedback within dwarfs it will be imperative to extend the BH -- galaxy scaling relations to lower masses. This will provide critical insights into the efficiency of BH feedback across the galaxy population as, within our model, only overmassive AGN are able to regulate star formation in dwarfs. If these overmassive BHs are not detected, this may hint that BHs could only play a minor role in the evolution of dwarf galaxies.

\subsection{DM properties} \label{subsec:dm_props}

The `cusp versus core problem' remains a hotly contested issue, with the inferred DM halo profiles of some observed dwarfs seemingly in disagreement with the cuspy DM profile predicted by the $\Lambda$CDM model \citep[e.g.][]{flores_observational_1994,moore_evidence_1994}. Various explanations have been offered including observational measurement uncertainties and/or biases, modifications to the DM model as well as baryonic processes. The latter has been mostly focused on stellar feedback, though the significant impact of AGN feedback that we find in our models (once BH accretion is allowed to be efficient) raises the question of whether AGN feedback could contribute to core transformations in dwarfs as well \citep[also see][]{silk_feedback_2017}.

Central DM densities can be used as a key diagnostic for core formation, see e.g. discussion in \citet{read_dark_2019}. In Fig.~\ref{fig:DMProperties}, we plot the central densities as a function of virial mass (left-hand panel) and as function of stellar mass to virial mass ratio (right-hand panel). The grey band in the left-hand panel assumes no cusp-core transformation, classic NFW profile \citep{navarro_universal_1997}, and the width of the band corresponds to $1 \sigma$ scatter in halo concentration ($\Delta \log(c) =0.1$). For comparison, we also show the estimated central densities and virial masses of some observed cuspy dwarf galaxies from \citet{read_dark_2019}.

We show the binned redshift evolution (bin midpoints at $z=6,5,4,3,2,1,0$) of our zoom-in simulations, with redshifts $z=2$, $z=1$ and $z=0$ marked by diamonds, stars and crosses, respectively. All of the simulation are in agreement with the predictions from the NFW profile at $z=0$, i.e. all simulations result in cuspy DM haloes. The central DM densities are not significantly affected by the variations in galactic feedback implementations trialled here -- however, there is a trend towards lower central densities with additional AGN feedback\footnote{We caution that at $r=150$~pc, our simulations are at the very edge of the resolution limit, barely exceeding the gravitational softening length for our fiducial resolution. However, we have verified that our simulations are also in good agreement with the NFW expectations at $r=500$~pc, and that there is good agreement between NFW's cuspy profile and the very-high-resolution \EddNoDutyFromZFourRes \ simulation.}.

We examine this trend more closely in the right-hand panel, where we plot the central densities as a function of stellar mass to virial mass ratio for $z=2$ to $z=0$. This plot demonstrates how more severe AGN feedback suppression of star formation (i.e. a lower stellar mass to virial mass ratio) is associated with lower central DM densities. Indeed, simulations with a sustained, efficient AGN (i.e. AGN feedback dominated set-ups) move towards lower central densities as their $M_\mathrm{stellar}/M_\mathrm{vir}$ ratio decreases. The simulations where the AGN is deactivated at $z=2$ (\EddNoDutyFromZFourToZTwo \ and \EddLowDutyLightFromZSixToZTwo) do not show any strong trends with $M_\mathrm{stellar}/M_\mathrm{vir}$ ratio. SN-only simulations (see \NoAGNHighSN \ and \NoAGN) or set-ups that only have a weak AGN (see \BondiFidDutyFromZFour) display the opposite behaviour, i.e. decreased central DM densities are associated with higher $M_\mathrm{stellar}/M_\mathrm{vir}$ ratios. The milder \EddNoDutyFromZFourRes \ is an interesting in between case, maintaining a near constant $M_\mathrm{stellar}/M_\mathrm{vir}$ ratio whilst the DM density drops between $z=2$ and $z=1$.

AGN feedback dominated dwarf simulations therefore interestingly show the opposite behaviour to the trend predicted for stellar-feedback-driven cusp-core transformation, where a higher stellar mass to halo mass ratio implies more star formation and therefore larger SN-driven feedback activity which results in lower central densities \citep[e.g.][]{read_dark_2019}. The difference between the AGN and no-AGN runs is more pronounced than between the different strengths of SN feedback. In particular, the runs with early AGN activity have notably lower densities.

In summary, with our set-ups the AGN feedback cannot dynamically heat the DM to transform the central region into a core, however, there is a small yet systematic effect where central densities are lowered with AGN feedback (and therefore lower stellar mass to halo mass ratios, as opposed to stellar feedback driven density suppression). As a final caveat, note that we find that our dwarfs do not have strong rotational support ($v_\mathrm{rot}/\sigma \sim 1$) so these systems are not directly analogous to the observed dwarfs presented in \citet{read_dark_2019}.

\subsection{Further comparisons with observations} \label{subsec:comp_obs}
In this section, we compare our dwarf simulations with some of the observed dwarf galaxies hosting AGN.

Particular interest has arisen from HI observations which found that dwarf galaxies with AGN signatures are associated with lower galaxy HI masses \citep{bradford_effect_2018,ellison_atomic_2019,guo_cold_2022}. With upcoming radio telescopes, such as SKA, set to map HI gas at much higher resolution and to higher redshifts, this is the ideal time to make predictions for the HI mass evolution of dwarfs (see Section~\ref{subsubsec:HImasses}).

Furthermore, X-ray surveys have detected hints of AGN activity down to the classical dwarf regime of $M_\mathrm{stellar} \sim 10^{7} \ \Msun$ \citep[e.g.][]{birchall_x-ray_2020} and we compare our simulated AGN luminosities to these local observations as well as make predictions for the high-redshift regime in Section~\ref{subsubsec:XrayLum}.

\subsubsection{HI gas masses} \label{subsubsec:HImasses}

\begin{figure}
    \centering
    \includegraphics[width=\columnwidth]{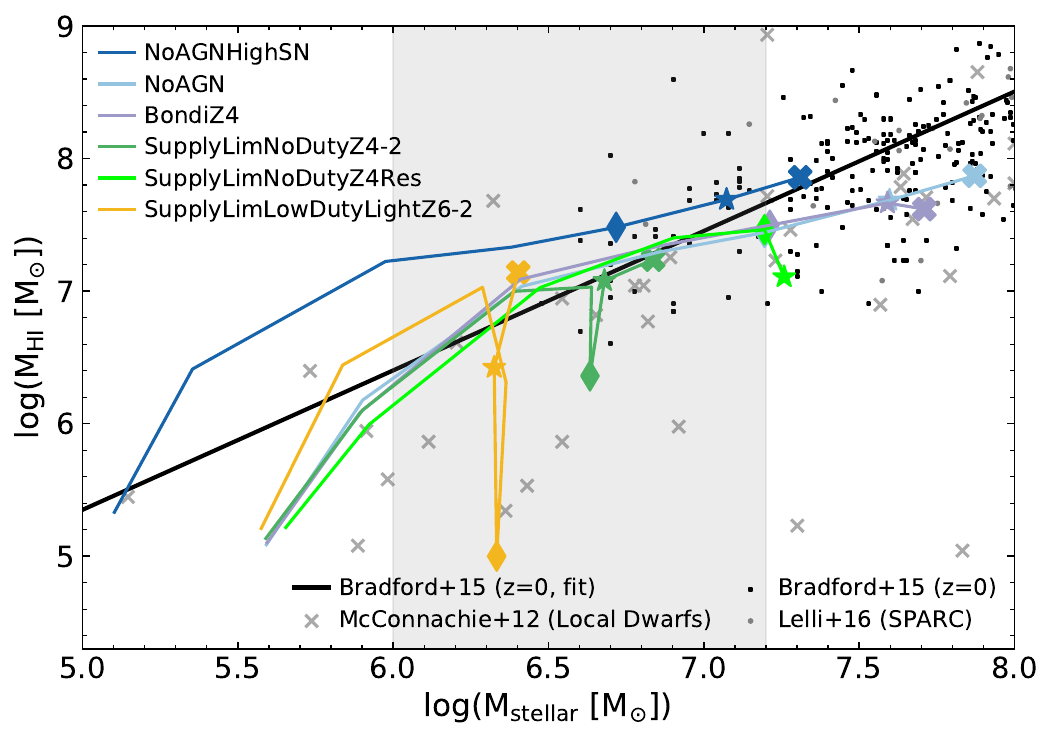}
    \caption{HI mass against stellar mass. We show the redshift evolution of the main dwarf zoom-in simulations that result in gaseous dwarfs from $z=6$~--~$0$, with the $z=2$, $z=1$ and $z=0$ data points highlighted by diamonds, stars and crosses, respectively. For comparison, we plot the $M_\mathrm{HI}$ -- $M_\mathrm{stellar}$ relation from \citet{bradford_study_2015} for isolated low-mass galaxies. We also plot the HI data used to derive this relation to indicate the scatter. Furthermore, we show the HI masses of the SPARC galaxies \citep{lelli_sparc_2016} and the local dwarfs \citet{mcconnachie_observed_2012}. The grey-shaded area indicates the stellar masses that would be expected for our simulated dwarf from the SMHM relations at $z=0$.}
    \label{fig:HIMassStellarMass}
\end{figure}

We compare the HI gas masses of our simulated dwarfs with observational constraints to assess whether our simulations would result in realistic gas properties. Note that, as in Illustris and \fable, the self-shielding of the non-star-forming gas\footnote{The UV background radiation field is not considered for the star-forming gas which is placed on the eEOS.} from the UV background is computed on-the-fly as a function of cosmological epoch and gas density following \citep{rahmati_evolution_2013}, see \citet{vogelsberger_model_2013} for details. We calculate the atomic fraction of the neutral hydrogen gas in the simulation in post-processing following \citet{bird_damped_2014} which is based on \citet{blitz_role_2006} and \citet{altay_through_2011} and then sum up the HI gas mass within twice the stellar half mass radius to obtain $M_\mathrm{HI}$. 

Fig.~\ref{fig:HIMassStellarMass} shows galaxy HI mass plotted against galaxy stellar mass. The colour-coded lines show the binned redshift evolution of the main zoom-in simulations (bin midpoints at $z=6,5,4,3,2,1,0$), with the $z=2$, $z=1$ and $z=0$ data points highlighted by diamonds, stars and crosses, respectively. We also show the HI mass evolution of the \EddNoDutyFromZFourRes \ simulation run (note that this high-resolution simulation was only performed until $z=1$). For three of the main zoom-in simulations (\BondiBoostFidDutyFromZFour, \EddNoDutyFromZFour, and \EddLowDutyLightFromZSix) the HI reservoir is completely depleted from $z\sim 3$, so we do not plot these runs here as all of the observational HI dwarf data is local. 

Two of the simulation runs which experience strong feedback at very high redshifts of $z > 4$ (\NoAGNHighSN \ and \EddLowDutyLightFromZSixToZTwo) initially have larger HI masses than the runs with weaker feedback, since the feedback leads to a less concentrated stellar distribution and hence larger half mass radii (not shown here). However, for $z < 4$, the half mass radii of the different runs converge again and the positive offset of the \NoAGNHighSN \ run is largely driven by the lower stellar masses.

For comparison, we also show the $M_\mathrm{HI}$ -- $M_\mathrm{stellar}$ relation from \citet{bradford_study_2015} for isolated low-mass ($M_\mathrm{stellar} < 10^{8.6} \ \Msun$) galaxies as well as the data points from their observational sample to indicate the scatter. For reference, the HI data from the SPARC sample \citep{lelli_sparc_2016} and the Local Dwarfs \citep{mcconnachie_observed_2012} is plotted as well.

\EddNoDutyFromZFourToZTwo \ and \NoAGNHighSN \ are in excellent agreement with the predicted relation, however, all other gaseous dwarfs are also within the scatter of the observed data. For the gaseous dwarfs, no simulations end up further than one dex away from the \citet{bradford_study_2015} relation, i.e. none of these would be categorised as HI poor. Indeed the offsets from the relation are largely driven by insufficient (e.g. \EddLowDutyLightFromZSixToZTwo) or excessive stellar mass growth (\NoAGN \ or \BondiFidDutyFromZFour). Though for the case of the mild AGN feedback in the high-resolution \EddNoDutyFromZFourRes \ simulation there is a clear vertical offset due to gas depletion that could become more significant at low redshifts.

We note that in observations, isolated dwarf galaxies without HI detections are rare, e.g. in the \citet{bradford_study_2015} study only 7 out of 144 isolated dwarfs are non-detections in HI, some of which would have been affected by insufficient signal-to-noise rather than being genuinely completely HI-deficient. This indicates that isolated dwarfs which are regulated by strongly ejective feedback are likely uncommon -- unless the feedback source only operates at higher redshift in which case the gas reservoir may recover so that the dwarf galaxies again fall on the observed relation as is the case for \EddLowDutyLightFromZSixToZTwo \ and \EddNoDutyFromZFourToZTwo.

Overall, we find that in our simulations gentle AGN activity does not result in significantly different local HI gas masses than the equivalent no-AGN run (see e.g. \NoAGN \ and \BondiFidDutyFromZFour), which could make it difficult to define a clear link between AGN and gas depletion. Indeed \citet{davis_radio_2022} find no significant difference in gas fractions for AGN versus no-AGN detections in nearby ($z<0.5$) observed dwarf galaxies.

However, other studies find that there is an apparent link between AGN activity and gas depletion in dwarf galaxies \citep{bradford_effect_2018,ellison_atomic_2019}. Though we note that because of either flux limits or the limited validity of the BPT diagram for identifying AGN in low-mass dwarfs \citep[e.g.][]{cann_limitations_2019}, these studies are currently restricted to the LMC-mass dwarf regime ($9.0 \lesssim \log(M_\mathrm{stellar}) \lesssim 9.5$), whilst no systematic studies of the impact of AGN activity on the neutral gas content have been carried out in the low-mass dwarf regime. Indeed the study by \citet{ellison_atomic_2019} focuses on HI stacks from the ALFALFA 100\% data release with the lowest mass bin ($9.0 < \log(M_\mathrm{stellar}) < 9.6$) containing 50 AGN galaxies (and 9951 non-AGN galaxies, allowing for efficient matching of the comparison sample in terms of stellar mass and SFR), which have on average significantly lower HI masses than their no-AGN counterparts. The \citet{bradford_effect_2018} study supplements the data from the ALFALFA 70\% data release with deeper HI data of dwarf galaxies with stellar masses $7.0 < \log(M_\mathrm{stellar}) < 9.5$, allowing them to focus on individual galaxies in the dwarf regime rather than comparing stacks. They find nine gas-depleted galaxies in their isolated sample with intermediate to large BPT distances, however, these galaxies are all in the mass range $9.2 < \log(M_\mathrm{stellar}) < 9.5$, which can at least partly be attributed to the issues with standard optical emission line ratios for identifying AGN in lower-mass dwarfs. Furthermore, they also note that due to the bias towards gas-rich galaxies in flux-limited surveys of HI emission, low-mass galaxies that have been significantly depleted of their HI gas may be missed by these surveys \citep[also see][]{koribalski_1000_2004,giovanelli_arecibo_2005}. Interestingly though \citet{bradford_effect_2018} find that the gas-depleted, low-mass, isolated galaxies with large BPT distances are red and compact with old stellar populations and no distinguishable signs of star formation -- inconsistent with the general population of isolated galaxies in the same stellar mass range.

Forthcoming surveys with LSST and SKA will significantly improve upon our knowledge of the AGN -- neutral gas connection in dwarf galaxies. Promisingly, LSST will also be able to identify dwarf AGN via their optical variability allowing for a BPT-independent method that can be extended down to low-mass dwarfs \citep[e.g.][]{baldassare_identifying_2018,davis_radio_2022}.

\subsubsection{X-ray luminosities} \label{subsubsec:XrayLum}

\begin{figure}
    \centering
    \includegraphics[width=\columnwidth]{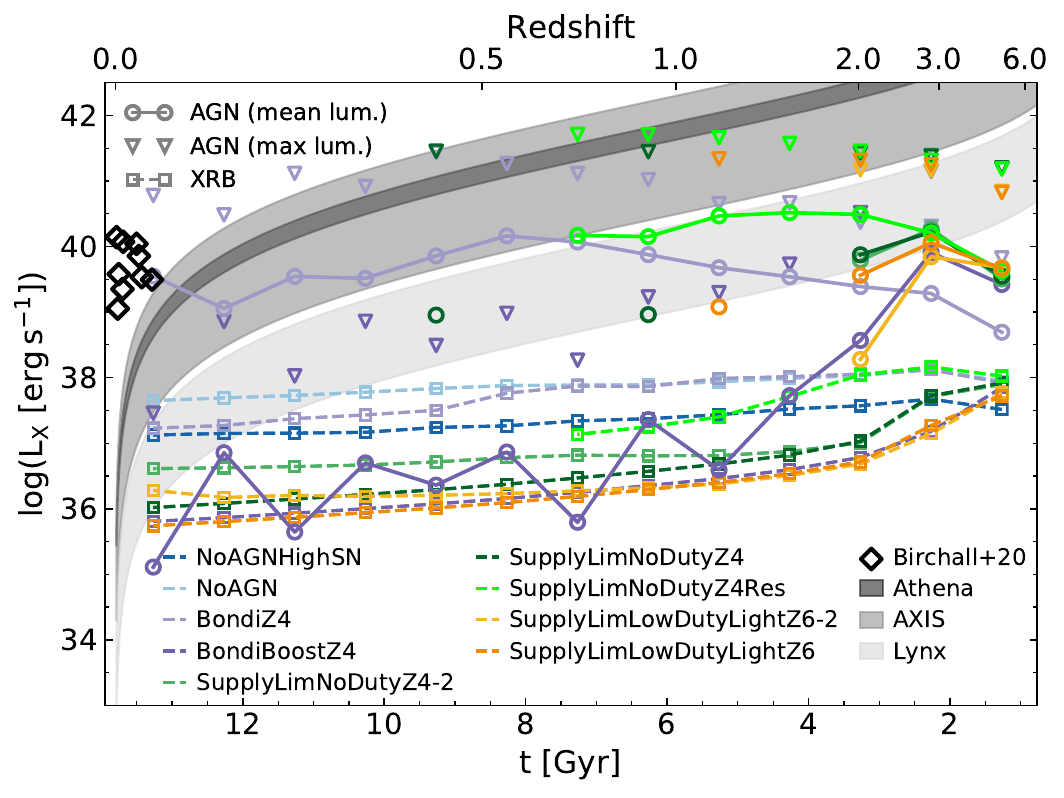}
    \caption{Evolution of X-ray luminosities as a function of cosmic time binned over $\Delta t = 1$~Gyr, with AGN luminosities shown as solid lines (mean bin values as circles, maximum bin values as triangles) and XRB luminosities, estimated based on observed scaling relations for the stellar mass and SFR, shown as dashed lines (mean bin values as squares). For comparison the observed dwarf galaxies from \citet{birchall_x-ray_2020} are also shown as black diamonds. Efficiently accreting AGN in dwarf galaxies are mostly restricted to high redshifts ($z \geq 2$) and should be easily distinguishable from XRBs in this regime, making them promising targets for future X-ray missions such as \textit{Athena}, \textit{Lynx} or \textit{AXIS}, as indicated by the grey shaded bands. The fiducial Bondi run has moderately high AGN luminosities throughout, matching the dwarfs observed by \citet{birchall_x-ray_2020} at $z \sim 0$.}
    \label{fig:XrayLum}
\end{figure}

In this section, we assess the (rest-frame) X-ray luminosities of our simulated dwarf galaxies and whether the AGN activity would be detectable by current and/or future surveys. 

We calculate the X-ray AGN luminosity as $L_\mathrm{X,AGN} = \kappa_\mathrm{X} \epsilon_\mathrm{r} \dot{M}_\mathrm{acc} \mathrm{c}^{2}$, where $\kappa_\mathrm{X}$ is the X-ray bolometric correction from \citet{shen_bolometric_2020}, $\epsilon_\mathrm{r}=0.1$ is the assumed radiative efficiency, $\dot{M}_\mathrm{acc}$ is the accretion rate and $\mathrm{c}$ is the speed of light. Here, we focus on the hard X-ray luminosity band ($2$~--~$10$~keV) as this is most promising for AGN detection due to less contamination than in the soft X-ray band.

To estimate the X-ray luminosities of X-ray binaries (XRBs), we use the observational scaling relations from \citet{lehmer_evolution_2016} for the low-mass and high-mass XRB components based on the galactic SFR and stellar mass, respectively. This closely follows the assessment criteria used in observational dwarf galaxy studies where these scaling relations are employed to establish whether the observed X-ray luminosity is significantly above the expected XRB luminosity. 

Fig.~\ref{fig:XrayLum} shows the redshift evolution of the BH X-ray luminosity for the main AGN simulations (solid lines) and the XRB luminosity (dashed lines) for all main simulations as well as for the high-resolution simulation \EddNoDutyFromZFourRes, which was performed until $z \sim 1$. These lines are based on the mean binned luminosity values (bin width $\Delta t = 1$~Gyr), with the mean BH luminosity values marked by circles and squares for AGN and XRB luminosities, respectively. Note that the AGN luminosities display large variability -- to indicate the peak luminosities that may be observed with these set-ups, we plot the maximum BH X-ray luminosity value in each bin as triangles (observations have a bias towards observing peak AGN luminosities). We also show the observed X-ray luminosities of the dwarf galaxies from the \citet{birchall_x-ray_2020} sample (only selecting the dwarf galaxies with $M_\mathrm{stellar} < 10^{8} \ \Msun$ to obtain a comparable mass range). Finally, we indicate the minimum observable luminosities based on the sensitivities of future X-ray telescopes (\textit{Athena}, \textit{AXIS} and \textit{Lynx}) as grey-shaded bands. The lower limit and upper limit of each band are set by the sensitivities for deep and wide surveys, respectively \citep[see][and references therein]{habouzit_supermassive_2022}.

At high redshifts ($z \gtrsim 2$), the AGN luminosities clearly exceed the XRB luminosities in all cases. At low redshifts, the AGN activity is significantly diminished for the majority of our simulations. The supply-limited-accretion runs (where the AGN has not been switched off by construction for $z<2$) only experience one to two high-luminosity bursts at low redshifts which are still clearly distinguishable from XRBs. The \BondiBoostFidDutyFromZFour \ run only has very low activity levels, with the mean luminosities from $z\sim 1$ at a similar level as the XRB population.

Overall, for our dwarf zoom-ins, the simulation set-ups that result in efficient AGN feedback have easily distinguishable high-luminosity AGN and significant levels of star formation at high redshifts, whilst at low redshifts these set-ups are associated with low SFRs and dormant AGN that would be difficult to detect either due to only very short duration bursts or low AGN luminosity levels.

This is in good general agreement with \citet{schirra_bringing_2021} who recently compiled the properties of the faint AGN and XRB populations in several large-scale cosmological simulations. Whilst the simulations that they consider have significantly lower resolution and are therefore restricted to more massive dwarf galaxies, it is still interesting to note that for dwarf galaxies at $z=0$ with low specific SFRs, the XRB emission always outshines AGN emission in all of these simulations. For Illustris and EAGLE this also applies to main-sequence galaxies and starbursts, whilst in TNG100 \citep[which has a significantly higher fraction of luminous dwarf AGN than Illustris or EAGLE, see e.g.][]{haidar_black_2022}, the AGN contribution tends to dominate in star-forming galaxies where AGN activity is also enhanced.

Whilst our high-luminosity dwarf AGN broadly follow these trends from TNG100, the exception to this is the AGN in the high-resolution \EddNoDutyFromZFourRes \ simulation which due to its more gentle impact is able to accrete continuously, with similar luminosities to the observed (local) dwarf AGN from \citet{birchall_x-ray_2020} even as star formation is significantly suppressed. The \BondiFidDutyFromZFour \ run also has relatively steady AGN luminosity levels throughout the simulation, with a slight increase in luminosity until $z\sim 0.5$. The mean $z=0$ luminosity falls well within the observed data from \citet{birchall_x-ray_2020}. 

This indicates that high-redshift ($z \geq 2$) dwarf AGN may be promising targets for X-ray searches, with e.g. \textit{Athena}, \textit{Lynx} or \textit{AXIS}. All of which would be able to observe the peak luminosity AGN bursts of (some of) the efficient AGN set-ups at high redshift. A deep X-ray survey with \textit{Lynx} may even be able to map low-level AGN activity out to intermediate redshifts. The Bondi set-up with the fiducial \fable \ parameters (\BondiFidDutyFromZFour) matches the observed dwarf AGN at low redshifts, however, the minimal impact of the AGN on star formation in this simulation demonstrates that X-ray detections of AGN in dwarfs do not necessarily imply that these AGN play an important role in star formation regulation in their dwarf hosts.

\section{Discussion} \label{sec:discussion}

\subsection{Comparison with previous theoretical work}

In recent years, there have been several theoretical investigations considering AGN feedback in dwarf galaxies within a cosmological environment -- so far yielding mixed results. Some find that accretion rates of AGN in dwarfs are generally low \citep[e.g.][]{bellovary_multimessenger_2019,trebitsch_escape_2018}, making electromagnetic detections of dwarf AGN extremely difficult, especially for off-centre BHs \citep[e.g.][]{bellovary_origins_2021,sharma_hidden_2022}. However, if AGN feedback is efficient, it can have a significant impact on its dwarf host in simulations, especially at high redshifts \citep[e.g.][]{barai_intermediate-mass_2019,sharma_black_2020,koudmani_little_2021}.

This diverse range of outcomes is also highlighted by \citet{haidar_black_2022} who compare the AGN fractions in dwarfs for several large-scale cosmological simulation projects and find that these span two orders of magnitude at fixed stellar mass. The differences between simulations are mainly driven by varying choices in the subgrid modelling of BH physics, in particular BH seed masses and BH accretion prescriptions. For example, some simulations, such as IllustrisTNG, power too many AGN in low-mass galaxies mostly due to too massive BH seeds. Similarly \citet{sharma_hidden_2022} find that the Romulus25 simulation (which also employs a relatively high seed mass of $10^{6} \ \Msun$) produces too many luminous AGN in dwarfs compared to X-ray constraints. Furthermore, they find that Romulus25 produces a high number of quenched isolated dwarf galaxies in tension with observational constraints \citep[][]{geha_stellar_2012}.

This demonstrates that reproducing the observed AGN population in dwarf galaxies whilst matching other constraints such as the SMHM relation is still very difficult for simulations -- compounded by observational uncertainties. Indeed, \citet{wellons_exploring_2022} recently investigated a plethora of AGN subgrid models and parametrizations in the FIRE cosmological zoom-in simulations and found that they need to employ a subgrid prescription for stellar feedback regulation of low-mass BH growth to avoid overquenching of dwarfs whilst still efficiently quenching the massive galaxy population.

Beyond models for stellar and BH physics, environment also shapes the dwarf AGN population in cosmological simulations \citep[e.g.][]{kristensen_merger_2021}. In this study, we only focus on one dwarf galaxy which lives in a relatively low-density environment and investigate the impact of varying the AGN subgrid models and stellar feedback parameters. In the future, it will also be critical to assess the role of the cosmological environment and merger histories. The role of merger histories is directly connected to the hotly debated topic of wandering BHs with various observational hints at off-centre AGN in dwarfs \citep[e.g.][]{reines_new_2020,mezcua_hidden_2020}. According to theoretical predictions, these off-centre AGN are driven by perturbations such as mergers \citep[e.g.][]{bellovary_origins_2021} and hence the merger history has a crucial impact on the efficiency of dwarf AGN.  

\subsection{BH model}

In our simulations the BH remains in the central region, as would be expected for a dwarf in a relatively quiet environment, due to the sufficiently high dynamical mass ($\geq 10^{4} \ \Msun$) and high resolution ($m_\mathrm{dm}=1536 \ \Msun$ and $\bar{m}_\mathrm{gas} = 287 \ \Msun$ for fiducial resolution and $m_\mathrm{dm}=192 \ \Msun$ and $\bar{m}_\mathrm{gas} = 35.9 \ \Msun$ for very high resolution). However, in high-density environments, we would need to include a formalism to properly model off-centre BHs and their decay towards the centre due to dynamical friction. Furthermore, independent of the cosmological environment, tracking the BH dynamics may be important in the presence of a clumpy ISM where the dynamical centre of the galaxy is ill-defined and BHs may not sink to the central region \citep[see e.g.][]{pfister_erratic_2019,ma_seeds_2021}.

Accurately modelling massive BH orbits for the whole galaxy population will be a key step towards more predictive galaxy formation simulations \citep[e.g.][]{bahe_importance_2022}, and in particular for dwarfs this could be a crucial component in the models for reproducing the diverse outcomes seen in simulations, with a small fraction of dwarfs in the local Universe hosting luminous AGN whilst the majority of dwarf galaxies only have low-luminosity or no detectable AGN activity \citep[also see][]{sharma_hidden_2022}. In this sense, our central dwarf AGN may represent an upper limit for AGN activity in dwarfs.

In addition to more sophisticated models for BH dynamics, it will also be important to investigate more realistic BH accretion models. Here, we increased the boost factor $\alpha$ in the Bondi prescription as a simple method for obtaining high BH accretion rates in dwarfs. Note that \citet{booth_cosmological_2009} find that for constant-$\alpha$ models the value of $\alpha$ effectively controls the ambient gas density at which the BH accretion rate becomes Eddington-limited -- and with more massive haloes reaching higher gas densities, this then sets the halo mass at which AGN feedback becomes effective. They conclude that rather than setting a constant $\alpha$ boost factor, $\alpha$ should be varied with the density as otherwise BH accretion rates (and consequently star formation suppression) may be overestimated in low-density gas. However, we emphasise here that our numerical experiments with increased $\alpha$ values are not be too regarded as within the Bondi framework but simply as a test of how AGN activity could affect dwarf galaxies if BH accretion were to be efficient (with the general Bondi dependencies on density and sound speed still providing some level of self-regulation).

We note that alternative accretion models based on gravitational torque-driven BH growth \citep[e.g.][]{hopkins_analytic_2011,angles-alcazar_black_2013,angles-alcazar_gravitational_2017} do not suffer from the suppression of low-mass BH growth found for the Bondi model, with simulations that have implemented these torque-driven models producing bright AGN in low-mass galaxies which can even exceed the Eddington limit \citep[e.g.][]{angles-alcazar_black_2017,dave_simba_2019,thomas_black_2019}.

We also explored supply-limited BH accretion schemes and found that for extremely high gas resolution ($m_\mathrm{gas} = 35.9 \ \Msun$), this scheme can provide efficient yet moderate AGN feedback that suppresses star formation without quenching the system. Whilst this is beyond the scope of this paper, in future work we plan to investigate more sophisticated accretion models taking advantage of the super-Lagrangian refinement technique \citep[e.g.][]{curtis_resolving_2015,beckmann_bondi_2018,angles-alcazar_cosmological_2021} and accretion disc based subgrid models \citep[e.g.][]{fiacconi_galactic_2018,cenci_black_2021}.

Disc-based models for BH accretion will also allow for the self-consistent modelling of AGN feedback phenomena such as disc winds \citep[e.g.][]{sala_non-isotropic_2021} or jets \citep[e.g.][]{talbot_blandford-znajek_2021}. In this study, we only include purely thermal, isotropic AGN feedback to reduce the number of free parameters and avoid further complicating the set-up. The exploration of more sophisticated AGN models is left for future work, though we note that employing different AGN feedback prescriptions may significantly impact our results. In particular, alternative injection geometries such as bipolar winds \citep[e.g.][]{curtis_resolving_2015,koudmani_fast_2019,wellons_exploring_2022} could allow for a dense gas component to survive in the central region and hence power more efficient BH growth.

Finally, TDEs could play a significant role in feeding AGN in dwarf galaxies \citep[e.g.][]{zubovas_tidal_2019} and it will be important to include this as a BH growth channel for future models.

Overall, we note that the modelling of AGN accretion and feedback in dwarf galaxies is still in its infancy and it will be crucial to investigate more sophisticated theoretical models that can reproduce observational constraints across the whole galaxy population. Indeed, one cannot conclusively determine whether AGN feedback can replace efficient stellar feedback in dwarfs until the same galaxy formation model is successfully tested against observations in higher-mass galaxies \citep[see e.g.][]{wellons_exploring_2022}.

\subsection{ISM model} \label{subsec:DiscuISMModel}

In addition to refining the modelling of BH physics, it will also be very important to improve the modelling of the ISM to accurately estimate AGN activity and its impact. In particular, capturing the cold, dense phase of the ISM could boost BH accretion rates \citep[provided that the BH can accrete these cold gas clumps rather than residing in a hot cavity, with a resolved multi-phase ISM leading to highly variable accretion rates, see e.g.][]{gabor_simulations_2013,angles-alcazar_cosmological_2021,sivasankaran_simulations_2022} and enhance the impact of AGN feedback as the gas is very likely much ``puffier'' with the eEOS model from \citet{springel_cosmological_2003}. On the other hand, multi-phase ISM models would also likely lead to more spatially concentrated star formation and stellar feedback which could be more efficient at ejecting the gas from the central region \citep[e.g.][]{angles-alcazar_black_2017,habouzit_blossoms_2017,trebitsch_escape_2018,hopkins_why_2022}. So whilst the explicit modelling of the cold gas phase may enhance AGN activity (and its variability), the inclusion of localised stellar feedback could suppress BH accretion rates. Hence the effect of moving towards more sophisticated star formation and stellar feedback models is not obvious and has to be explored.

The AGN efficiency may also be hampered by a resolved multiphase ISM, with dense gas clumps potentially resistant to AGN feedback and the hot gas escaping along the polar regions without significantly affecting star formation in the galaxy \citep[e.g.][]{gabor_active_2014,bourne_resolution_2015,koudmani_fast_2019,torrey_impact_2020}.

Localised stellar feedback coupled with a multi-phase ISM may also significantly alter the properties of SN-driven outflows. Indeed, the differences between the SN- and AGN-driven outflow are likely amplified by the Illustris/\fable \ galactic wind model. Even with \fable's modifications to the thermal content of wind particles, galactic winds produced by this scheme are still comparatively cold as the hydro-decoupled wind scheme does not allow for the simultaneous generation of a hot, fast phase with low mass loading and a cold, slow, high mass-loaded phase. In principle, the former phase could be accelerating out of the halo while the latter is a fountain flow, all in the same galaxy, which is not possible with the galactic wind model employed here.

Though we note that even with the explicit modelling of multi-phase outflows, we would still expect to see a significant difference in outflow properties between SNe and AGN along the lines demonstrated in this work. Indeed in previous work \citep[][]{koudmani_fast_2019}, we compared the properties of SN- and AGN-driven outflows using the multi-phase ISM model from \citet{smith_supernova_2018} in isolated dwarf galaxy simulations, and found that AGN-boosted outflows reach significantly higher temperatures and velocities than their SN-only counterparts.

Beyond outflow and star formation properties, the explicit modelling of a cold and hot phase within the ISM and localised stellar feedback could also affect our results with respect to cusp-to-core transformations, in particular with bursty star formation likely contributing to core formation \citep[e.g.][]{pontzen_how_2012}.

In the future, we will aim to model the interplay between SNe and AGN in dwarfs more accurately by employing a multi-phase ISM model that injects energy and momentum in accordance with the stage of the SN remnant evolution resolved \citep[e.g.][]{kimm_feedback-regulated_2017,hopkins_how_2018,smith_supernova_2018,marinacci_simulating_2019,gutcke_lyra_2021}. This would also potentially reduce the number of free parameters in the star formation and stellar feedback model.

Lastly, it will also be important to investigate non-thermal components of the ISM that could play a role for AGN in dwarfs such as magnetic fields or cosmic rays \citep[e.g.][]{uhlig_galactic_2012,booth_simulations_2013,hanasz_cosmic_2013,salem_cosmic_2014,pakmor_galactic_2016,farber_impact_2018,holguin_role_2019,dashyan_cosmic_2020,martin-alvarez_how_2020}. In particular, \citet{wellons_exploring_2022} find cosmic rays to be a crucial ingredient for efficient AGN feedback.

\subsection{Observational context}
Finally, we discuss our results in the context of observational constraints on the quenched fraction of field dwarf galaxies and how future observations of AGN in dwarfs may break the present degeneracies between simulations to guide the modelling of AGN feedback across the whole galaxy population.

\subsubsection{Quiescent dwarf galaxies in the field}

In our dwarf simulations, we find that efficient AGN can easily quench the system. However, with our dwarf galaxy living in a low-density environment, this system would generally be expected to be star-forming at $z=0$ \citep[e.g. see][who probe this with SDSS dwarf galaxies in the field]{geha_stellar_2012}. Whilst SDSS is incomplete for low surface brightness galaxies and the \citet{geha_stellar_2012} study can only provide upper limits on the quenched fraction below stellar masses of  $10^{9} \ \Msun$ (with quenched field dwarfs below $10^{9} \ \Msun$ not detected by SDSS), we would expect the trend of a decreasing quenched fraction of field galaxies towards lower galaxy masses to continue and we need to interpret our results in the context of these observations.

First of all, we emphasise again that our efficient dwarf set-ups, including the boosted Bondi runs and supply-limited accretion at the Eddington limit, are best-case scenarios for AGN accretion in dwarfs and likely \textit{overestimate} the typical impact AGN feedback would have on a dwarf system (for example an angular momentum barrier is not taken into account for the accretion schemes considered here). We also investigated more moderate set-ups, including restricting the AGN phase to high redshifts ($z \geq 2$) or increasing the resolution so that the BH is accreting from and injecting feedback into a much smaller region. Both of these more moderate set-ups lead to star formation suppression compared to the strong SN feedback equivalent set-up but do not quench the dwarf and manage to maintain a neutral gas reservoir.

Nevertheless, these results also demonstrate that rare conditions allowing for very efficient AGN accretion in dwarf galaxies could lead to quenching of the whole system. \citet{dickey_agn_2019} investigated quiescent dwarf galaxies for LMC-type dwarfs ($M_\mathrm{stellar}=10^{9.0 - 9.5} \ \Msun$). They found that 16 out of 20 quiescent dwarfs contain AGN-like line ratios, suggesting a possible link between AGN activity and self-quenching in dwarfs for this mass range.

For lower dwarf galaxy masses with $M_\mathrm{stellar} \lesssim 3 \times 10^{7} \ \Msun$ only very few examples of quenched field dwarfs are known, including the five Local Group dwarfs Cetus \citep{whiting_new_1999}, Tucana \citep{lavery_local_1990}, KKR25 \citep{makarov_unique_2012}, KKS3 \citep{karachentsev_new_2015}, and recently discovered Tucana B \citep{sand_tucana_2022}. Furthermore, \citet{polzin_recently_2021} reported an additional serendipitous discovery of a low-mass quenched dwarf galaxy COSMOS-dw1 in the COSMOS field. Despite their current isolated position, some of these dwarfs might have still been quenched by environmental effects \citep[also see][]{fillingham_environmental_2018}, e.g. Cetus and Tucana are believed to be splashback galaxies. Furthermore reionization suppression could have quenched dwarf galaxies such as Tucana B \citep{sand_tucana_2022}. COSMOS-dw1, KKR25, and KKS3 on the other hand all have complex stellar populations, indicating that star formation was suppressed and rejuvenated several times. This makes an environmental cause or reionization suppression unlikely and suggests that internal feedback could have quenched these dwarfs \citep[see discussion in][]{polzin_recently_2021}.

We note that from the theoretical side, overquenching at the low-mass end is a common problem in cosmological simulations \citep[e.g.][]{dickey_iq_2021}. Whilst general trends of red dwarf galaxies in group and cluster environments versus blue dwarf galaxies in the field are reproduced by the latest efforts \citep[see e.g.][]{joshi_cumulative_2021}, the quenched field fraction is still too high for dwarf galaxies in cosmological simulations.

To sum up, quenched field dwarf galaxies are rarely observed in the Local Universe, though the mass range we consider here ($M_\mathrm{stellar} \lesssim 3 \times 10^{7} \ \Msun$) remains underexplored. Efficient dwarf AGN should be able to quench systems in this mass range provided that the BH stays in the central region and is able to accrete the ambient gas near the Eddington limit (for limited periods of time). We note that in reality this should be a rare occurrence as we did not consider additional limiting factors such as an angular momentum barrier in this investigation so that the impact of the efficient AGN in our study should be considered as an upper limit.

\subsubsection{Gravitational waves and multi-messenger signatures}

LISA will be sensitive to the detection of massive BHs with masses in excess of a few thousand solar masses out to high redshifts \citep{amaro-seoane_astrophysics_2022} and therefore provide crucial and completely independent constraints on BH formation and BH growth in dwarfs via mergers across cosmic time. In the near future, LIGO and Virgo will provide complementary constraints on the low-mass end of the IMBH regime with total binary masses of $50$ -- $500 \ \Msun$ expected to be well within the constraining capabilities of the fourth observing run \citep[e.g.][]{mehta_observing_2022}. Moreover, atom interferometer observatories, such as the proposed experimental programme AION, will map gravitational waves in the mid-frequency range between LIGO and LISA \citep{badurina_aion_2020}.

As discussed in previous sections, our simulated dwarf galaxy is in a relatively low-density environment and for our redshift range of interest ($z \leq 6$), there is only one merger with stellar mass ratio $> 1/100$ which occurs at $z=4.0$. This merger has a stellar mass ratio of $0.02$ and a gas mass ratio of $0.04$ (with the mergee stellar and gas mass being $5 \times 10^{4} \ \Msun$ and $4 \times 10^{6} \ \Msun$, respectively). Note that in our simulations, we only seed one BH, into the central dwarf galaxy, by construction, so we do not model BH mergers. However, we can consider whether this merger would be detectable by LISA if the mergee also hosted a massive BH.

There are currently virtually no observational constraints on central BH masses in low-mass dwarfs (see Section~\ref{subsec:bh_assembly}), but we may take the extrapolated NSC mass -- stellar mass relation to obtain an upper limit for the mass of the central BH, which yields $M_\mathrm{BH} \sim 3 \times 10^{4} \ \Msun$. Extrapolating the BH mass -- stellar mass scaling relations is even trickier given that, independent of the seeding model assumed, we would expect BH -- galaxy scaling relations to flatten in this mass regime \citep[e.g.][]{greene_intermediate-mass_2020}, however, if we extrapolate the BH mass -- stellar mass scaling relation with the highest normalisation \citep[early-type relation from][]{greene_intermediate-mass_2020}, then we obtain $M_\mathrm{BH} \sim 10 \ \Msun$, giving an expected mass range for the secondary BH spanning more than three orders of magnitude. 

We note that the L3-proposed LISA configuration covers a consistent signal-to-noise parameter space in total binary mass for mass ratios between $\sim 0.05$ and 1 \citep[though there is an increase in signal strength for more equal mass binaries, see][for details]{kaiser_sensitivity_2021}. Therefore, we may employ the equal-mass LISA sensitivity curves for a secondary BH mass down to $\sim 500 \ \Msun$ (for an assumed primary mass of $10^{4} \ \Msun$ which matches the seed mass of the BHs seeded at $z_\mathrm{seed}=4$) or to  $\sim 100 \ \Msun$ (for an assumed primary mass of $2 \times 10^{3} \ \Msun$ which matches the BH mass at $z=4$ of the main simulations runs with $z_\mathrm{seed}=6$). For even lower mass ratios $\lesssim 10^{-2}$, we would be in the intermediate-mass ratio inspiral (IMRI) regime, where the waveforms are much harder to predict and LISA detectability would depend on the orbital parameters \citep[e.g.][]{amaro-seoane_detecting_2018}.

However, assuming that the secondary BH has a mass of at least $100$~--~$500 \ \Msun$, we can estimate the signal-to-noise ratio using the equal-mass LISA sensitivity curves for a total binary mass of $\sim 2 \times 10^{3}$~--~$10^{4} \ \Msun$ at $z=4$. With these mass ranges for the secondary BH mass and total binary mass, we would expect this merger event to be observable by LISA with an expected signal-to-noise ratio of $\gtrsim 10$ given the sensitivity of the L3-proposed LISA configuration \citep[e.g.][]{kaiser_sensitivity_2021}.

As a final caveat we note that, given the potentially long dynamical friction timescales for low-mass BHs, the dynamics of the secondary BH would be crucial in determining whether this putative BH -- BH merger could actually take place. Again this would also significantly depend on the BH mass of the secondary. In future work, we aim to explore the merger rates of BHs in dwarf galaxies for different seed masses more self-consistently using large cosmological volumes covering a range of environments.

In terms of electromagnetic counterparts to this possible BH merger, we can see in the right-hand panel of Fig.~\ref{fig:SfProperties} that shortly after $z=4$, the BH accretion rates of most\footnote{The \EddNoDutyFromZFourRes \ set-up also has high BH accretion rates at this point, yet does not reach its peak luminosity until $z \sim 1.5$, and overall displays much steadier accretion behaviour.} efficient AGN runs reach their peak value (even if seeded at $z=6$). This then also corresponds to a peak in X-ray luminosities as can be seen in Fig.~\ref{fig:XrayLum}. Excitingly, these bright X-ray AGN are above the \textit{Lynx} sensitivity limit, as discussed in Section~\ref{subsubsec:XrayLum}, so that an X-ray counterpart to this gravitational-wave event may be observable with future facilities.

\subsubsection{Future observational constraints}

In the near future, the Rubin Observatory should significant improve upon our knowledge of low surface brightness galaxies, especially in the dwarf regime \citep[e.g.][]{jackson_origin_2021,wright_formation_2021}. And in the long term, together with \textit{RST}, this will determine how common quenched field dwarfs are in this mass range. In addition to Rubin and \textit{RST} delivering valuable constraints on the low-surface brightness galaxies, it will also be crucial to map AGN in dwarf galaxies to lower luminosities and to higher redshifts with forthcoming X-ray telescopes such as \textit{Athena}, \textit{AXIS} and \textit{Lynx} and radio telescopes such as SKA playing a key role. \textit{RST} will also play an important role in pushing into the IMBH regime and identifying targets for dynamical BH mass measurements \citep[e.g.][]{greene_intermediate-mass_2020}. Rubin will also directly constrain the IMBH population by mapping TDE events \citep[e.g.][]{gezari_tidal_2021} and with AGN variability searches \citep[e.g.][]{baldassare_identifying_2018,ward_variability-selected_2021}.

Furthermore \textit{JWST} will be important for constraining star formation and feedback processes at the low-mass end both by mapping outflow properties with NIRSpec and by detecting faint emission lines with NIRISS to map (stochastic) star formation and extend the mass-metallicity relation towards lower masses and higher redshifts, see e.g. the WDEEP Survey using NIRCam and NIRISS to constrain feedback in low-mass galaxies from  $z \sim 1$~--~$12$ \citep{finkelstein_webb_2021} and the JADES survey using NIRCam and NIRSpec to discover and characterise the first galaxies as well as study the formation and evolution of galaxies more generally from $z \geq 12$ to $z \sim 2$ \citep{rieke_jwst_2019}. With our set-up we find that AGN feedback is more ejective than SN feedback and the main difference between the two feedback mechanisms lies in the outflow temperatures and velocities, which are significantly boosted by the AGN, and \textit{JWST} may be able to constrain these outflow features. 

Moreover, we find that overmassive BHs (compared to extrapolated scaling relations), low SFRs, and bright X-ray AGN at high redshifts are salient features of efficient AGN activity in the low-mass regime with our model, providing succinct predictions for future observational facilities.

\section{Conclusions} \label{sec:conclusion}

There is increasing observational evidence for AGN activity in dwarf galaxies, yet in most cosmological simulations AGN play a negligible role in dwarf galaxy evolution by construction. The reason for weak AGN feedback in the low-mass regime with these models is twofold, with both strong SN feedback and the suppression of low-mass BH growth in the Bondi rate contributing.

We have performed a series of high-resolution ($\bar{m}_\mathrm{gas} = 287 \ \Msun$) zoom-in simulations of a dwarf galaxy ($M_\mathrm{vir}(z=0) \sim 10^{10} \ \Msun$) in a relatively low-density environment, building on the \fable \ galaxy formation model \citep[see][]{henden_fable_2018}, but relaxing the assumption of strong SN feedback in low-mass galaxies. The goal is to determine whether more moderate SN feedback combined with an efficient AGN could have the same success in regulating star formation at the low-mass end as the strong SN feedback set-up. 

To reach the efficient AGN regime we used two strategies. Firstly, we increased the boost factor in the Bondi rate to compensate for the $\dot{M}_\mathrm{Bondi} \propto M_\mathrm{BH}^{2}$ dependency suppressing the growth of low-mass BHs. These simulations (labelled as \textit{BondiBoost}) should then be regarded as numerical experiments rather than within the `theoretical' Bondi framework. Secondly, we implemented a gas supply limited BH accretion scheme (whereby the BH accretes at the Eddington limit if there are at least 16 gas cell neighbours within the central resolved region set by three times the gravitational softening length, these set-ups are labelled as \textit{SupplyLim}). 

We performed a large suite of zoom-in simulations varying seeding times, seeding masses and the length of the AGN duty cycle (see Table~\ref{tab:zoomruns}) down to $z=1$. We also continued a selected set of representative zoom-in simulations to $z=0$. Furthermore, we repeated three of these representative set-ups at even higher resolution ($\bar{m}_\mathrm{gas} = 35.9 \ \Msun$) to $z=1$ (labelled as \textit{Res}). Our main findings from this investigation are summarised below.

\begin{enumerate}
    \item Importantly, we find that there are sufficient amounts of gas to power brief, Eddington-limited accretion episodes in cosmological simulations of dwarf galaxies. However, with the standard Bondi model, AGN accretion in dwarfs is inefficient especially for lighter seeds ($M_\mathrm{seed} \lesssim 10^{3} \ \Msun$), indicating severe limitations of such an accretion model for low-mass galaxies.
    \item AGN feedback in dwarfs can lead to a variety of outcomes in terms of star formation regulation including no additional suppression, moderate suppression and catastrophic quenching.
    \item Powerful AGN-boosted outflows can deplete the gas reservoir of their hosts via ejective feedback and then maintain a quiescent state through heating the circumgalactic medium, thereby preventing cold gas accretion from the cosmic web. More moderate AGN-driven outflows (as in our very high resolution simulation) never completely deplete the gas reservoir, and stellar mass suppression is at a comparable level to the strong SN feedback set-up.
    \item For our simulated dwarfs and with the feedback models adopted here, we find that the AGN-driven outflows are accelerating whilst SN-driven outflows are of a decelerating nature, these kinematic features should in principle be distinguishable by observations \citep[e.g.][]{steidel_structure_2010}.
    \item Comparing our simulations with observational constraints on the SMHM relation, galaxy colours and metallicities does not disfavour AGN feedback in dwarf galaxies. Dwarfs with efficient AGN feedback tend to have lower stellar masses as well as be redder and more metal-poor. We note that the quenched galaxies produced by the most extreme AGN cases are likely to be rare and represent an upper limit for the impact of AGN on dwarfs.
    \item Efficient AGN are overmassive compared to extrapolations of observed $M_\mathrm{BH}$~--~$M_\mathrm{stellar}$ scaling relations (but are in good agreement with extrapolations of observed $M_\mathrm{BH}$~--~$\sigma_\mathrm{stellar}$ scaling relations), with future {\it direct} observational constraints in this low-mass regime being crucial. 
    \item All of our simulations are consistent with cuspy DM profiles, even for maximally efficient AGN. However, there is a small yet systematic trend towards lower central DM densities with AGN feedback, which becomes stronger with decreasing $M_\mathrm{stellar}/M_\mathrm{vir}$ ratio (opposite trends from stellar feedback induced DM depletion).
    \item The HI gas masses of our dwarfs are either completely suppressed or within the scatter of the observed $M_\mathrm{HI}$~--~$M_\mathrm{stellar}$ relations for field dwarfs.
    \item The X-ray activity of most efficient AGN set-ups is mainly restricted to high redshifts ($z>2$) with only a few occasional bursts at low redshifts, rendering these dwarfs difficult to detect with current facilities, but with exciting prospect for next generation X-ray missions such as \textit{Athena}, \textit{AXIS} and \textit{Lynx}.  
    \item Predicted X-ray luminosities of AGN at low redshifts, which either have very little impact on star formation or which efficiently suppress star formation, are in good agreement with the dwarf X-ray AGN luminosities observed by \citet{birchall_x-ray_2020} in a similar stellar mass range. This implies that detections of these moderately bright local AGN do not necessarily allow for conclusions with regards to the impact on the host dwarf galaxy.
\end{enumerate}

To conclude, AGN can have a significant impact on dwarf galaxies provided the BH growth is efficient -- however, the jury is still out as to whether these comparatively high BH masses could be excluded by future dynamical measurements.  According to our simulations, if the AGN is to significantly affect the evolution of its dwarf host galaxy, it has to largely operate in the ejective regime, driving strong outflows at high luminosities. If we do not find these bright AGN at higher redshifts, then BH feedback likely plays a less relevant role in dwarfs. On-going and upcoming observations with e.g. MUSE, \textit{JWST}, Rubin, \textit{RST}, SKA, ngVLA, \textit{Athena}, \textit{Lynx} and LISA will be able to constrain these theoretical possibilities and help us unravel the mysterious role of AGN in the low-mass regime.

\section*{Acknowledgements}
The authors would like to thank the anonymous referee for a comprehensive and insightful review which improved this work. The authors are also grateful to Martin Haehnelt, Roberto Maiolino, Sandro Tacchella and Sylvain Veilleux for valuable comments, suggestions, and discussions during the development of this manuscript. Part of this work was completed during the 2022 Galaxy Formation Workshop at Ringberg and benefited from stimulating discussions with participants. SK is supported by a Junior Research Fellowship from St Catharine's College, Cambridge. SK and DS acknowledge support by the STFC and the ERC Starting Grant 638707 `Black holes and their host galaxies: co-evolution across cosmic time'. MCS is supported by the Deutsche Forschungsgemeinschaft (DFG, German Research Foundation) under Germany’s Excellence Strategy EXC 2181/1-390900948 (the Heidelberg STRUCTURES Excellence Cluster). This work was performed using the Cambridge Service for Data Driven Discovery (CSD3), part of which is operated by the University of Cambridge Research Computing on behalf of the STFC DiRAC HPC Facility (www.dirac.ac.uk). The DiRAC component of CSD3 was funded by BEIS capital funding via STFC capital grants ST/P002307/1 and ST/R002452/1 and STFC operations grant ST/R00689X/1. This work used the DiRAC@Durham facility managed by the Institute for Computational Cosmology on behalf of the STFC DiRAC HPC Facility (www.dirac.ac.uk). The equipment was funded by BEIS capital funding via STFC capital grants ST/P002293/1 and ST/R002371/1, Durham University and STFC operations grant ST/R000832/1. DiRAC is part of the National e-Infrastructure.

\section*{Data Availability}
The data underlying this article will be shared on request to the corresponding author.



\bibliographystyle{mnras}
\bibliography{sk_2022} 




\appendix

\section{Bondi accretion prescription} \label{appsec:BondiAccretionRates}

In this appendix, we present a detailed analysis of the accretion rates for the Bondi-based zoom-in simulations. With this accretion prescription, we tested two different seed masses, $M_\mathrm{seed}=10^{3} \ \Msun$ and $M_\mathrm{seed}=10^{4} \ \Msun$. Given the difficulty we found with growing the $M_\mathrm{seed}=10^{3} \ \Msun$ BHs, we did not perform any Bondi-based runs with the light BH seeds with $M_\mathrm{seed}=10^{2} \ \Msun$. For all of the Bondi-based zoom-in simulations, we set the AGN duty cycle to $25$~Myr, as in the original \fable \ implementation, which should enhance both BH growth and feedback as it allows for the gas in the central region to recover in between feedback episodes and reduces overcooling.

The Bondi prescription was used for the original \fable \ simulations as well as in several other prominent cosmological simulation projects, such as Horizon-AGN \citep{dubois_dancing_2014}, EAGLE \citep{schaye_eagle_2015}, Illustris \citep{sijacki_illustris_2015} or IllustrisTNG \citep{weinberger_simulating_2017}. However, these galaxy formation models were developed with mainly AGN feedback from supermassive BHs (SMBHs, $M_\mathrm{BH} \gtrsim 10^{6} \ \Msun$) in mind and the quadratic dependency on $M_\mathrm{BH}$ in the Bondi rate (see Equation~\ref{eq:ZoomsBondiRate}) makes it very difficult for IMBHs ($10^{2} \leq M_\mathrm{BH} \leq 10^{5} \ \Msun$) to accrete. In fact, with the Bondi rate scaling as $\dot{M}_\mathrm{Bondi} \propto M_\mathrm{BH}^{2}$ and the Eddington rate scaling as $\dot{M}_\mathrm{Edd} \propto M_\mathrm{BH}$ with BH mass, the Eddington fraction consequentially scales as $f_\mathrm{Edd} \propto M_\mathrm{BH}$, i.e. lower AGN activity levels for lower-mass BHs (for the same environmental conditions).

One way to circumvent the suppression of BH growth in the low-mass regime is then to set a higher boost factor $\alpha$ for the Bondi rate (see Equation~\ref{eq:ZoomsBondiRate}). In the \fable \ simulations (as in Illustris), the boost factor is set to $\alpha = 100$ and the BH seed mass is $M_\mathrm{seed}=10^{5} \ \Msun$. However, we found evidence that this parameter combination still does not lead to sufficient BH growth in dwarf galaxies, with the BHs in \fable's low-mass galaxies undermassive compared to most observed scaling relations, and a scarcity of high-luminosity AGN dwarfs in \fable \ compared to constraints from X-ray observations \citep{koudmani_little_2021}.

Here we therefore also test alternative values, increasing the $\alpha$ boost factor to $\alpha=10^{3}$ (labelled as `\textit{Boost}') for both seed masses as well as to $\alpha=10^{4}$ (labelled as `\textit{ExtraBoost}') for the $10^{3} \ \Msun$ seed. Furthermore, we test accretion-only set-ups to assess the impact of AGN feedback on BH growth.

\begin{figure}
    \centering
    \includegraphics[width=\columnwidth]{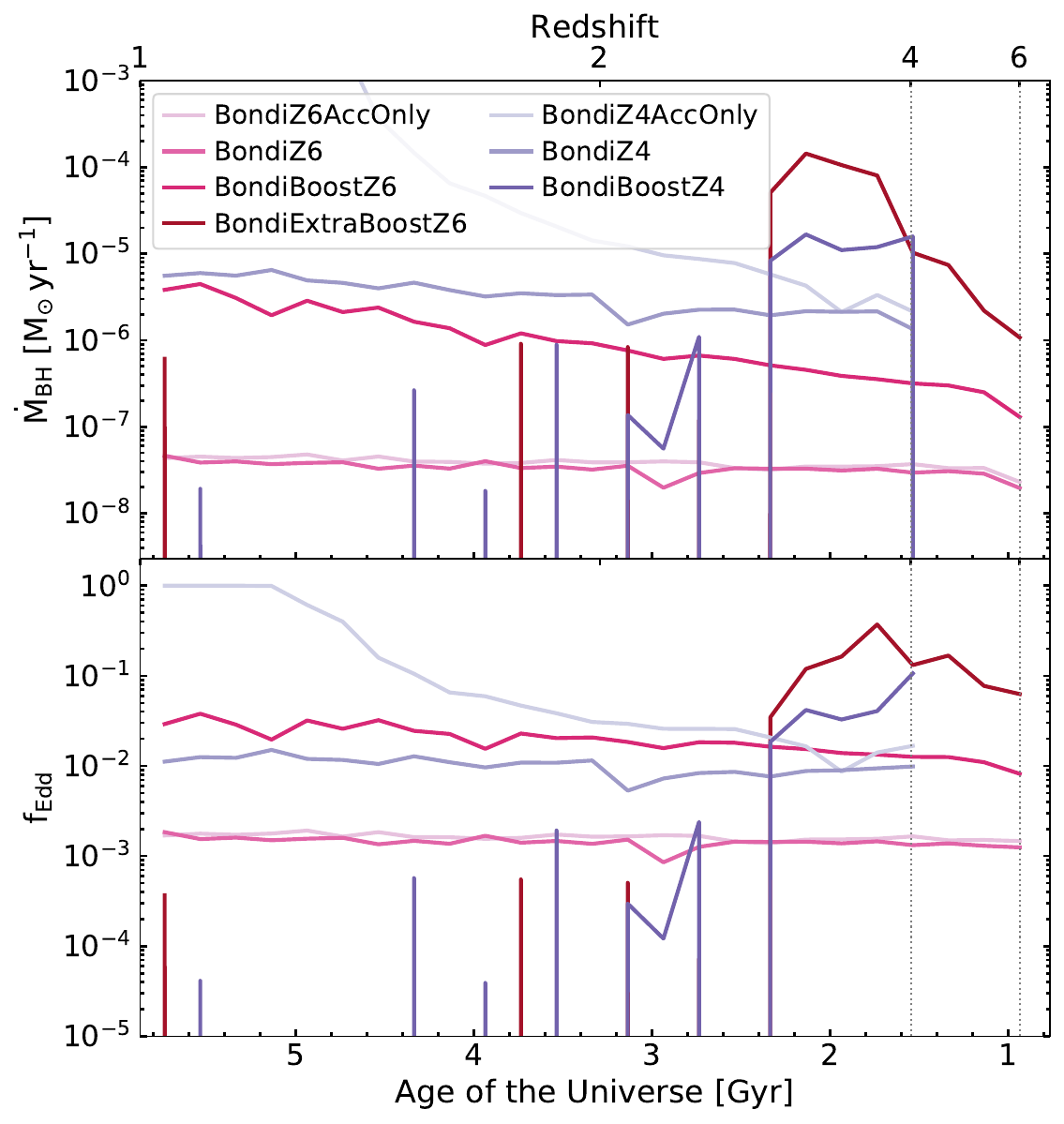}
    \caption{BH growth as a function of cosmic time for the Bondi-based dwarf zoom-in simulations. The upper panel shows the BH accretion rate and the lower panel shows the Eddington fraction. With the Bondi model, the accretion rates in the low-mass regime are strongly degenerate with the assumed BH seed mass and the boost factor.}
    \label{fig:BHAccPropertiesBondi}
\end{figure}

The BH accretion rates and Eddington ratios (binned over 200 Myr for clarity) of the Bondi-based zoom-in simulation runs are shown in Fig.~\ref{fig:BHAccPropertiesBondi}, also see Table~\ref{tab:zoomruns} for a list of all zoom-in simulations including details on the different set-ups. Note that apart from the main simulation runs \BondiFidDutyFromZFour \ and \BondiBoostFidDutyFromZFour \ (which were performed to $z=0$), these zoom-in simulations were generally only performed to $z=1$, so we focus our analysis on $z\geq 1$ here. 

Firstly, we inspect the simulations using the fiducial $\alpha=100$ boost factor, \BondiFidDutyFromZFourAccOnly \ and \BondiFidDutyFromZFour \ (BH seeding with $M_\mathrm{seed}=10^{4} \ \Msun$ at $z=4$) and \BondiFidDutyFromZSixAccOnly, \BondiFidDutyFromZSix \ (BH seeding with $M_\mathrm{seed}=10^{3} \ \Msun$ at $z=6$).

With the accretion-only set-up \BondiFidDutyFromZFourAccOnly, the BH growth is rapid, reaching the Eddington limit around $t=5.2$~Gyr. The equivalent set-up with an order of magnitude less massive BH seeded at $z=6$, \BondiFidDutyFromZSixAccOnly, only has negligible BH accretion rates due to the suppression of low-mass BH growth by the Bondi model. In fact, comparing the Eddington ratios at $z=4$ for these two accretion-only runs, they are offset almost exactly by an order of magnitude ($f_\mathrm{Edd}=10^{-2}$ versus $f_\mathrm{Edd}=10^{-3}$), which can be entirely attributed to the order of magnitude difference in BH masses at this point in time.

With AGN feedback added to this set-up, the growth of the more massive $10^{4} \ \Msun$ seed is markedly suppressed. The AGN feedback regulates the BH growth and the Eddington ratio stays around $10^{-2}$ until $z=1$ for the \BondiFidDutyFromZFour \ set-up, resulting in much lower growth rates than for the equivalent accretion-only set-up. As discussed in Section~\ref{ResultsSubSec:CosmicEvMainRuns}, we found that in the \BondiFidDutyFromZFour \ simulation the BH has no significant impact on star formation or gas content of its host galaxy until $z=1$. 

For the lighter seed with $M_\mathrm{seed}=10^{3} \ \Msun$, the BH feedback (which is directly coupled to the BH growth) is extremely weak so that the accretion-only run \BondiFidDutyFromZSixAccOnly \ and feedback run \BondiFidDutyFromZSix \ have virtually the same growth histories with only negligible feedback suppression.

To explore the efficient AGN regime, we therefore performed additional simulations where we increase the boost factor to $\alpha=1000$. For the $M_\mathrm{seed}=10^{4} \ \Msun$ set-up (see \BondiBoostFidDutyFromZFour) this results in high accretion rates ($f_\mathrm{Edd} \sim 0.1$) at early times followed by a rapid decline in BH accretion rates at $t = 2.3$~Gyr as the resultant high AGN feedback activity drives the gas out of the dwarf galaxy and halts BH growth. After this rapid shutdown, the BH only experiences a few short accretion episodes with $f_\mathrm{Edd} \lesssim 10^{-3}$.

On the other hand, for the lighter seed with $M_\mathrm{seed}=10^{3} \ \Msun$ at $z=6$, whilst there is a marked increase in the BH accretion rate from \BondiFidDutyFromZSix \ to \BondiBoostFidDutyFromZSix, these levels of BH activity are still not sufficient to have a significant impact on the host galaxy. Interestingly, the \BondiBoostFidDutyFromZSix \ set-up leads to very similar Eddington ratios as the \BondiFidDutyFromZFour \ run (albeit slightly higher), demonstrating the degeneracy between the $\alpha$ boost factor and the seed mass: with the Bondi prescription the Eddington ratio scales as $f_\mathrm{Edd} \propto \alpha M_\mathrm{BH}$ so that $M_\mathrm{seed}=10^{4} \Msun$ and $\alpha=100$ give very similar outcomes to  $M_\mathrm{seed}=10^{3} \Msun$ and $\alpha=1000$.

We investigate this further by performing an additional boosted run for the $M_\mathrm{seed}=10^{3} \ \Msun$ set-up with $\alpha=10^{4}$, so that \BondiBoostFidDutyFromZFour \ and \BondiExtraBoostFidDutyFromZSix \ have the same value for the $\alpha M_\mathrm{BH}$ product (though slight differences are to be expected due to the different seeding times). Indeed we again obtain a very similar evolution history for the Eddington ratios, with the early seeding allowing for an even higher accretion peak of approximately 40 per cent of the Eddington rate (the initial Eddington ratios are both very similar at around 10 per cent). Similarly to the \BondiBoostFidDutyFromZFour \ set-up, the efficient accretion phase is halted at 2.3~Gyr due to gas depletion, and then the BH only experiences small accretion bursts for the remainder of the simulation.

Our analysis of the Bondi-based dwarf zoom-in simulations demonstrates two important points. Firstly, with the Bondi model, BH accretion rates at the low-mass end are strongly degenerate with the assumed seed mass and boost factor. Secondly, the boost factors required to obtain efficient AGN accretion in dwarf galaxies lie between $10^{3}$ and $10^{4}$ (depending on the seed mass) which is difficult to motivate physically, in particular the boost factors required for the lighter seeds. It will be crucial to develop novel, more sophisticated BH accretion models that produce bright dwarf AGN naturally. Gas supply based accretion schemes or accretion disc based schemes could be a promising avenue -- provided the simulation resolution is high enough for these schemes to be implemented on a physical basis.

\section{Outflow properties in the inner halo regions} \label{appsec:Outflows}

\begin{figure*}
    \centering
    \includegraphics[width=0.985\textwidth]{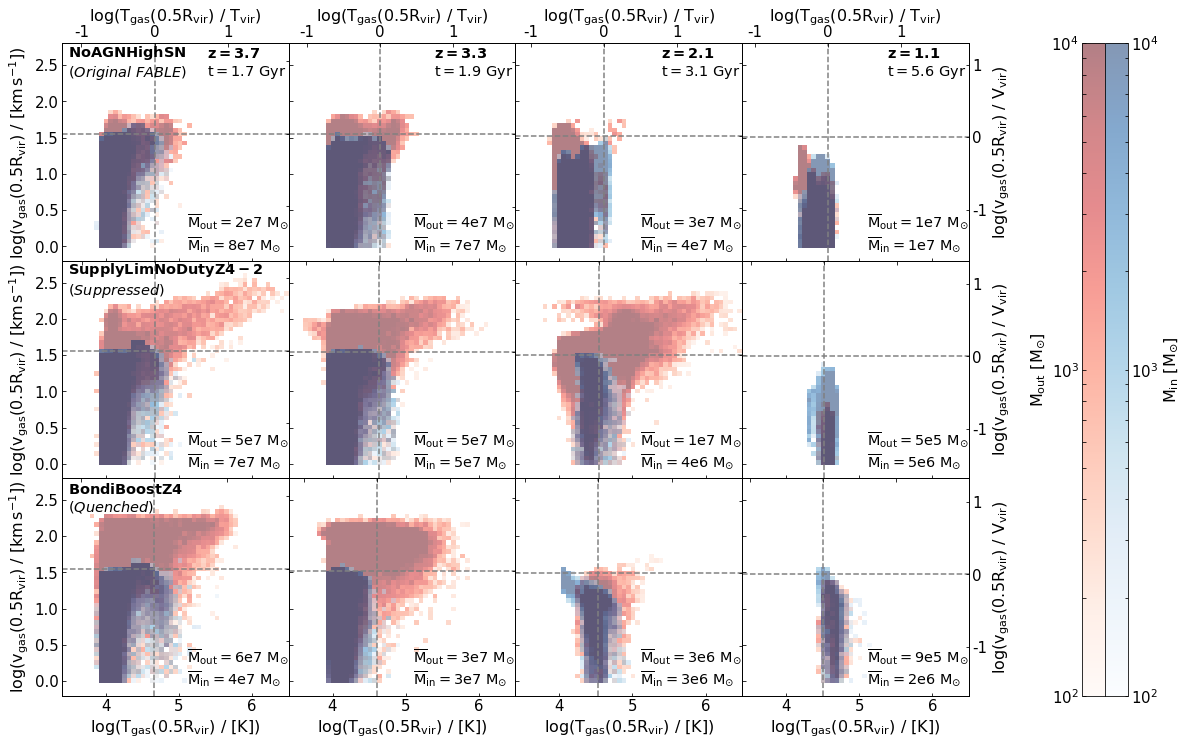}\\
    \includegraphics[width=0.985\textwidth]{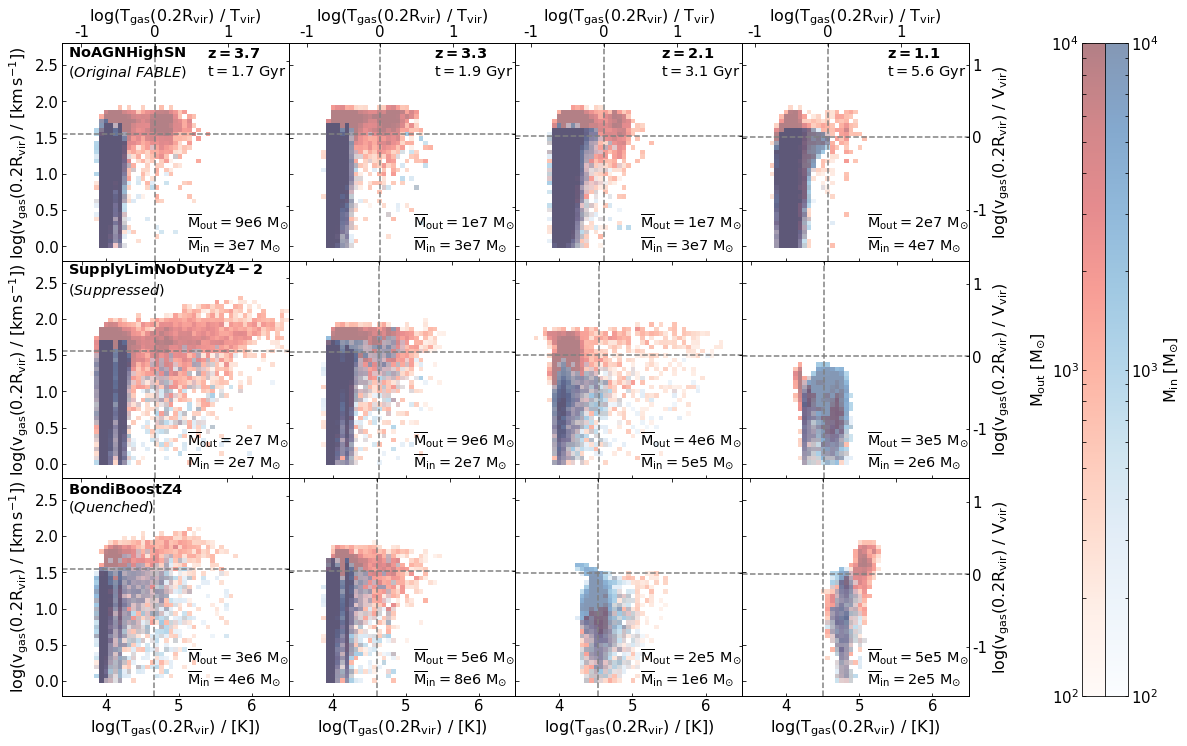}
    \caption{Outflow properties of representative dwarf zoom-in simulations. The blue-shaded and red-shaded histograms show the distribution of the inflowing and outflowing gas, respectively, in temperature -- velocity space. The gas properties are calculated within spherical slices placed at 50 per cent of the virial radius with slice width $2$~kpc (upper panel) and at 20 per cent of the virial radius with slice width $1$~kpc (lower panel). The total outflowing and inflowing mass within the slices is indicated in the lower right-hand corners.}
    \label{fig:OutflowProperties0_5-0_2Rvir}
\end{figure*}

Here we present the distributions of the outflowing and inflowing gas in temperature -- velocity space at $0.5 R_\mathrm{vir}$ (slice width $2$~kpc) and $0.2 R_\mathrm{vir}$ (slice width $1$~kpc) in Fig.~\ref{fig:OutflowProperties0_5-0_2Rvir}. These plots are analogous to Fig.~\ref{fig:OutflowProperties} in Section~\ref{subsec:outflow_props}, which shows the outflows and inflows at the virial radius. 

As in Section~\ref{subsec:outflow_props}, we focus on the \NoAGNHighSN \ simulation as examples of the fiducial FABLE set-up (which would not seed a BH into a halo of this mass range), the \EddNoDutyFromZFourToZTwo \ AGN set-up which evacuates gas from dwarf momentarily quenching star formation but is then replenished as AGN is switched off, and finally the \BondiBoostFidDutyFromZFour \ set-up where the AGN is active throughout the whole simulation, maintaining a quenched state.

The salient points regarding outflow properties both at small and large scales are discussed in Section~\ref{subsec:outflow_props} and here we just re-emphasise the most important conclusions and present the other two plots. The most notable difference at smaller radii pertains to the SN-only run \NoAGNHighSN \ which has a fast and hot component at $0.2 R_\mathrm{vir}$, which is completely absent at halo scales. At $0.5 R_\mathrm{vir}$, the fast-flowing hot gas is already markedly reduced, demonstrating the deceleration and cooling of the outflow towards larger scales, resulting in a galactic fountain.

In addition to investigating the small-scale outflows, the $0.2 R_\mathrm{vir}$ slices also provide us with useful insights into the general ISM gas conditions for the different runs. The most significant differences here can be seen at $z=2$ where the gas (both inflowing and outflowing) in the AGN runs is offset towards higher temperatures than the \NoAGNHighSN \ set-up. This offset is most significant for the \BondiBoostFidDutyFromZFour \ run demonstrating that the remaining galactic gas is sufficiently heated to maintain the quenched state. The smaller offset for the \EddNoDutyFromZFourToZTwo \ run, on the other hand, still allows for residual warm gas so that star formation is merely suppressed rather than completely shut down.

\section{High-resolution simulations: convergence of stellar feedback} \label{appsec:HiresStellar}

In Section~\ref{subsec:ResDependence}, we presented the outflow properties of the very-high-resolution ($\bar{m}_\mathrm{gas}=35.9 \ \Msun$) zoom-in simulation runs compared to the equivalent set-ups at the fiducial resolution ($\bar{m}_\mathrm{gas} = 287 \ \Msun$). Apart from the differing BH accretion and feedback histories (and therefore outflow rates), there are also clear differences between the SN-only runs at different resolutions.

\begin{figure}
    \centering
    \includegraphics[width=\columnwidth]{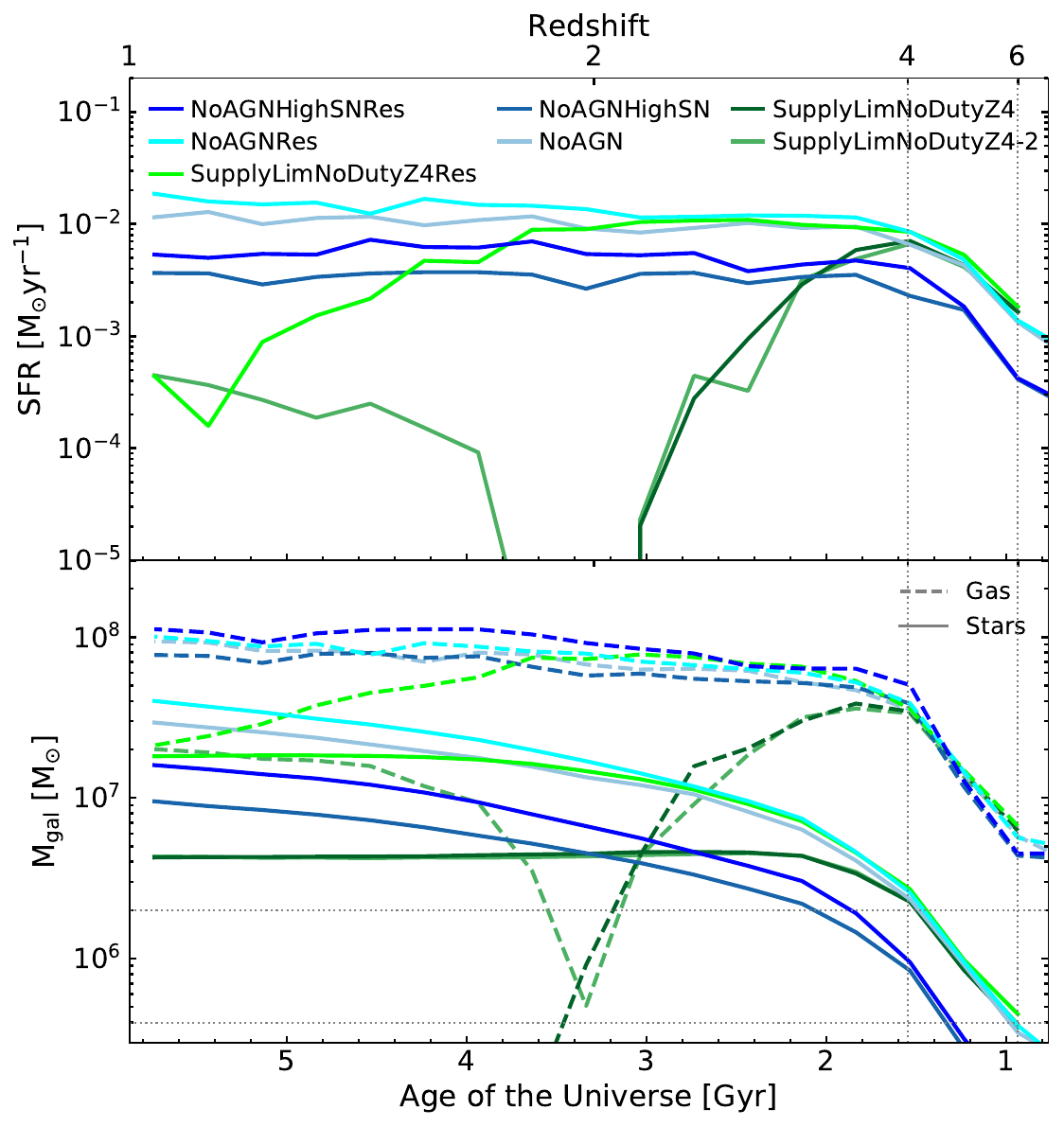}
    \caption{Star formation properties as a function of cosmic time. Here we contrast a selection of the main simulation runs at the fiducial resolution ($\bar{m}_\mathrm{gas} = 287 \ \Msun$) with equivalent zoom-in simulations with eight times higher mass resolution ($\bar{m}_\mathrm{gas} = 35.9 \ \Msun$). The upper panel shows the SFRs within twice the stellar half-mass radius and the lower panel shows stellar masses (solid lines) and gas masses (dashed lines) within twice the stellar half mass radius. The high-resolution simulations result in higher SFRs due to the increased sampling of high-density regions.}
    \label{fig:SFHiResRuns}
\end{figure}

These differences mainly stem from different levels of star formation activity as can be seen in Fig.~\ref{fig:SFHiResRuns}, which shows the SFRs (upper panel) as well as stellar and gas masses (lower panel) as a function of cosmic time. We show the \NoAGNHighSNRes, \NoAGNRes, and \EddNoDutyFromZFourRes \ simulations as well as their fiducial-resolution counterparts. For comparison we also show the results from the \EddNoDutyFromZFourToZTwo \ set-up at the fiducial resolution.

\NoAGNHighSNRes \ and \NoAGNRes \ both have slightly higher gas masses and SFRs than their fiducial-resolution counterparts. Note that we have checked that the smaller smoothing length is not responsible for this discrepancy by repeating the \NoAGNRes \ simulation with a resolution-adjusted neighbour number.

Indeed, higher galactic stellar masses are commonly found with increasing resolution, e.g. see discussion in \citet{pillepich_simulating_2018} of this effect in the IllustrisTNG model. The main reason for this lies in the increased sampling of high-density regions at higher resolution resulting in higher SFRs. Whilst this will at least in part be balanced by correspondingly stronger feedback, this balance can be disrupted by any resolution dependencies inherent to the feedback implementation itself, such as interactions with the circumgalactic medium.

\bsp	
\label{lastpage}
\end{document}